\documentclass[fleqn,usenatbib]{mnras}
\usepackage{newtxtext,newtxmath}
\usepackage[utf8]{inputenc}

\DeclareRobustCommand{\VAN}[3]{#2}
\let\VANthebibliography\thebibliography
\def\thebibliography{\DeclareRobustCommand{\VAN}[3]{##3}\VANthebibliography}

\usepackage{graphicx}	
\usepackage{amsmath}	
\usepackage{amssymb}	
\usepackage{multicol}
\usepackage{algorithm,algpseudocode}

\title[KLT performance for astronomical signals]{Performance analysis of the Karhunen-Lo\`{e}ve Transform for artificial and astrophysical transmissions: denoising and detection}

\author[M. Trudu et al.]{
Matteo Trudu,$^{1,2}$\thanks{E-mail: matteo.trudu@inaf.it}
Maura Pilia,$^{1}$
Gregory Hellbourg,$^{3}$
Pierpaolo Pari,$^{4}$
Nicol\`{o} Antonietti,$^{4,5}$
\newauthor 
Claudio Maccone,$^{1,5}$
Andrea Melis,$^{1}$
Delphine Perrodin$^{1}$
and Alessio Trois$^{1}$
\\
$^{1}$INAF-Osservatorio Astronomico di Cagliari, via della Scienza 5, I-09047, Selargius (CA), Italy\\
$^{2}$Universit\`{a} degli Studi di Cagliari, Dipartimento di Fisica, SP Monserrato-Sestu km 0.7, I-09042 Monserrato  (CA), Italy\\
$^{3}$California Institute of Technology, Pasadena, California 91125, USA\\
$^{4}$INAF-Istitituto di Radio Astronomia, via Gobetti 101, 40129 Bologna, Italy\\
$^{5}$IAA SETI Permanent Committee\\
}

\date{Accepted XXX. Received YYY; in original form ZZZ}

\pubyear{2015}

\begin{document}
\label{firstpage}
\pagerange{\pageref{firstpage}--\pageref{lastpage}}
\maketitle

\begin{abstract}
In this work, we propose a new method of computing the Karhunen-Lo\`{e}ve Transform (KLT) applied to complex voltage data for the detection and noise level reduction in astronomical signals. We compared this method with the standard KLT techniques based on the Toeplitz correlation matrix and we conducted a performance analysis for the detection and extraction of astrophysical and artificial signals via Monte Carlo simulations. We applied our novel method to a real data study-case: the Voyager 1 telemetry signal.
We evaluated the KLT performance in an astrophysical context:  our technique provides a remarkable improvement in computation time and Monte-Carlo simulations show significant reconstruction results for signal-to-noise ratio (SNR) down to -10 dB and comparable results with standard signal detection techniques. The application to artificial signals, such as the Voyager 1 data, shows a notable gain in SNR after the KLT.
\end{abstract}


\section{Introduction}
\label{sec:introduction}
The possibility of using the  Karhunen-Lo\`{e}ve Transform \citep{karhunen1947ueber,loeve1978probability} (KLT)   in order to recover a Signal Of Interest (SOI) buried in noise was proposed during the 1980s by Biraud \citep{BIRAUD1983759}, and was further explored by Maccone \citep{d2012noise} and Dixon \citep{1993ASPC...47..129D}  in the context of the Search for Extra-Terrestrial Intelligence (SETI). More recently, the KLT has triggered the renovated interest in the astronomical community, since it has proved to be particularly effective in areas such as the Cosmic Microwave Background power spectrum estimation \citep{Gjerlow:2015mqa}, the filtering out of 21 cm fluctuations \citep{Shaw:2014khi}, astronomical imaging \citep{lauer2002, Shaw:2013wza}, cosmological parameter extraction \citep{dba88a22d0d94e5e9a764017ad5bb1db}, as well as
spectra classification \citep{1999AJ....117.2052C}.
The KLT technique involves the decomposition of a stochastic process in a Hilbert space using orthonormal functions, which can in principle have any shape, unlike Fourier or Wavelet bases. It is also known as the Principal Component Analysis (PCA) in the finite dimensional case. The KLT statistically adapts to the data in order to extract an embedded pattern, by maximising the data covariance. For this reason the KLT is, at least in principle, an ideal operator for performing blind adaptive filtering, and offers a  better separation between the  deterministic components within  the  received signals  and  the  stochastic  ones. 

The aim of this work is to study the applicability and detection and extraction performance of the KLT in the interstellar telecommunication and astronomical context, and introduce a new method which permits a fast implementation of the KLT based on a  variant of the autocovariance matrix. 

The  paper is organised as follows. In section \ref{sec:sig_types} we introduce the artificial and astrophysical SOIs used in this analysis and we discuss their standard detection techniques. In section \ref{sec:klt} we discuss the main mathematical equations for the KLT techniques used in this paper. In section \ref{sec:reconstruction} we show the reconstruction results of the KLTs for artificial signals. In section \ref{sec:montecarlo} we discuss the Monte Carlo simulations for both denoising and detection. In section \ref{sec:voyager} we present our results for Voyager 1 data. In section \ref{sec:conclusion} we summarise our main results and conclude. 

\section{Signals of Interest}
\label{sec:sig_types}

\subsection{Astrophysical Signals}

Most astrophysical emissions present sparsity either in the spectral, temporal, or spatial domains, or in a combination of those.
The present analysis focuses on single receiver radio astronomy instruments that deliver a single spatial sampling of the electromagnetic field in any given sky direction.
Therefore, only spectral and temporal energy sparsity are considered.
Hereafter we describe the sample of SOIs that we analysed and the standard techniques used for their detection.

\subsubsection{Spectral lines}

Spectral sparsity is a feature of spectral line emissions, generated within molecular clouds and gases across the universe, and originating from molecular recombinations or atomic radiative transfers \citep{ep51}.
The proposed model for such a signal consists in the convolution between a stationary white Gaussian noise and a narrow bandpass filter:
\begin{equation}
\label{eq:spec_line}
s_{\text{line}}(t) = h(t) \circledast x(t),
\end{equation}
where $x(t) \sim \mathcal{NC}\left(\mu=0,\sigma_x^2\right)$ is the realisation of a complex white Gaussian noise with mean $\mu = 0$ and variance $\sigma_x^2$ at time $t$, $h(t)$ is the (finite) impulse response of a narrow bandpass filter describing a Gaussian bell curve in the frequency domain, and $\circledast$ stands for the convolution operator.

\subsubsection{Astrophysical transients}

The term ``transients" in astrophysics refers to wide-band and temporally sparse bursts of energy. These pulses are either unique or repetitive, such as Fast Radio Bursts \citep{lorimer07,spitler14}, or even periodic like pulsars \citep{lk04}. Their emissions experience a hyperbolic dispersion in the time-frequency domain due to their propagation in the interstellar medium (ISM).

Single pulses are modelled as an amplitude-modulated, complex white Gaussian noise that is associated with a frequency-dependent time delay following an ISM dispersion measure (DM), as follows:
\begin{equation}
\label{eq:transient}
s_{\text{pulse}}(t) = a(t) \cdot x(t) \circledast d(t)
\end{equation}
where $a(t)$ is a temporal envelope describing a Gaussian Bell curve envelope in the time domain, and $x(t) \sim \mathcal{NC}\left(\mu=0,\sigma_x^2\right)$.
$d(t) = \mathcal{F}^{-1} \left\{ \text{exp}\big( -\frac{ 2 j \pi f^2 k_{\text{DM}} \text{DM}}{f_0^2(f+f_0)} \big) \right \}$ models the influence at frequency $f$ -- relatively to the central frequency $f_0$ of the observed data -- of the ISM on the transient emission, $\mathcal{F}^{-1} \left\{.\right\}$ is the inverse Fourier transform, $j = \sqrt{-1}$, $k_{\text{DM}}^{-1} = 2.41\times10^{-4}$s$\cdot$MHz$^2$ is a constant of proportionality, and $\text{DM}$ is the transient dispersion measure.

Additionally, the ISM affects the distribution of energy across frequencies of the received transient through scintillation. This effect is neglected in this work, since we are assuming a narrow-enough processed bandwidth.

\subsubsection{Narrowband frequency-drifting transmissions}

A common target signal in the Search for Extra-Terrestrial Intelligence (SETI) is an engineered pure sine wave transmitted as a signalling beacon from a potential technologically advanced, non-terrestrial civilisation. This signal type presents the advantage of maximising the detection potential in the Fourier domain and minimising the impact of the ISM in the transmission. Accounting for the Doppler drift due to the dynamical environment of the Earth (Earth rotation, orbit, solar orbit in the galaxy, etc...), such a signal is modelled as a linear chirp on the receiver side of the communication channel:
\begin{equation}
\label{eq:sine_chirp}
s_{\text{nfd}}(t) = A \cdot \text{exp} \left(j 2 \pi \left(f_0 + \frac{k}{2} t \right) t \right)
\end{equation}
where $A$ is an amplitude factor (assumed constant over short periods of time), $f_0$ is the intrinsic transmission signal frequency, and $k$ is the frequency drift rate embedding the Doppler effect of the transmission as perceived from Earth.

\subsubsection{Binary Phase-Shift Keying transmission}

An information-bearing transmission is often regarded as a possible SETI target. A commonly-used modulation scheme for terrestrial transmissions  is Binary Phased-Shift Keying (BPSK), expressed as:
\begin{equation}
s_{\text{bpsk}}(t) = A \cdot m(t) \cdot \text{exp}\big(i 2 \pi f_0 t \big)
\end{equation}
where $m(t) = \sum_{k=-\infty}^{+\infty}\epsilon[k]\delta[t - k \cdot T_B] \circledast h_{T_B}(t)$ is a message signal, $\epsilon[k] = \pm 1$, $T_B$ is a bit-period, $h_{T_B}(t)$ is a pulse shaping window of length $T_B$, $\delta[t]$ is the Dirac delta function, $A$ is a constant amplitude factor, and $f_0$ is the central frequency of the transmission.

\subsection{Signal Detection in Astronomy and SETI}

Modern observatories run specialised detection pipelines when looking for particular signal types. This section briefly describes the classic detection pipelines employed to detect the various signals presented in section \ref{sec:sig_types}.

\subsubsection{Spectral line detection}

Spectral lines are detected using standard spectroscopic methods that involve the production of power spectra with ideally matching frequency resolution to the spectral line width ($\approx$ 100s of kHz frequency resolution), and threshold excesses of energy at given frequency bins \citep[see][and references therein]{koribalski12}. Power spectral density (PSD) estimation requires time integration, usually of the order of seconds to minutes, to reach the appropriate sensitivity to detect faint spectral lines. The sensitivity of a radio telescope to spectral line detection follows the radiometer equation defined as:
\begin{equation}
\rho(\tau, \Delta f) = \frac{T_{\text{source}}}{T_{\text{system}}} \sqrt{\tau \cdot \Delta f}
\label{eq:radiometer}
\end{equation}
where $\rho(\tau, \Delta f)$ is the apparent signal-to-noise ratio (SNR) of a given source in the field-of-view of the telescope after a time integration $\tau$ (in s) and over a frequency bandwidth $\Delta f$ (in Hz), $T_{\text{source}}$ is the source brightness temperature (in K), and $T_{\text{system}}$ is the system temperature (in K). This equation assumes a unit gain in the direction of the astronomical source.

\subsubsection{Astrophysical transient detection}

The sparsity in time of astrophysical transients prevents the utilisation of spectroscopic methods and long time integrations for improving the SNR of the emission. The common approach for detecting such signals involves a match-filtering process known as de-dispersion, and integration over large frequency bandwidths. The de-dispersion process consists in cancelling the effect of ISM dispersion by aligning the transient emission in time. This procedure requires the knowledge of the transient's DM; blind transient searches usually involve multiple de-dedispersion trials over a given range of DMs. Two types of de-dispersion are employed:
\begin{itemize}
\item incoherent de-dispersion : the telescope data are channelised and spectra are produced at regular time periods (typically every few ms). Incoherent de-dispersion consists in aligning each frequency bin according to a given DM.
\item coherent de-dispersion : incoherent de-dispersion assumes that the SNR of the transient emission is sufficiently high in each frequency channel to enable a confident detection after averaging all channels together. Signal smearing is however experienced over the individual channels, and can be detrimental to the detection when either the individual channel bandwidths are large, the DM is large, or the observation is conducted at low frequencies. In that case, coherent de-dispersion consists in cancelling the ISM response on the voltage data by applying the appropriate phase inversion for a given DM.
\end{itemize}
For periodic transients (pulsars), unless they are particularly bright, this approach has to be complemented by ``folding'' the de-dispersed transient profile, i.e. averaging time windows of the de-dispersed data. This technique also requires the knowledge of the transient's period; otherwise a search has to be performed like in the case of DM.

\subsubsection{Narrowband extra-terrestrial transmission detection}

Similarly to spectral line detection, the search for narrowband emissions aims at detecting sparse excesses of energy in the frequency domain. The frequency resolution of typical narrowband SETI searches is much higher ($\approx$ 1 Hz resolution) than the one for astrophysical spectral lines to match the narrow frequency bandwidths of these transmissions. The classic Fourier transform acts as a matched filter for pure sine waves.

A frequency-drift search for such transmissions is usually employed to cancel the Doppler effect experienced by a potential transmission (mostly due to the Earth's rotation, typically up to a few Hz/s), and therefore improve the detection performance after time integration. The detection sensitivity also follows the radiometer equation \ref{eq:radiometer}.

\section{The Karhunen-Lo\`{e}ve Transform}
\label{sec:klt}
\subsection{Mathematical Formulation}
Considering a complex valued stochastic process $X(t)$ where $t \in [0,T]$, the KLT of $X(t)$ consists in the following series expansion \citep{maccone2012mathematical}:
\begin{equation}
    \label{2.1}
    X(t) =  \sum_{m = 0}^{+ \infty} \zeta_{m} \phi_{m}(t) +\mu(t) \ ,
\end{equation}
where $\mu(t) = \mathbb{E} \left[X(t) \right]$ and $\mathbb{E}[.]$ is the expectation value operator, $\zeta_m$ are statistically independent complex random variables and $\phi_m(t)$ are complex basis functions, the eigenfunctions of the operator $R(t,s)$, defined as the autocovariance operator of $X(t)$
\begin{equation}
    \label{2.2}
    R(t,s) = \mathbb{E} 
    \left[
    \left( X(t) - \mu(t) \right) 
    \left( X(s) - \mu(s) \right)^{*} 
    \right] \ ,
\end{equation}
where $(.)^{*}$ stands for the complex conjugate operation. 
The expansion coefficients $\zeta_{m}$ are obtained by projecting the process $X(t)-\mu(t)$ over the corresponding eigenfunction $\phi_{m}(t)$, as
\begin{equation}
    \label{2.4}
    \zeta_{m} = \int_{0}^{T} \left( X(t) -\mu(t) \right) \phi_{m}^{*}(t) dt .
\end{equation}
The eigenfunctions $\phi_{m}(t)$ obey the equation \citep{maccone2012mathematical}:
\begin{equation}
    \label{2.3}
    \int_{0}^{T} R(t,s) \phi_{m}(s) ds = \lambda_{(m)} \phi_{m}(t) \ ,
\end{equation}
where $\lambda_{(m)}$ are the eigenvalues of the operator $R(t,s)$. $R(t,s)$ acts as the kernel of the integral equation (\ref{2.3}). The eigenfunctions $\phi_{m}(t)$ will form a complete orthonormal set in the Hilbert space. 

From definition (\ref{2.2}), $R(t,s)$ is a Hermitian operator and therefore its eigenvalues $\lambda_{(m)}$ are always real. By combining equations (\ref{2.2},\ref{2.4},\ref{2.3}), we obtain  $ \mathbb{E} \left[ \zeta_m \zeta_n^{*} \right] = \lambda_{(m)} \delta_{m n} $ (where $\delta_{m n}$ is the Kronecker symbol) which ensures that the eigenvalues are also always positive. From Mercer's Theorem \citep{1909RSPTA.209..415M} the sum of all  eigenvalues converges to the total variance $\sigma^2$ of the stochastic process
\begin{equation}
    \label{totalvariance}
    \sigma^2 = \int_{0}^{T} \sigma^2(t) dt = \sum_{m=0}^{+ \infty} \lambda_{m} \ ,
\end{equation}
where $\sigma^2(t)$ is the variance of the process at the fixed time $t$ computed according to (\ref{2.2}) with $t=s$.
Since series (\ref{totalvariance}) is convergent, the eigenvalues can be arranged in decreasing order $\lambda_{(0)} \geq \lambda_{(1)} \geq ...  0  \ $  and, therefore, there is a finite number of linearly independent eigenfunctions for each eigenvalue.

The sequence of the eigenvalues sorted in decreasing order is commonly referred to as the Eigenspectrum and it plays a key role for the KLT when used as a noise filter \citep{2010AcAau..67.1427M}.

No generic analytical closed-form expression of (\ref{2.3}) exists \citep[though see][for possible results]{elkaroui2008,6262495}. However, when the process is discretised, as in the case of a digitised output signal of a radio receiver, equation (\ref{2.3}) will reduce to a linear system of equations.

\subsection{Autocovariance Operator}
Considering a fixed time $t_i \in [0,T]$, the variable $X(t_i) = x_i$ is a random variable that is characterised by a probability density function (PDF) $\rho(x_i,t_i)$. The expectation value $\mathbb{E} \left[X(t_i) \right]$ is
\begin{equation}
    \label{2.5}
    \mu(t_i) = \mu_i = \mathbb{E} \left[X(t_i) \right]  =
    \int\limits_{\Omega(x_i)}  x_i \rho \left (x_i,t_i \right) d x_i \ ,
\end{equation}
where $\Omega (x_i)$ denotes the probability space of the random variable $x_i$.  Equation (\ref{2.5}) defines a function of time that represents the mean value of the random variable $x_i$ at each time $t_i$. Similarly, from \citet{leon2008probability}, the autocovariance operator $R(t,s)$ for a complex stochastic process  is defined as
\begin{equation}
    \label{2.6}
    R(t,s) = \iint\limits_{ \Omega(x_t, x_s) }
    \left( x_t -\mu_t \right)
    \left( x_s -\mu_s \right)^{*} 
    \rho(x_t,x_s,t,s) d x_t d x_s \ ,
\end{equation}
where $\rho(x_t,x_s,t,s) $ is the joint PDF of the random variables $x_t$ and $x_s$ and  $\Omega(x_t, x_s)$ is the joint probability space of the two random variables.

When a stochastic process is Wide-Sense Stationary (WSS), that is when it has a constant average  $\mu(t) = m$ and its autocovariance is dependent only on the time difference $R(t,s) = R(t-s)$, expression (\ref{2.6}) can be computed in the time domain by only considering a single realisation. In this case, the expectation operator is computed in the following way:
\begin{equation}
    \label{timemean}
    m =
    \mathbb{E}_{T} \left[X(t) \right] =
    \langle X(t) \rangle_{T} =
    \frac{1}{T} \int_{0}^{T} X(t) dt \ .
\end{equation}
If we define $\tau = t-s $, expression (\ref{2.6}) assumes the following form for a WSS process:
\begin{equation}
    \label{acfunction}
    R(\tau) =
    \mathbb{E}_{T} 
    \left[ 
    \left(X(t)-m \right) \left(X(t+\tau)-m \right)^{*}
    \right] \\
\end{equation}
Equation (\ref{acfunction}) is commonly called the autocorrelation function. An important property of (\ref{acfunction}) is that, for a zero mean signal, its average total energy $\mathcal{E}$ corresponds to the autocorrelation function $R(\tau=0)$ at zero time-lag
\begin{equation}
    \mathcal{E} = \mathbb{E}_{T} \left[X(t)^{2} \right] = R\left( \tau = 0 \right) \ .
\end{equation}

\subsection{Estimators of the Autocovariance Operator}

In the discrete case, the stochastic process $X$ will in general be characterised by two indices $X = x_{\alpha i}$, where $\alpha$ labels the realisation of $X$ we are considering, while $i$ labels the specific time sample of the realisation we are considering. 

Consider a discrete complex stochastic process $X = x_{\alpha i}$ where $\alpha = 0,1,...,M-1$ and $i = 0,1,...,N-1$. The unbiased estimator for (\ref{2.6}) is the sample covariance matrix \citep{mardia1979multivariate,chatfield1981introduction} :
\begin{equation}
    \label{2.8}
    R_{ij} = \frac{1}{M-1} \sum_{\alpha=0}^{M-1} 
    \left(x_{\alpha i}-\mu_i \right)
    \left(x_{\alpha j}-\mu_j \right)^{*} \ ,
\end{equation}
where $\mu_i = \mathbb{E}[x_i] = \frac{1}{M} \sum_{\alpha = 0}^{M-1} x_{\alpha i} $ .

A major limitation of this estimator arises in the case $ M < N$, leading to the singularity and non-invertibility of the matrix \citep{doi:10.1198/jasa.2011.tm10155}.
The KLT based on the assumption that more realisations of the process are available and that the KLT kernel is computed according to (\ref{2.8}), will be referred to as multiple realisations KLT (MRKLT). 

The proposed MRKLT extends the classic KLT\citep{1993ASPC...47..129D, 2010AcAau..67.1427M} to the case where multiple independent observations of the same signal set is available. In section \ref{sec:reconstruction}, we will investigate the MRKLT under various signal scenarios. The single realisation (or observation) case will then be addressed under the framework of the MRKLT, as discussed at the end of this section.

When the stochastic process is WSS, we can define the N-dimensional vector $R_i$ ($i = 0,1,...,N-1$) as an estimator for (\ref{acfunction}):
\begin{equation}
    \label{2.9}
    R_i = \sum_{k=i+1}^{N} 
    \left(x_k -m \right) \left(x_{i-k} -m \right)^{*} \ , 
\end{equation}
where $m = \frac{1}{N} \sum_{k = 0}^{N-1} x_k $. The autocovariance matrix $T_{ij}$ $(i,j=1,0,...,N-1)$ for a WSS process depends only on the autocorrelation vector (\ref{2.9}), and it assumes the form of a Toeplitz matrix \citep{press1992numerical,1993ASPC...47..129D}:
\begin{equation}
    \label{2.10}
    T_{ij} =
    \left(
    \begin{matrix}
     1 & r_1 & r_2 & r_3 & \cdots & \cdots & r_{N-1} \\
     r_1 & 1 & r_1 & r_2 & \cdots & \cdots & r_{N-2} \\
     r_2 & r_1 & 1 & r_1 & \cdots & \cdots & r_{N-3} \\
     \vdots & \vdots & \vdots & \vdots & \vdots & \ddots & \vdots \\
     r_{N-1} & \cdots & \cdots & r_3 & r_2 & r_1 & 1 \\ 
    \end{matrix} 
    \right) \ ,
\end{equation}
where $r_i$ is the autocorrelation vector (\ref{2.9}) normalised with respect to the autocorrelation vector itself at time-lag zero
\begin{equation}
    \label{2.11}
    r_i = \frac{R_i}{R_0} \ .
\end{equation}
When no prior regarding the stationarity of the process is available, expression (\ref{2.6}) (more precisely its estimator (\ref{2.8})) has to be used. This fact constrains the applicability of the KLT, since not all processes come in multiple realisations.  In the case of raw voltage data, (\ref{2.8}) is not suitable since our input is an $ N$ dimensional complex vector.

In order to address the single realisation case (e.g. signals), we will follow two approaches. The first one is based on the assumption that the signal is WSS, therefore equation (\ref{2.10}) can be used for the KLT kernel. This variant of the KLT will be referred to as Toeplitz KLT (TKLT), and has been suggested to filter out classic SETI target signals like sinewaves or chirps in noisy measurements \citep{1993ASPC...47..129D, 2010AcAau..67.1427M}.

A novel approach based on MRKLT is proposed in this work, as discussed earlier, extending the WSS assumption to periodic signals.
Suppose we have an $N$ dimensional complex vector $x_i$ $i=0,1,...,N-1$, which we split into several sub-vectors of length $W$. We define $W$ as KLT Window. The total number $K$ of sub-vectors  contained in $x_i$ is
\begin{equation}
    \label{2.12}
    K = \text{Floor} \left(\frac{N}{W} \right) \ ,
\end{equation}
where $\text{Floor(.)}$ is the floor function. Our vector $x_{i}$ is now a $K \times W$ matrix 
\begin{equation}
    \label{vmatrix}
    v^{\beta}_{l} =
    \left(
    \begin{matrix}
     x_{0} & x_{1}   & \cdots & x_{W-1}  \\
     x_{W}& x_{W+1} & \cdots & x_{2W-1}  \\
     \vdots  & \vdots & \ddots & \vdots \\
     x_{KW-W} & \cdots & \cdots&  x_{KW-1}\\ 
    \end{matrix} 
    \right) \ ,
\end{equation}
where, for the sake of clarity, the upper index is the row index and the lower index is the column index. Similarly to (\ref{2.8}), we can build an autocovariance matrix $\Sigma_{lm}$ ($l,m=0,1,...,W-1$) according to the following expression:
\begin{equation}
    \label{2.13}
    \Sigma_{l m} = \frac{1}{K-1} \sum_{\beta=0}^{K-1}
     \left(v^{\beta}_{l} - \xi_{l}\right)
    \left(v^{\beta}_{m} - \xi_{m}\right)^{*} \ ,
\end{equation}
where we defined $\xi_{l} = \frac{1}{K} \sum_{\beta = 0}^{K-1} v^{\beta}_{l} $.  We will call the KLT based on this approach covariance KLT (CKLT). The idea behind CKLT is that when the SOI is periodic, each sub-vector acts as a \textquotedblleft realisation\textquotedblright for the matrix (\ref{2.13}). 

\begin{figure*}
    \centering
    \includegraphics[width = 0.8 \textwidth]{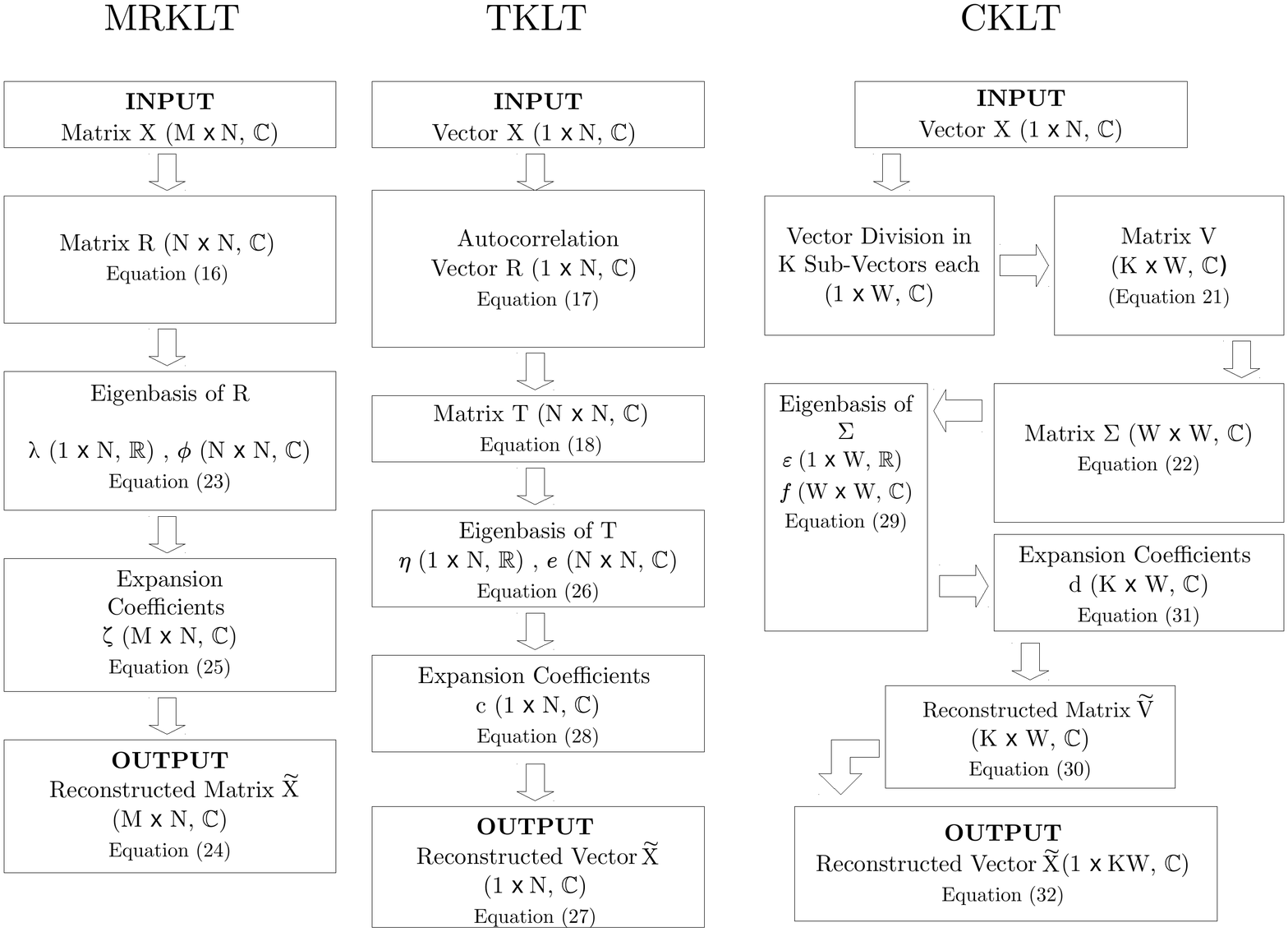}
        \caption{Block diagram for the three presented KLTs. For each of them, we indicate the type of data considered and we show all the mathematical operations (with reference equations) that are necessary in order to have the produced output.}
    \label{fig:klt_diagrams}
\end{figure*}

\subsection{KLT for Discrete Processes}

Once we have computed the autocovariance matrices, we can define the KLT expansions. Equation (\ref{2.3}) becomes a standard secular equation for all of these approaches, and it allows us to compute the eigenvalues and the eigenvectors of autocovariance matrices. For MRKLT equation (\ref{2.3}) becomes: 
\begin{equation}
    \label{2.15}
   \sum_{j = 0}^{N-1} R_{ij} \phi_{j m}  =\lambda_{(i)} \phi_{mi} \ .
\end{equation}
If we compute the eigenbasis of $R_{ij}$, the MRKLT for the process assumes the form:
\begin{equation}
    \label{2.16}
    \tilde{x}_{\alpha i} = \sum_{m = 0}^{N-1} \zeta_{\alpha m} \phi_{mi} + \mu_i \ , 
\end{equation}
where 
\begin{equation}
    \label{2.17}
    \zeta_{\alpha m} = \sum_{j = 0}^{N-1}  \left( x_{\alpha j} -\mu_j \right) \phi_{j m}^{*} \  .
\end{equation}
In the case of MRKLT, the M realisations of the stochastic process $X$ will share the same eigenbasis computed from matrix (\ref{2.8}). Each realisation will be reconstructed differently because they will have different expansion coefficients computed using (\ref{2.17}). 

In the case of TKLT, equation (\ref{2.3}) assumes the form:
\begin{equation}
    \label{2.18}
    \sum_{j=0}^{N-1} T_{ij} e_{j m}  =\eta_{(i)} e_{mi} \ ,
\end{equation}
where $\eta_{(i)}, e_{mi}$ are respectively the eigenvalues and the eigenvectors of $T_{ij}$. The TKLT reconstructed vector is
\begin{equation}
    \label{2.19}
    \tilde{x}_i = \sum_{l=0}^{N-1} c_{l} e_{li} + m \ , 
\end{equation}
where
\begin{equation}
    c_{l} = \sum_{j = 0}^{N-1}  \left( x_{j} -m \right) e_{j l}^{*} \  .
\end{equation}
Lastly,  for CKLT reconstruction equation (\ref{2.3}) is
\begin{equation}
    \label{2.20}
    \sum_{j=0}^{W-1} \Sigma_{ij}  f_{j m}  =\varepsilon_{(i)} f_{mi} \ ,
\end{equation}
where $\varepsilon_{(i)}, f_{mi}$ are respectively the eigenvalues and the eigenvectors of $\Sigma_{ij}$. The CKLT reconstructed sub-vectors are
\begin{equation}
    \label{2.21}
    \tilde{v}^{\beta}_{ i} = \sum_{m = 0}^{W-1} d^{\beta}_{ m} f_{mi} \ ,
\end{equation}
where
\begin{equation}
    \label{2.22}
    d^{\beta}_{ m} = \sum_{j = 0}^{W-1}  \left( v^{\beta}_{ j} -\xi_j \right) f_{j m}^{*} \  .
\end{equation}
The reconstructed initial vector $ \tilde{x_i}$ is the rearrangement of the K reconstructed sub-vectors $\tilde{v}^{\beta}_{ i}$, that is:
\begin{equation}
    \label{2.23}
    \tilde{x_i} = 
    \left(
    \tilde{v}^{0}_{0},\cdots, \tilde{v}^{0}_{W-1},
    \tilde{v}^{1}_{0},\cdots, \tilde{v}^{1}_{W-1},
    \cdots,
    \tilde{v}^{K-1}_{0},\cdots,\tilde{v}^{K-1}_{W-1} 
    \right) \ .
\end{equation}
We point out that the reconstructed vector $\tilde{x}_i$ (\ref{2.23}), as opposed to the other KLTs, might have fewer samples than the original signal. This is because of the way we build our input matrix (\ref{vmatrix}) for the CKLT, that is by dividing the initial signal into K sub-vectors.
An optimal KLT Window W is chosen such that the number of rejected samples is not high, but also such that $ K \leq W$   to make sure that matrix (\ref{2.13}) is not singular, to prevent the loss of information due to samples exclusion. Therefore if the initial signal has length N, the KLT Window W should not be greater than $\sqrt{N}$.

In figure \ref{fig:klt_diagrams} we show the block diagrams for the three types of KLT discussed. For each of them we describe the type of input data processed by the KLTs and what steps are necessary to produce the output.

\subsection{KLT for Signal Detection Theory}
We propose to use the KLT as a signal detector. The classical signal detection problem can be formulated as a binary hypothesis testing problem. Supposing we have a $(1 \times N)$ dimensional complex vector $x_i$, the two possible hypotheses $\mathbb{H}_0$ and $\mathbb{H}_1$ are
\begin{equation}
    \label{3.3.1}
    x_i = 
    \begin{cases} 
    n_i : \mathbb{H}_0 \\ 
    s_i + n_i : \mathbb{H}_1 
    \end{cases} \ ,
\end{equation}
where in the first hypothesis, only the noise is present, while in the second one, both SOI and noise are present. After computing the KLT kernel according to (\ref{2.10}) or (\ref{2.13}) and its eigenspectrum, we define the following quantity as a decision statistic parameter for the KLT:
\begin{equation}
    \label{3.3.2}
    \Lambda = \frac{\lambda_{(0)}}{\sum_{i} \lambda_{(i)}} \ .
\end{equation}
According to the Random Matrix Theory \citep{tao2012topics,2017arXiv171207903L}, the decision parameter (\ref{3.3.2}) follows a Marchenko-Pastur distribution \citep{1967SbMat...1..457M,pastur2011eigenvalue}; if we consider  $W/K \leq 1 $, this distribution for $\Lambda$ is well defined \citep{1967SbMat...1..457M}. 

In order to evaluate the KLT performance as a signal detector, we will compare it with standard detectors as Energy-based detectors, Fast Fourier Transform (FFT) detectors, and more recent Autocorrelation detectors \citep{5073595,8059467} . 
Energy detection is widely employed in signal processing for the detection of unknown signals. Considering the discrete signal $x_i$, the energy detector is expressed as the total sum of the modulus square of its samples
\begin{equation}
    \label{3.3.3}
    E = \sum_{i=0}^{N-1} \left|x_{i} \right|^2 .
\end{equation}
The FFT detector is the maximum value of the PSD of $x_i$
\begin{equation}
    \label{3.3.5}
    \Phi = \max \left( 
    \left|
   \mathcal{F}\left[ x_{i} \right]  
    \right|^2 \right) \ ,
\end{equation}
where $ \mathcal{F}\left[ x_{i} \right]$ is the FFT of $x_i$ 
\begin{equation}
    \label{3.3.6}
    \mathcal{F}\left[ x_{i} \right] 
    = \sum_{m=0}^{N-1}x_{m} \exp\left(-2 \pi j \frac{m i}{N} \right) \ .
\end{equation} 
Lastly, the autocorrelation detector, follows a similar definition proposed by \citet{5073595}
\begin{equation}
    \label{3.3.4}
    A = R_{0} + R_{1}  \ ,
\end{equation}
where $R_0,R_1$ are the autocorrelation vectors (\ref{2.9}) computed respectively at 0 and 1 sample time-lag. 

\subsection{KLT Implementation}
\begin{figure}
    \centering
    \includegraphics[width = 0.8 \columnwidth]{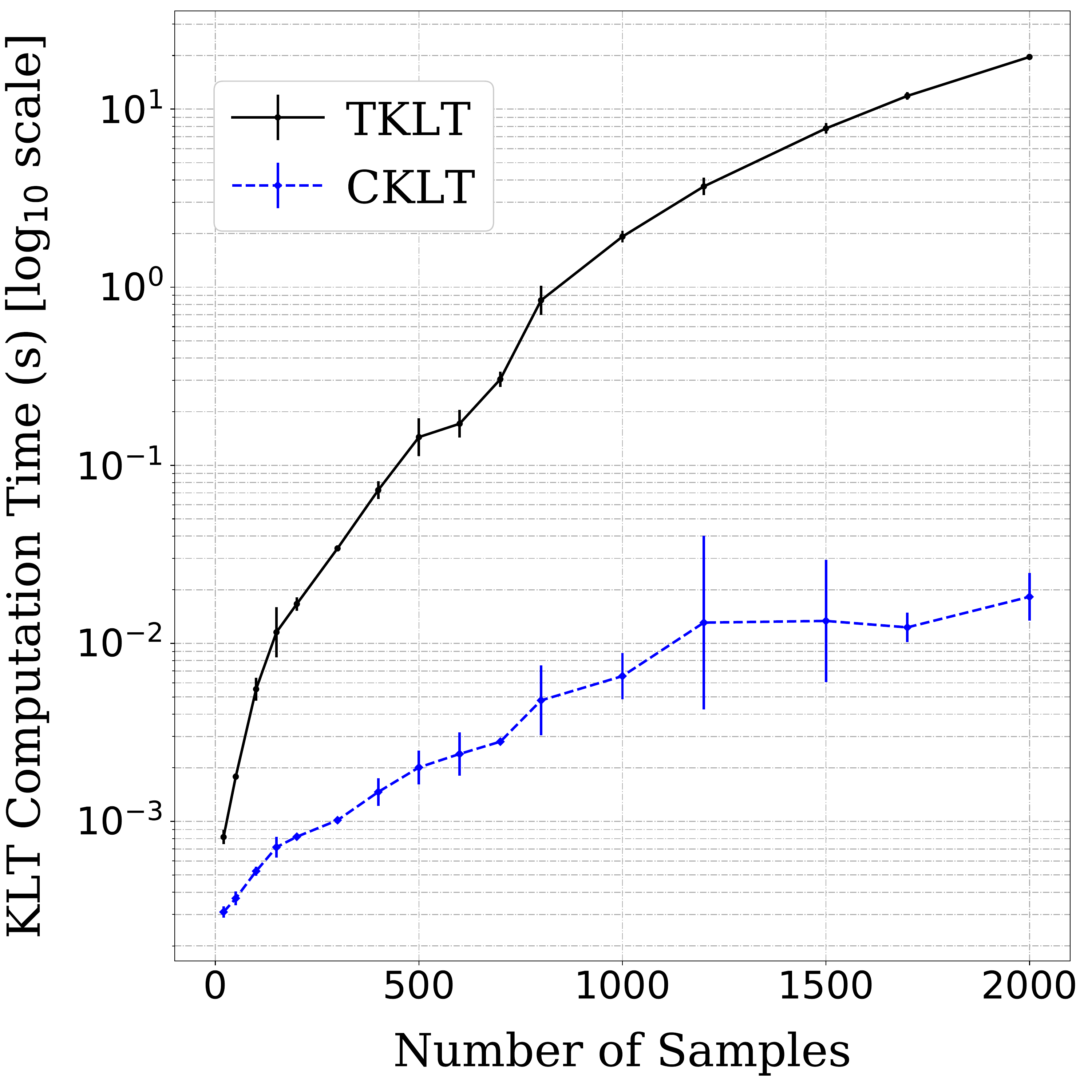}
    \caption{Computation time comparison between CKLT and TKLT. The KLT Window W used for each number of samples is 1/10 of the number of samples themselves.}
    \label{fig:time_analysis}
\end{figure}
The KLT has the main disadvantage to possess a high computational complexity, mainly do to the need to diagonalise the autocovariance matrix in order to compute the eigenbasis. More details concerning the KLT computational complexity are provided in \ref{app:matrix}. 
Figure \ref{fig:time_analysis} shows a comparison in computation time between CKLT and TKLT with their respective error bars. The y axis is in logarithmic scale. For both KLTs, we considered the same generated signal: a complex sinewave plus complex white noise. For different signal lengths, we performed 100 CKLTs and TKLTs. The KLT Window for CKLT was select at 1/10 of the number of samples considered. The CKLT remarkably outperforms TKLT in computation time. For the case of 2000 samples, we have a difference of 3 orders of magnitude between the two algorithms. This is because, as we already mentioned, the most computational heavy part of the algorithm consists in the computation of the eigenbasis: the autocovariance of the TKLT always needs the same length of the received signal, while the autocovariance of the CKLT has the same length as the selected KLT Window W. This is a huge advantage for CKLT, and can make it a more suitable instrument for processing real data, which tend to have a considerably high number of samples. This simulation and the following ones were done using a Linux platform Ubuntu 18.04 running on a Intel\textsuperscript{\textregistered}  Core\textsuperscript{TM} i7 i7-6700HQ, 4$\times$2.60GHz CPU, RAM memory 16GB DDR4.

\section{Interstellar Telecommunication Signal Reconstructions}
\label{sec:reconstruction}
\subsection{MRKLT Reconstruction}
\label{sec:mrklt}
In order to understand how the different KLTs recover a SOI, we consider interstellar telecommunication signals as a first test, since it is simpler understand their features. For the MRKLT we consider a sinewave with normalised  frequency $f_0 = 0.6f_s $, where $f_s$  is the sampling frequency, and a linear chirp with normalised starting frequency $f_0 = 0.6 f_s$, and normalised drift rate $k=0.2f_s/N$, where N are the samples of the SOI.

The input for the MRKLT was a complex matrix $ x_{\alpha i}  = s_{\alpha i} + n_{\alpha i} $ with $ 10^4 \times 10^3 $ entries.
The matrix $s_{\alpha i}$ is the matrix the considered SOI, while $n_{\alpha i}$ is a matrix that contains the noise. In each realisation, the phase of both sinewave and linear chirp was randomly generated with a uniform distribution $\mathcal{U}(0,2 \pi)$. The noise matrix contains complex coloured noise generated using white noise convolved with a Hanning window. 

The SOIs are generated according to (\ref{eq:sine_chirp}), where the sinewave is the case with $k=0$. The SNR is defined as the ratio between the total energy of the SOI (which is 1 for the way we generate the SOI) and the total energy of the noise.
\begin{figure}
\begin{multicols}{2}

        \includegraphics[width= \columnwidth]{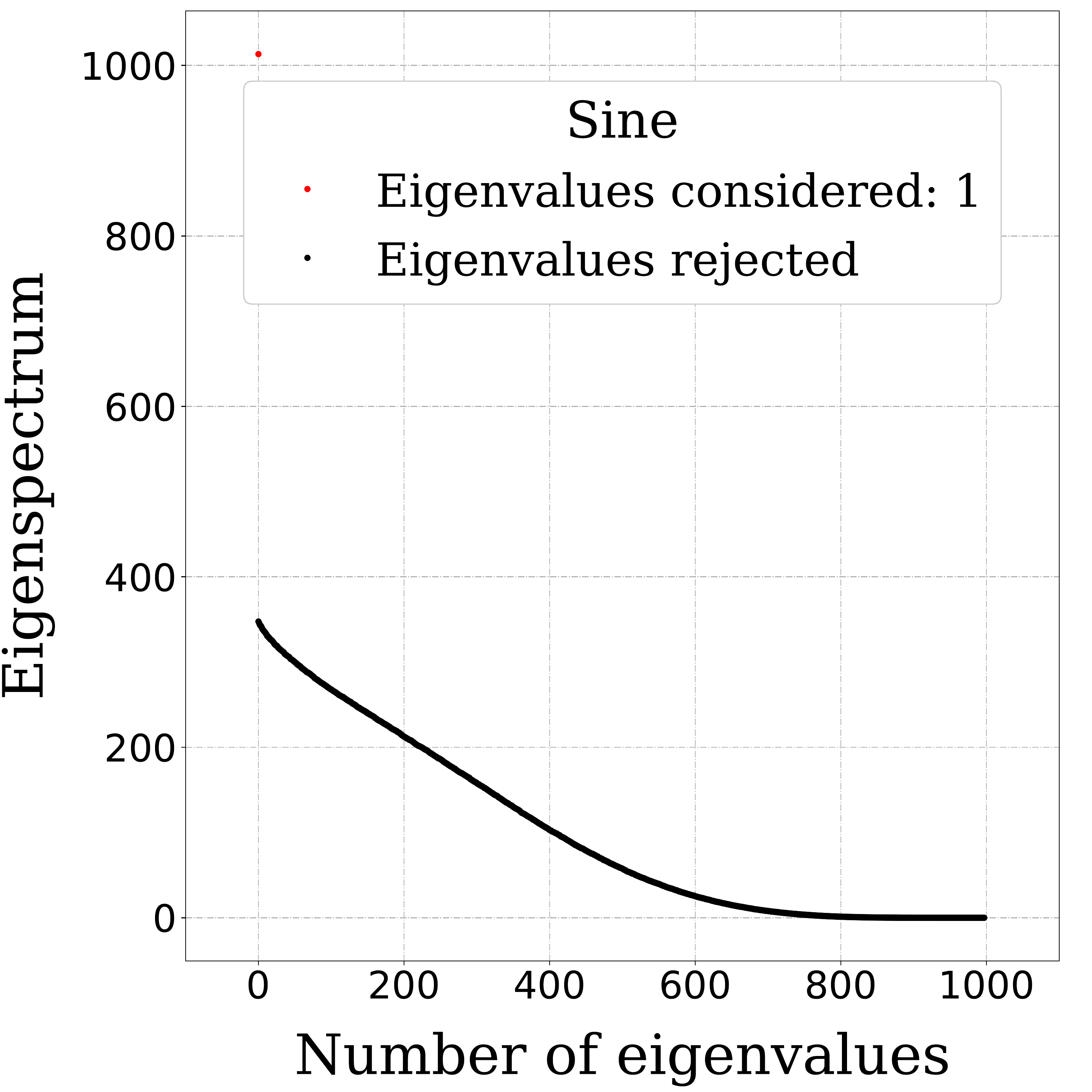}
        
        \includegraphics[width= \columnwidth]{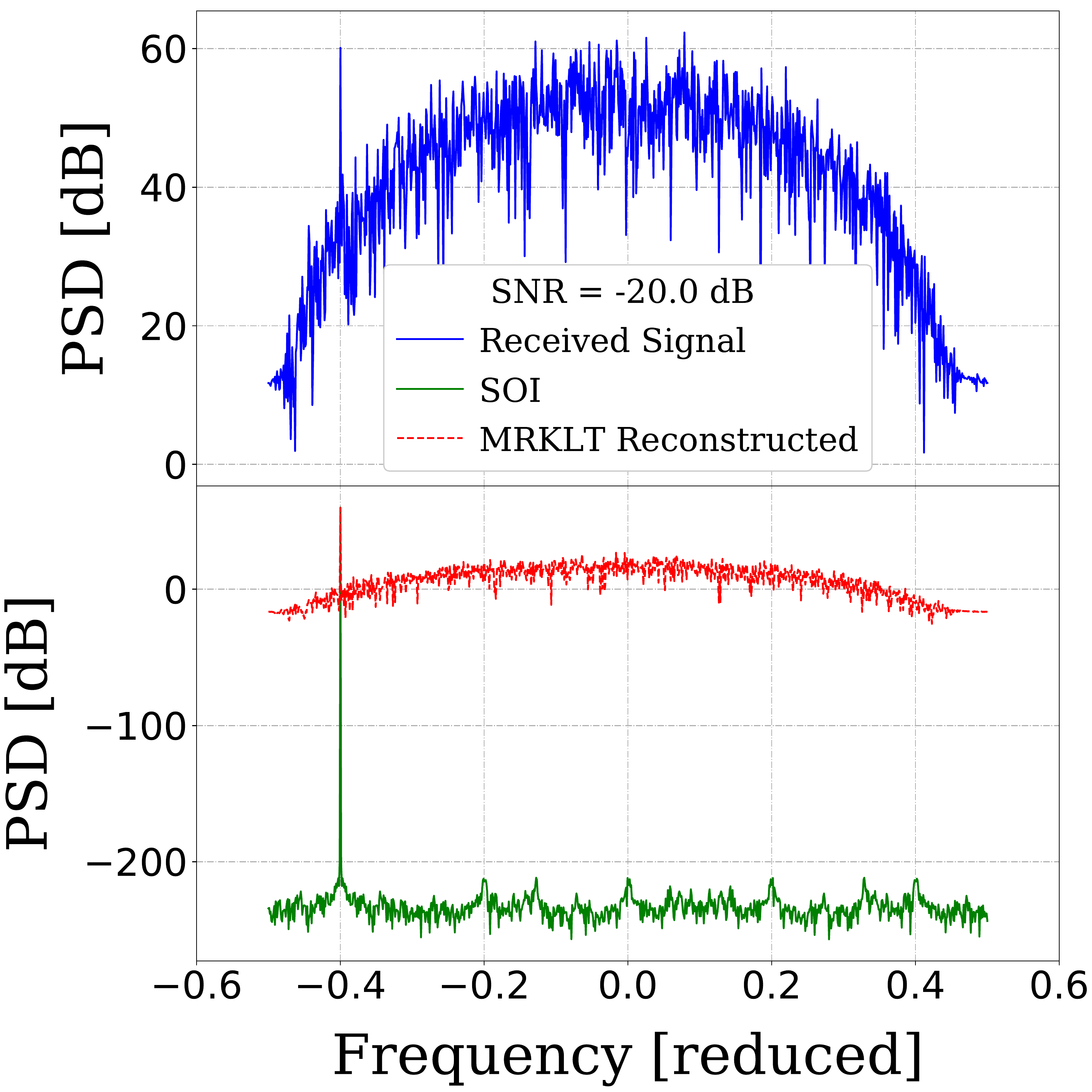}

\end{multicols}
\begin{multicols}{2}

        \includegraphics[width= \columnwidth]{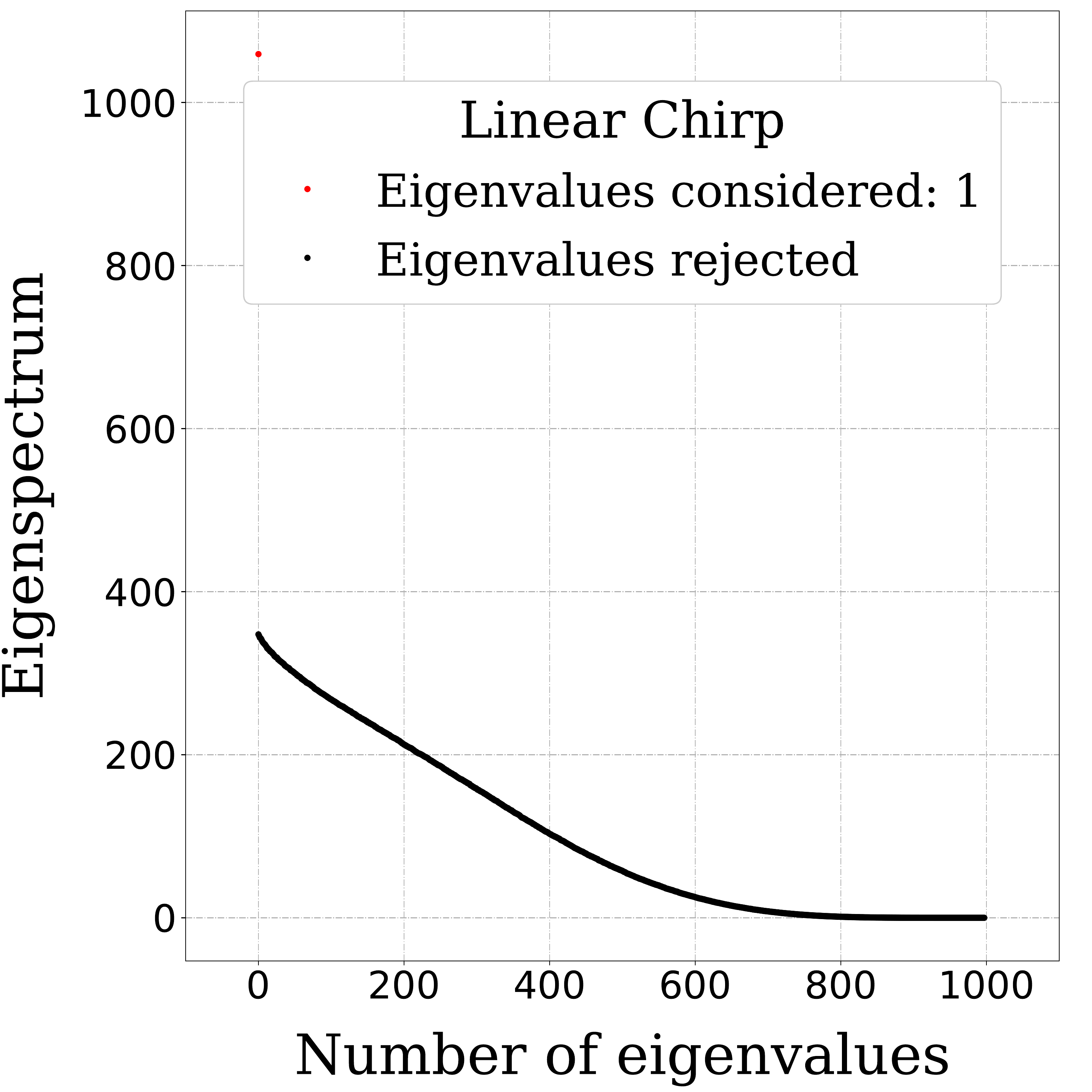}

        \includegraphics[width= \columnwidth]{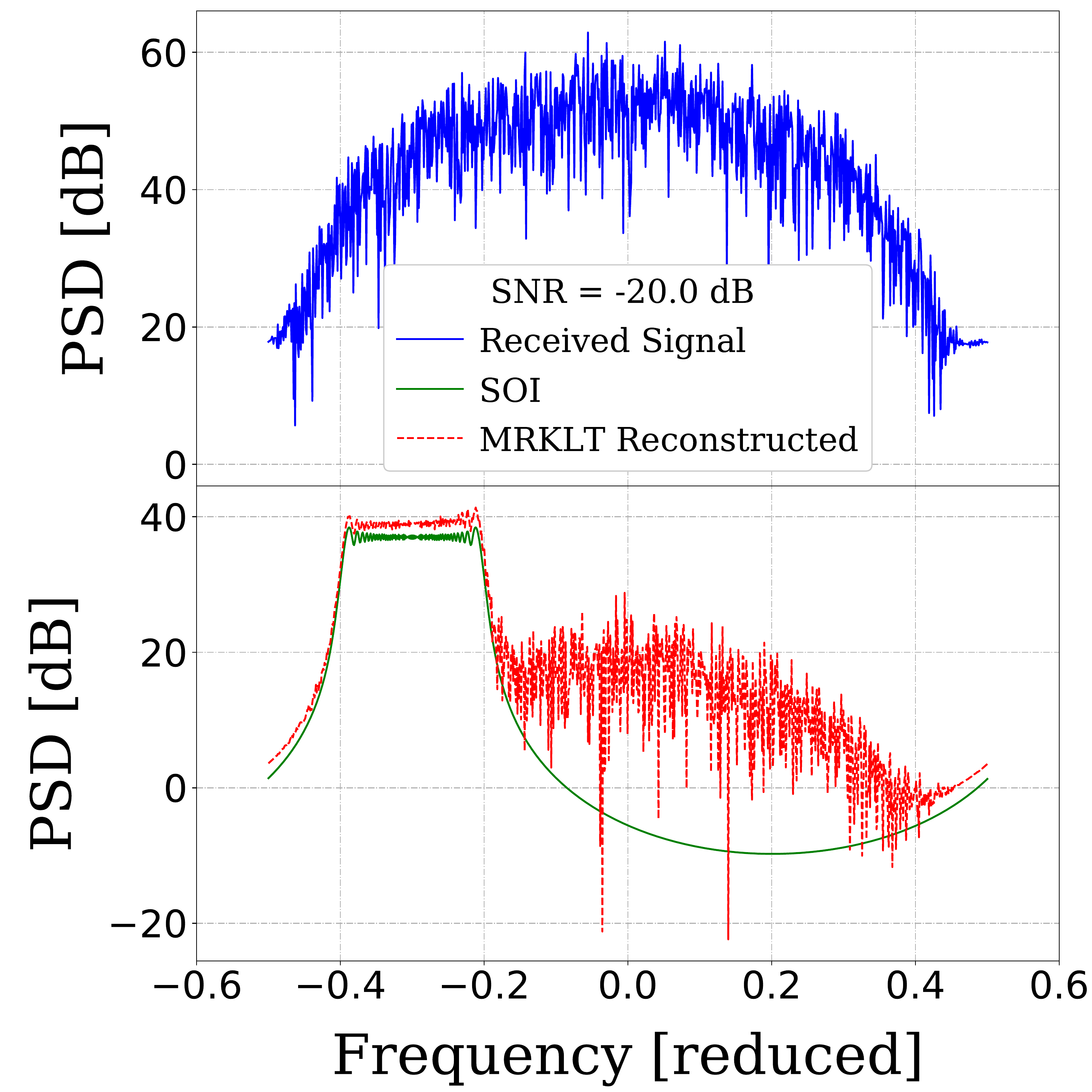}

\end{multicols}
    \caption{MRKLT reconstruction for a sinewave and a linear chirp. The total number of realisations is 10000. The figure shows the realisation number 3973 which was chosen randomly. SNR is -20 dB. Top-left panel: eigenspectrum for the sinewave obtained with MRKLT. Top-right panel: PSD of the received signal (blue);   the original SOI (green) and the MRKLT reconstructed (red). Bottom-left and bottom-right: the same as top-left and top-right respectively for the chirp.}
    \label{fig:mrklt_reconstruction}
\end{figure}

Figure \ref{fig:mrklt_reconstruction} shows the MRKLT results for the sinewave and the linear chirp. The plots on the left show the eigenspectrum of the autocovariance matrix (\ref{2.8}), while the plots on the right show the PSD of the SOI and MRKLT reconstructed. The plots show the results for  one (randomly chosen) of the 10000 realisations generated.

Both eigenspectra clearly show that there is only one dominant eigenvalue, which advocate for the choice of only one expansion term for the reconstruction. The PSDs of the MRKLT reconstructed signal are good representations of the PSDs of both SOIs despite the very low SNR level (-20 dB). There is a loss in reconstruction quality at high frequencies. This result is consistent with the injected coloured noise, which has a frequency dependence. 
MRKLT provides good reconstructions even for very low SNRs, its drawback being that real data do not always come in multiple realisations.

\subsection{TKLT and CKLT Reconstruction}
We next consider the case where only a single realisation is available. We show the results for the TKLT and CKLT by again considering a sinewave and a linear chirp as described in section \ref{sec:mrklt}. In addition, we consider a complex BPSK with normalised frequency $f_0 = 0.6f_s $.  In this case, our input is a complex vector $x_i = s_i + n_i  $ of N samples.
The number of samples $N$ is $10^3$ for the TKLT, while, for the CKLT, we consider $N=10^4$ and $W=10^2$ samples. The bit-period $T_B$ of the BPSK is 10 samples for TKLT and 100 samples for CKLT, in order to have the same relative length respect to the length of the input vector. coloured noise is added to the SOI, as in the previous case.

\begin{figure}
\begin{multicols}{2}
        \includegraphics[width= \columnwidth]{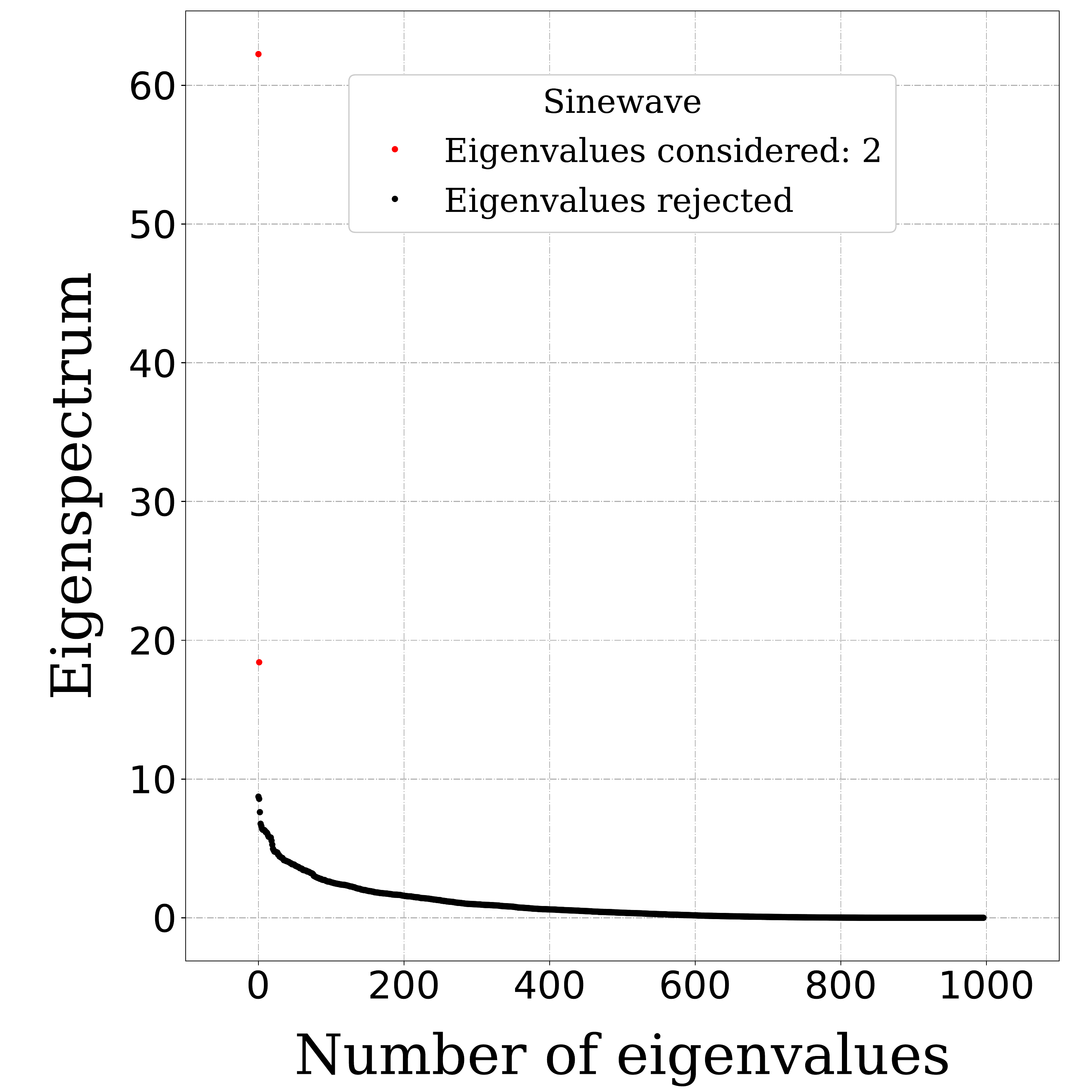}

        \includegraphics[width= \columnwidth]{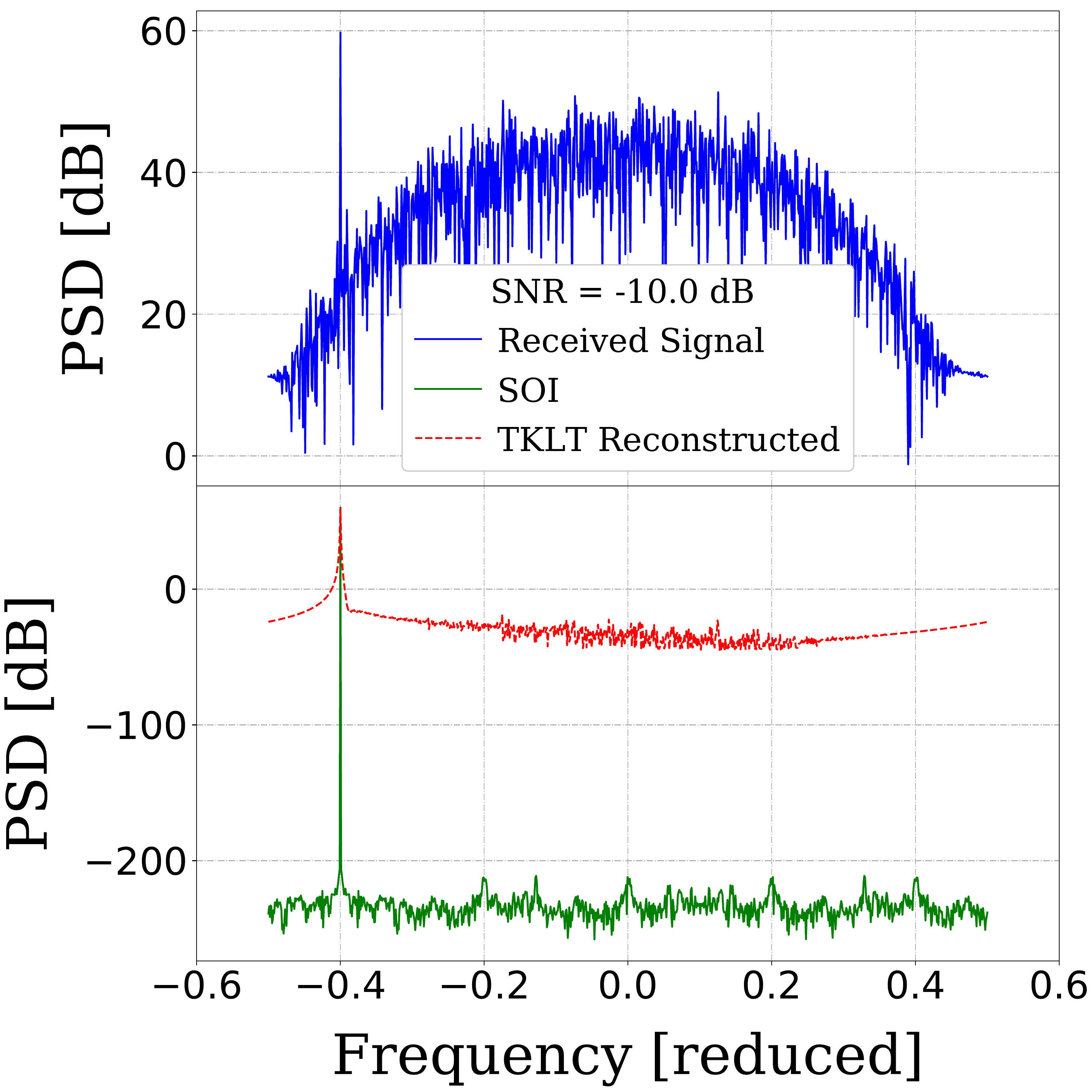}
        
\end{multicols}
\begin{multicols}{2} 
        \includegraphics[width= \columnwidth]{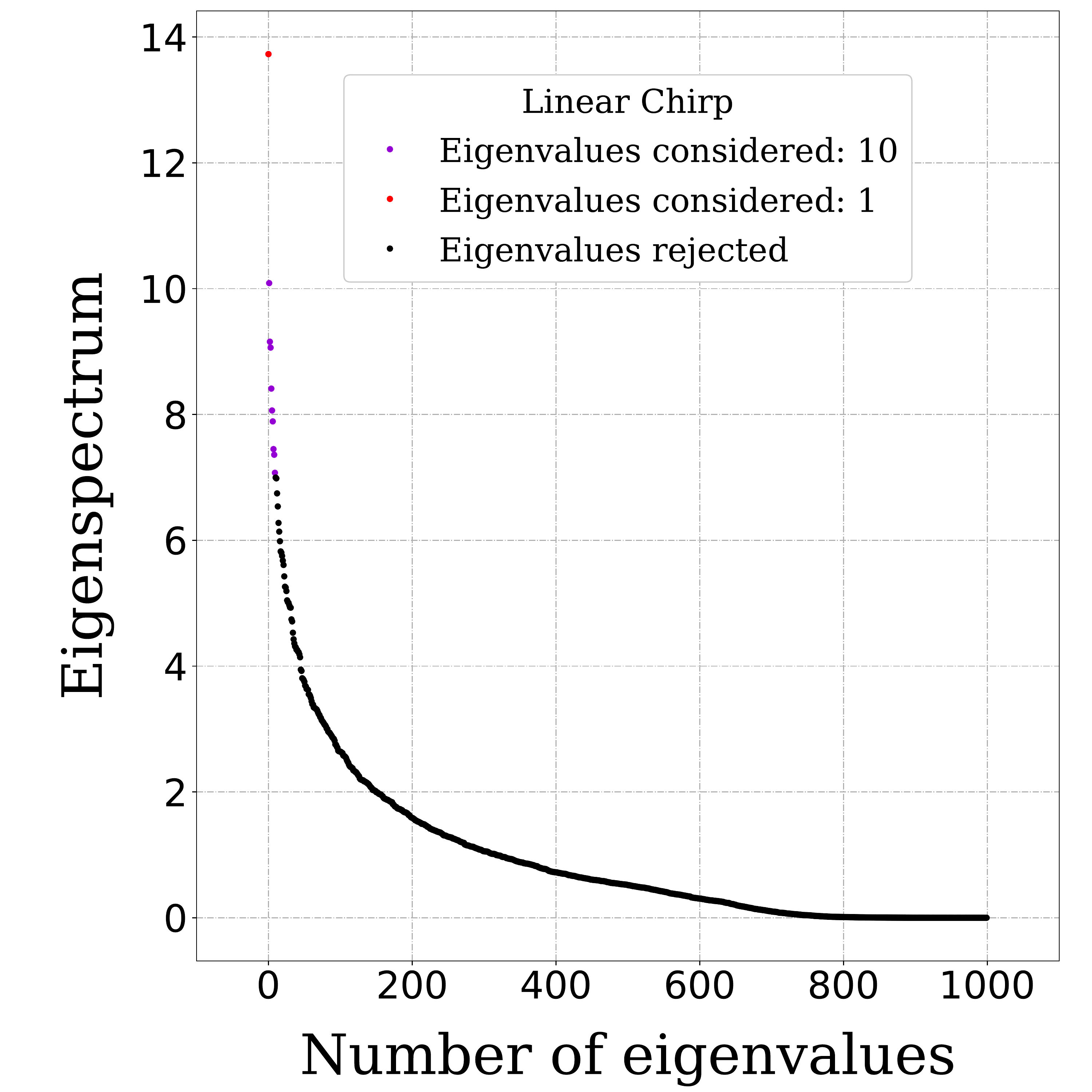}

        \includegraphics[width= \columnwidth]{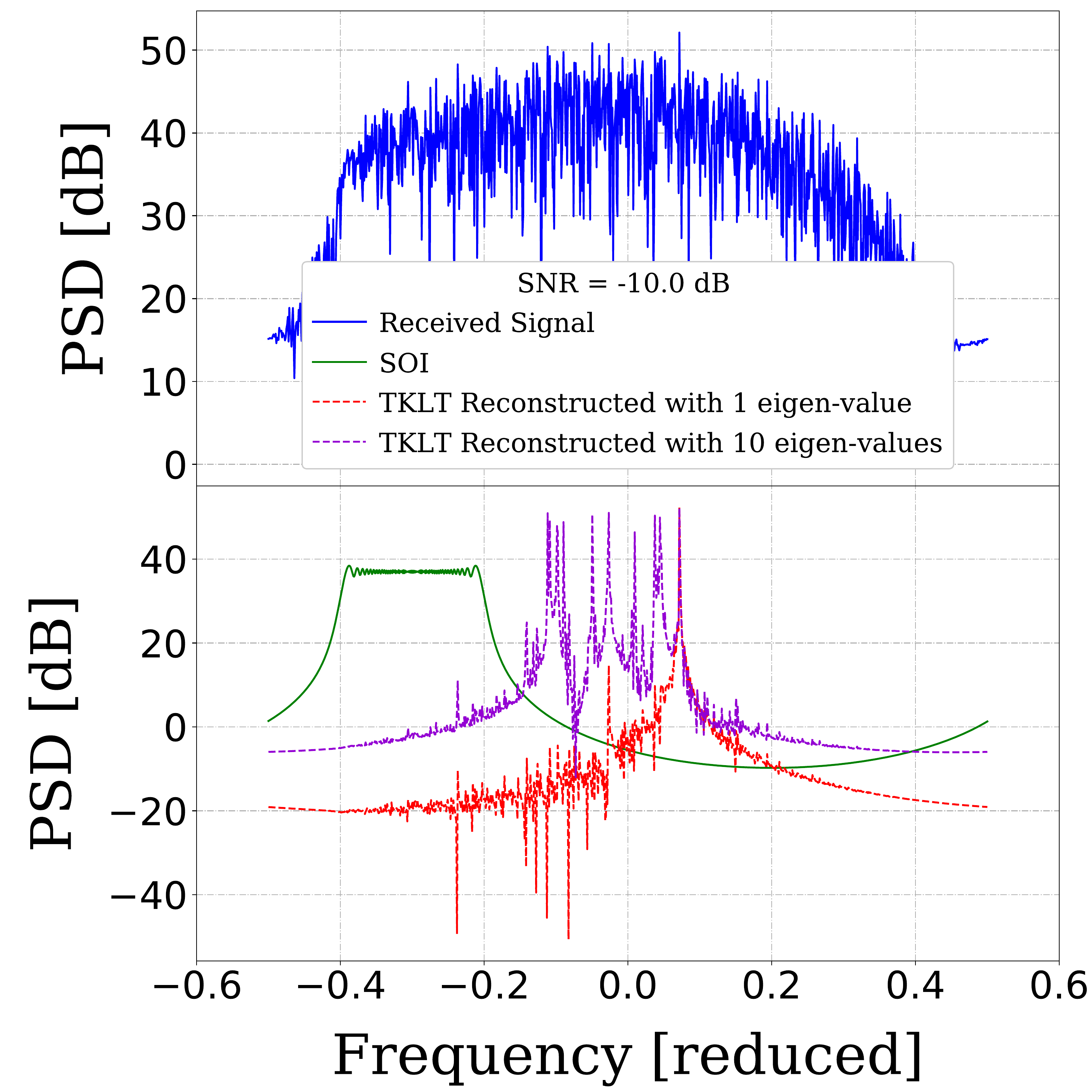}
        
\end{multicols}
\begin{multicols}{2}
        \includegraphics[width= \columnwidth]{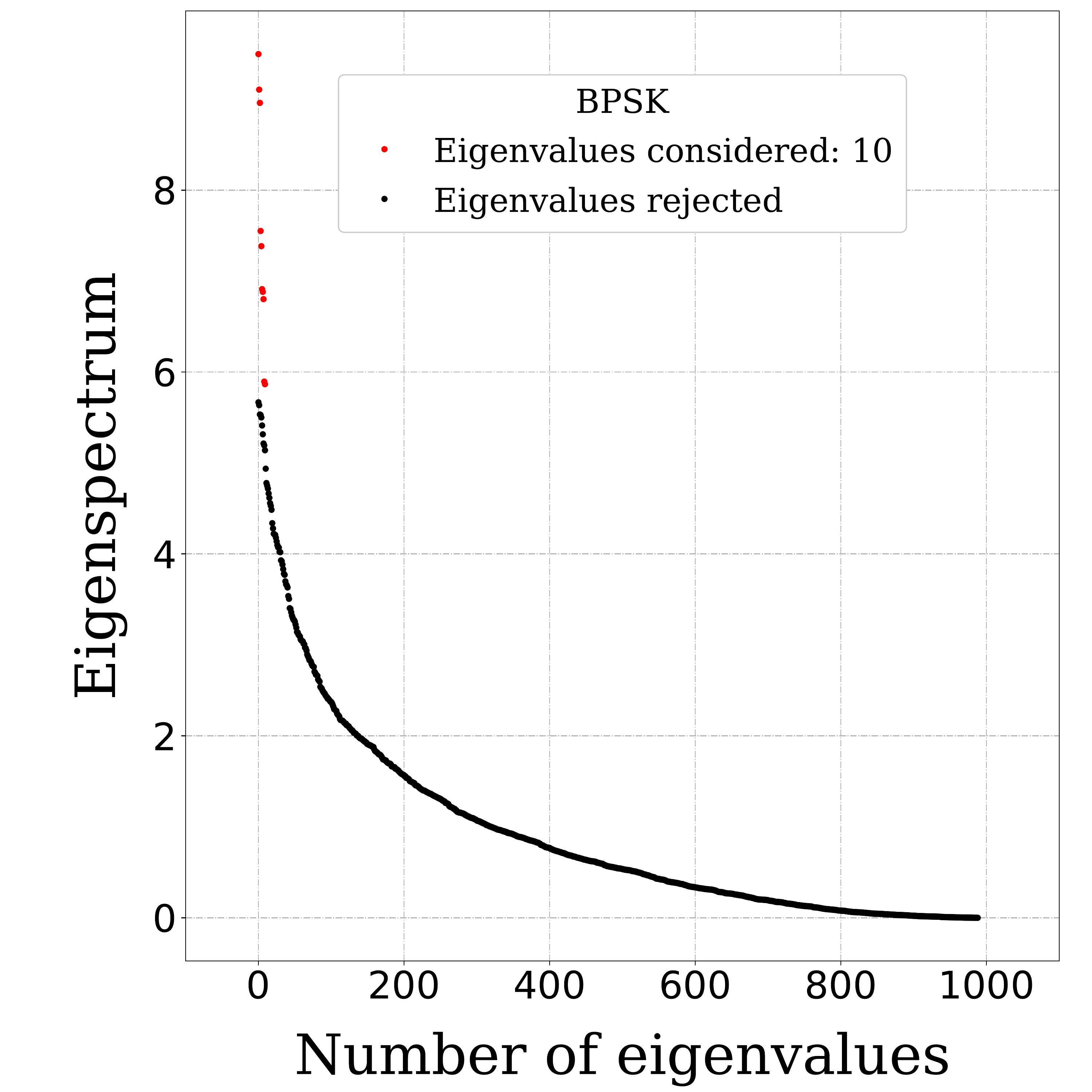}

        \includegraphics[width= \columnwidth]{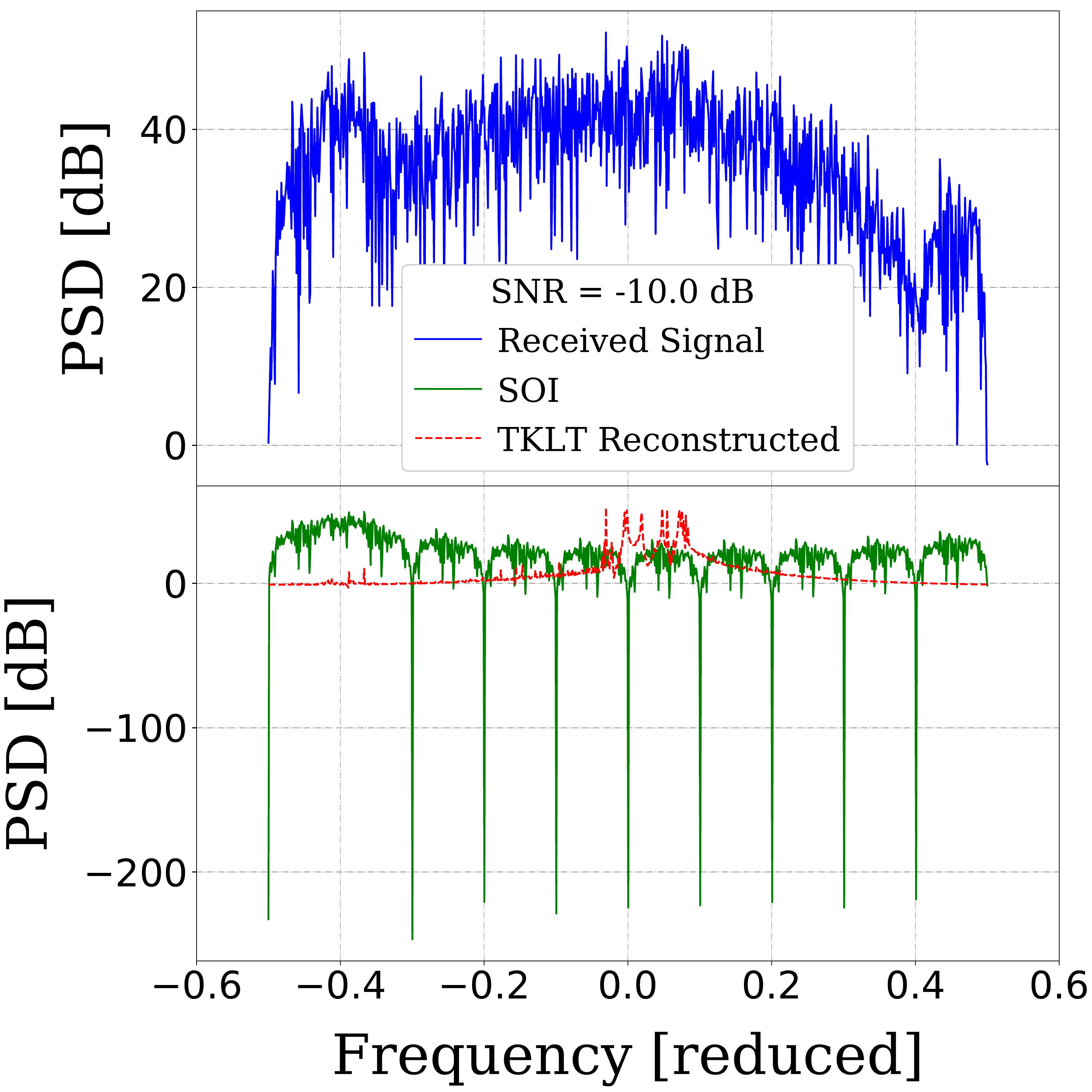}
        
\end{multicols}    
    \caption{TKLT reconstruction for a sinewave, a linear chirp and a BPSK.  The SNR is -10 dB. Top-left panel: eigenspectrum for the sinewave obtained with the TKLT. Top-right panel: PSD of the received signal (blue), the original SOI (green) and the TKLT reconstructed (red).  Middle-left and middle-right: the same as top-left and top-right respectively for the chirp. Bottom-left and bottom-right: the same as top-left and top-right respectively for the BPSK. }
    \label{fig:tklt_reconstruction}
\end{figure}

Figure \ref{fig:tklt_reconstruction} shows, on the left, the eigenspectra for the sinewave, the chirp and the BPSK while on the right the compared PSDs between the received signal, the SOI and the TKLT reconstructed. For the sinewave, the eigenspectrum in Fig.\ref{fig:tklt_reconstruction} (top-left panel) shows two dominant eigenvalues and therefore we considered two coefficients in the expansion, for the reconstruction. Fig.\ref{fig:tklt_reconstruction} (top-right panel) shows the PSD for the reconstructed signal. The TKLT reconstructed PSD shows a peak in the same position as the SOI, while the noise is notably reduced, showing that the TKLT filtered the injected noise.

For the chirp, Fig.\ref{fig:tklt_reconstruction} (middle-left panel) shows that the eigenspectrum does not display a clear break: in this case the TKLT does not manage to separate the SOI space and the noise space. Conservatively,  we attempted reconstruction of the chirp using only one eigenvalue (red line in Fig.\ref{fig:tklt_reconstruction}, middle-right panel) and 10 eigenvalues (violet line in Fig.\ref{fig:tklt_reconstruction}, middle-right panel).
When we consider only one eigenvalue, the reconstructed PSD shows a single peak which is unrelated to the starting frequency of the chirp. 
The PSD of the reconstructed signal, when considering 10 eigenvalues, shows several single peaks. There is a hint that only some specific components are being reconstructed, using a base of sinewaves. Further analysis will be needed in order to understand how a number of significant eigenvalues can be extracted in cases like this.
 
Analogously to the chirp, the eigenspectrum of the BPSK corrupted by the noise does not show a clear separation (Fig.\ref{fig:tklt_reconstruction}, bottom-left panel) and the PSD of the reconstructed signal with 10 eigenvalues (Fig.\ref{fig:tklt_reconstruction}, bottom-right panel) shows several single peaks. 

\begin{figure}
\begin{multicols}{2}
        \includegraphics[width= \columnwidth]{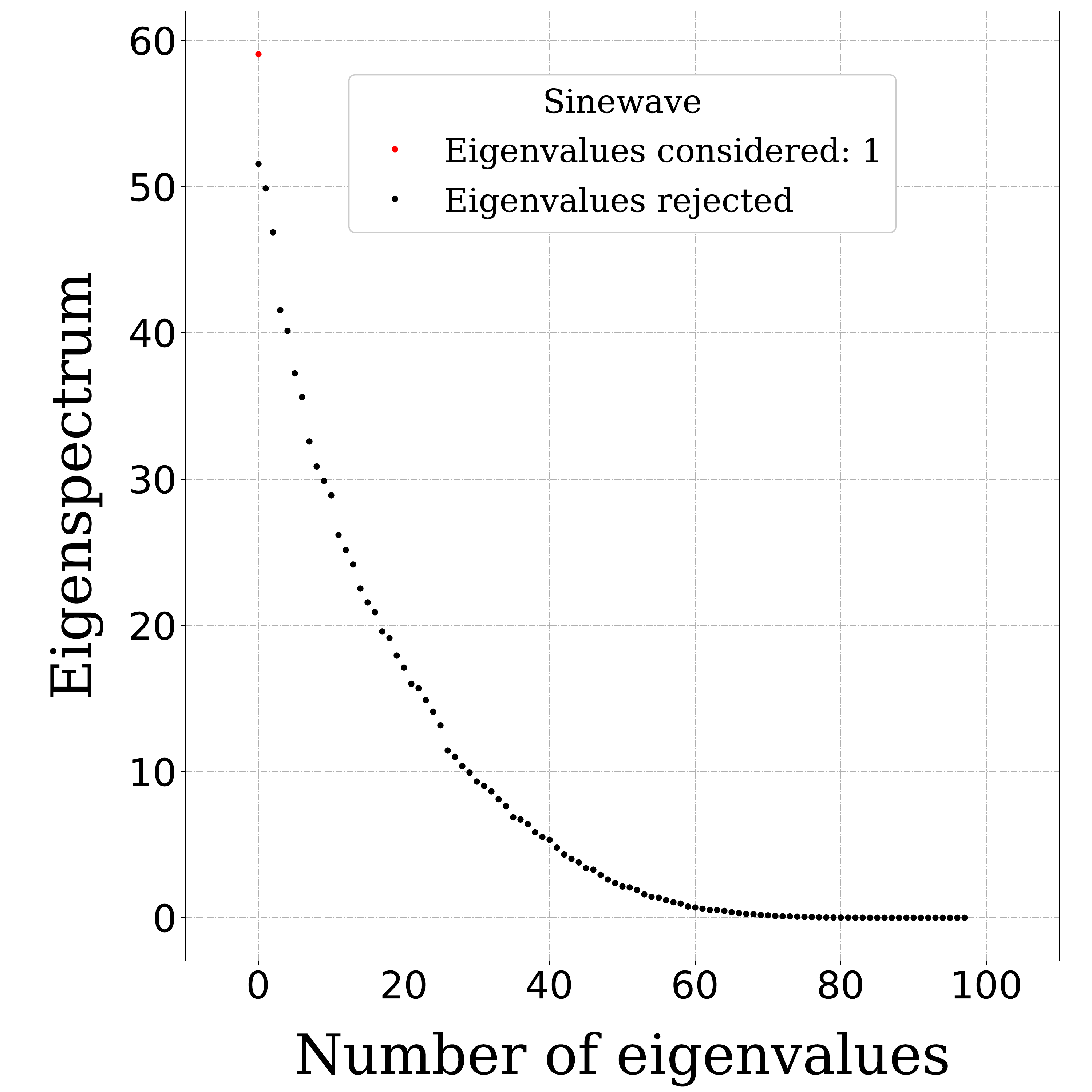}

        \includegraphics[width= \columnwidth]{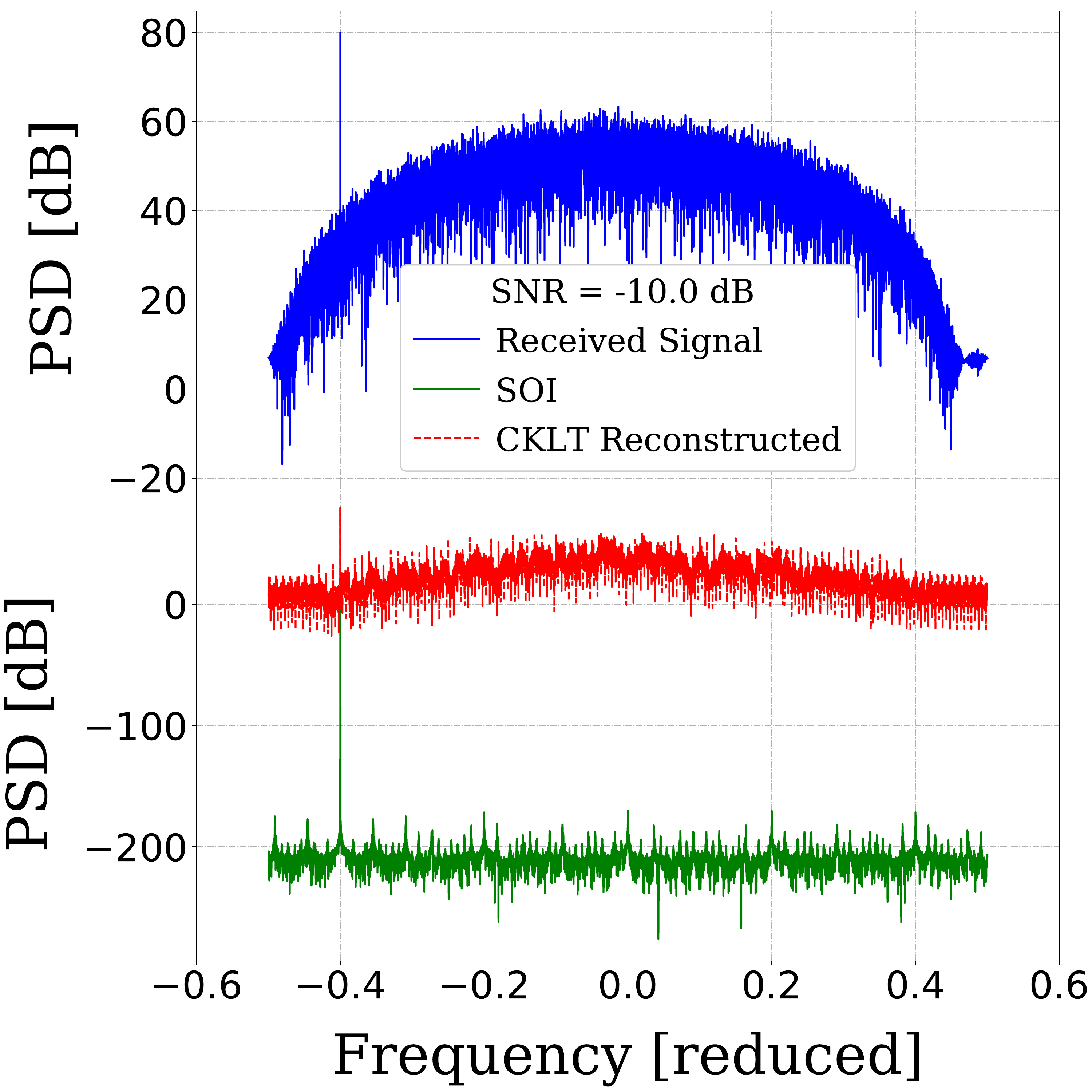}
        
\end{multicols}        
\begin{multicols}{2}
        \includegraphics[width= \columnwidth]{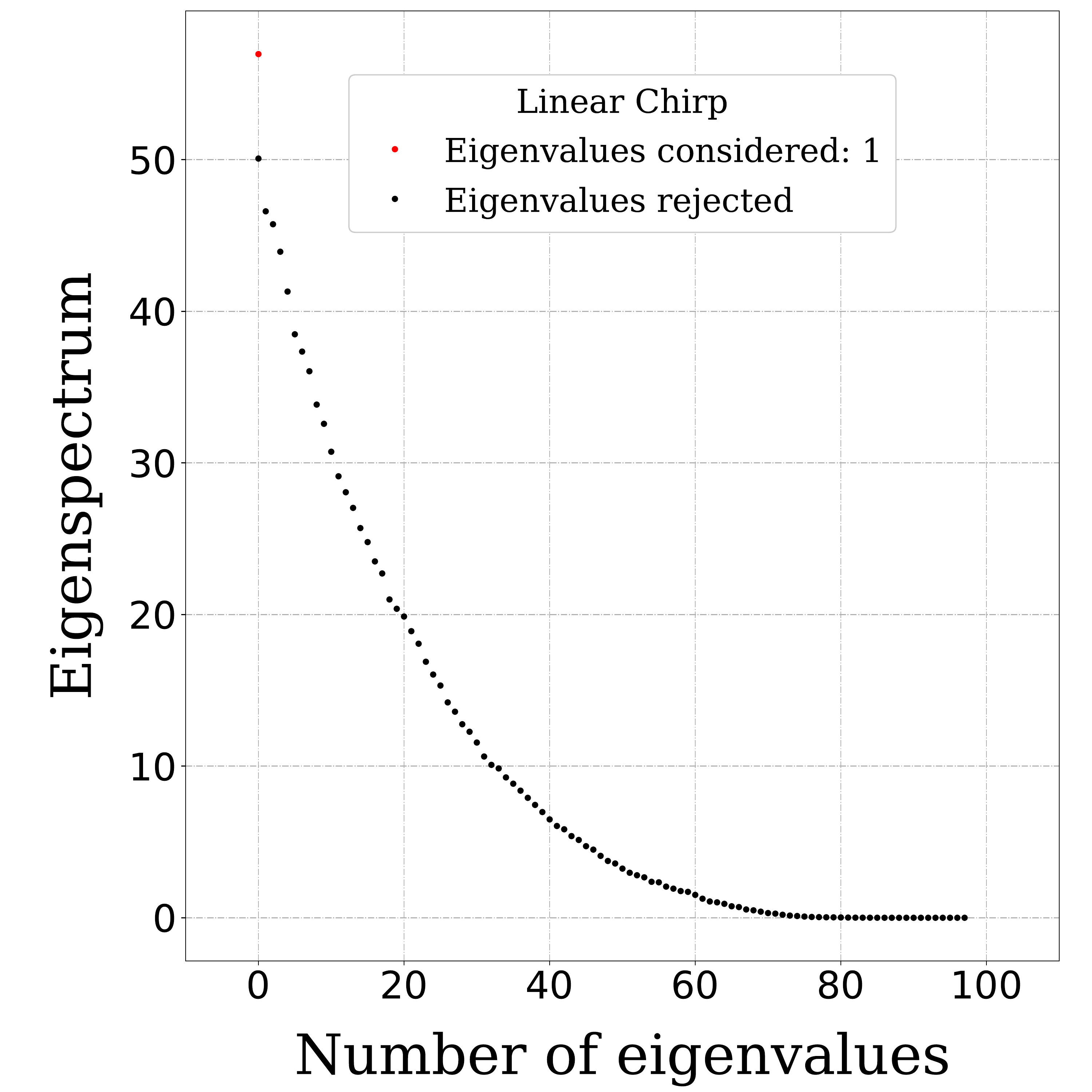}

        \includegraphics[width= \columnwidth]{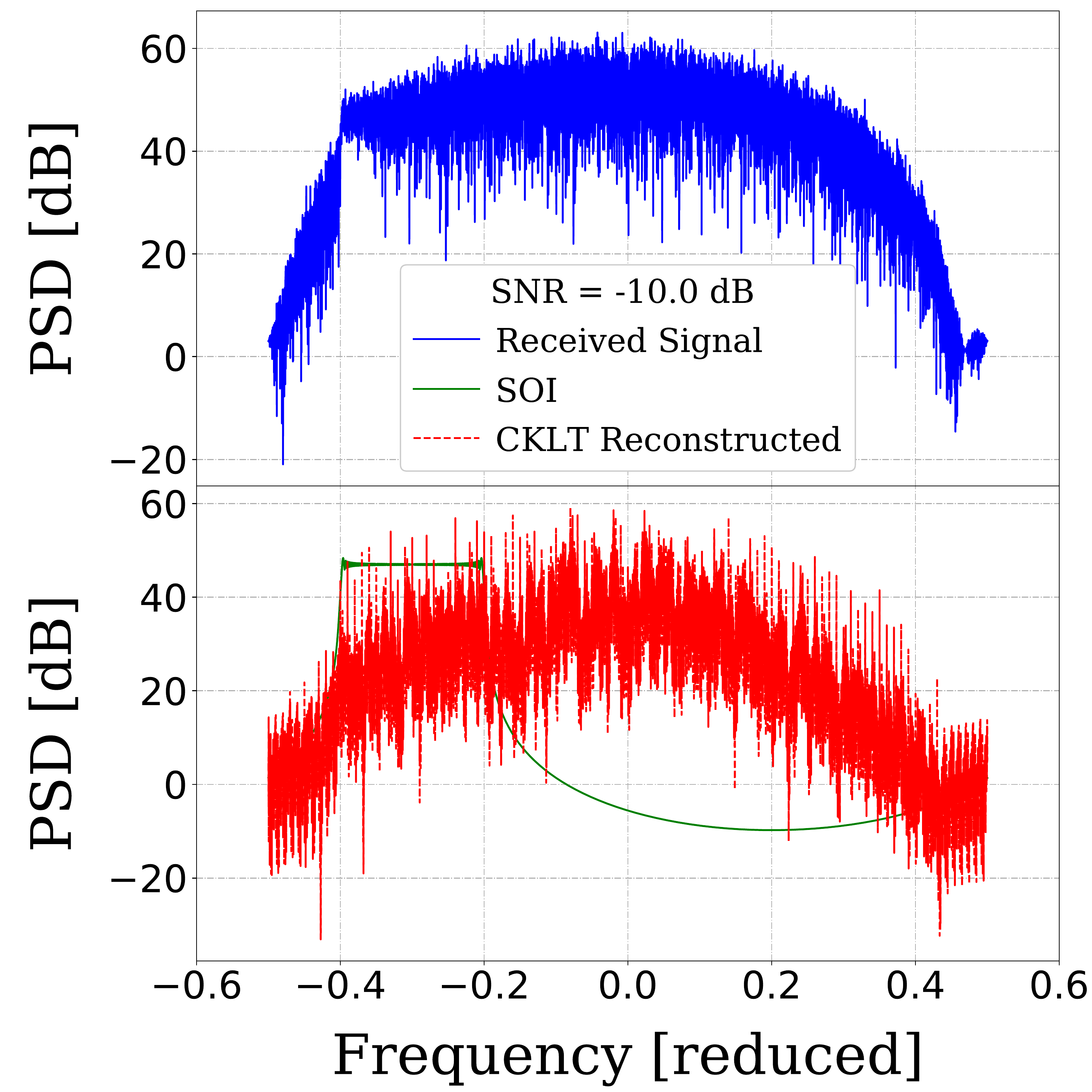}
        
\end{multicols}
\begin{multicols}{2}
        \includegraphics[width= \columnwidth]{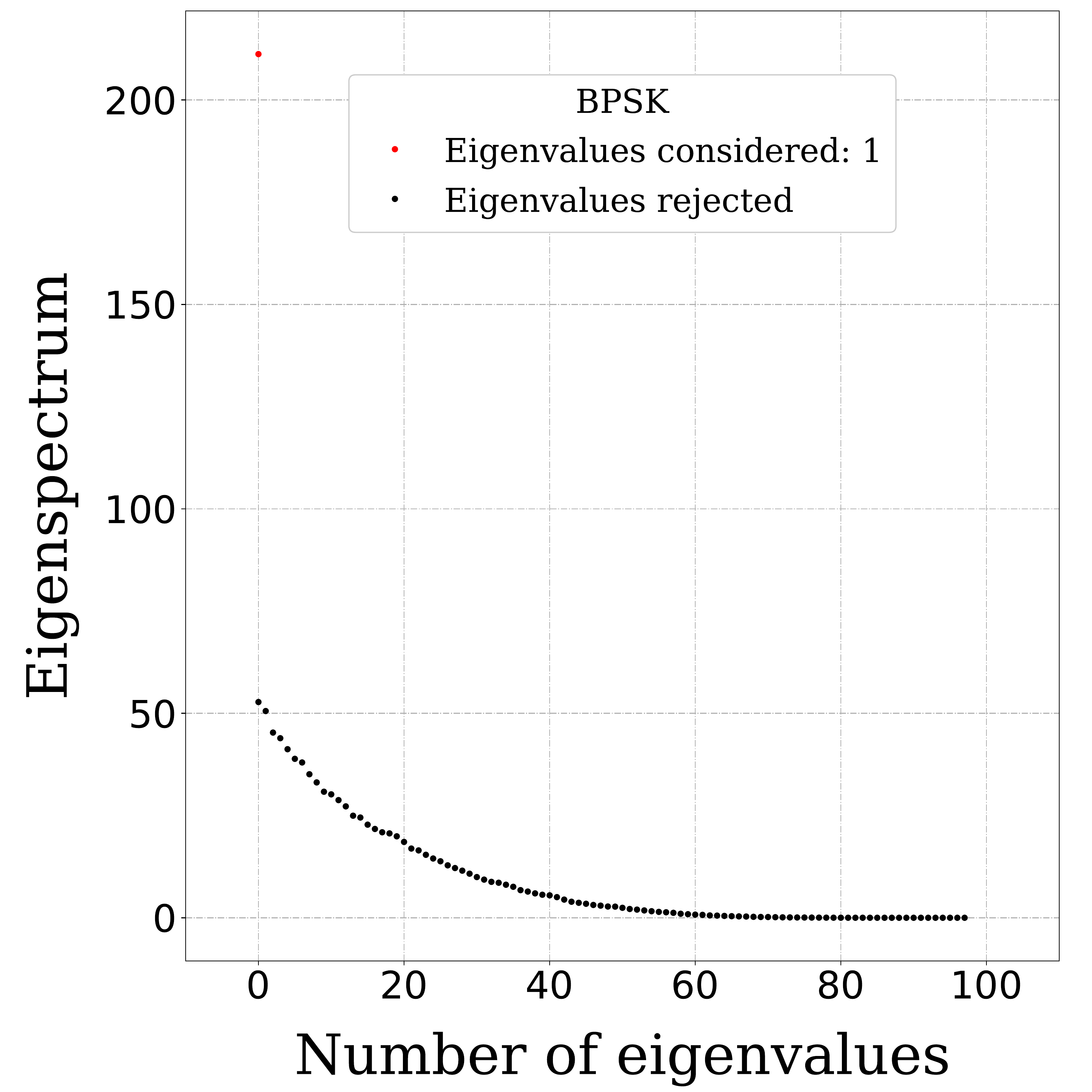}

        \includegraphics[width= \columnwidth]{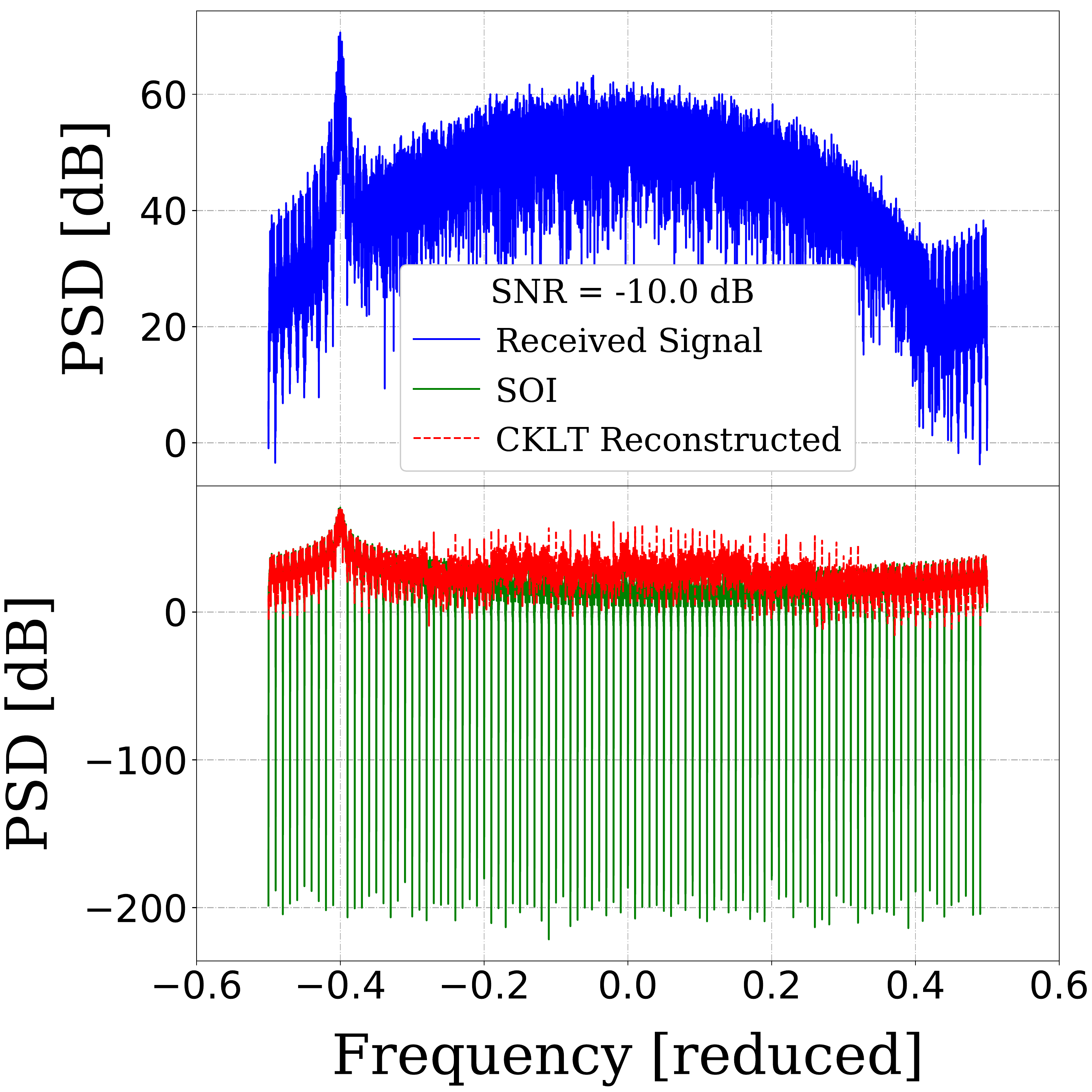}
        
\end{multicols}
    \caption{CKLT reconstruction for a sinewave, a linear chirp and a BPSK.  The SNR is -10 dB. Top-left panel: eigenspectrum for the sinewave obtained with the CKLT. Top-right panel: PSD of the received signal (blue), the original SOI (green) and the CKLT reconstructed (red).  Middle-left and middle-right: the same as top-left and top-right respectively for the chirp. Bottom-left and bottom-right: the same as top-left and top-right respectively for the BPSK. }
    \label{fig:cklt_reconstruction}
\end{figure}

Figure \ref{fig:cklt_reconstruction} shows the corresponding results for CKLT. For the sinewave and the chirp, neither eigenspectrum shows a break, hence the CKLT does not achieve a separation between signal and noise subspaces. Using only 1 eigenvalue for each SOI, the CKLT is able to find the correct position of the peak for the sinewave, despite a more noisy reconstruction
compared to the TKLT.
For the chirp, even when considering only 1 eigenvalue, the reconstructed signal does not show any feature of the SOI, and it contains a significant amount of noise.
In the BPSK case, as shown in Fig.\ref{fig:cklt_reconstruction} (bottom-left panel), the eigenspectrum possesses a significantly dominant eigenvalue with respect to the others. 
As opposed to the previous two cases, the PSD of the signal reconstructed by the CKLT (Fig.\ref{fig:cklt_reconstruction}, bottom-right panel) is describe similar to the PSD of the SOI. There is a considerable discrepancy between the SOI and the CKLT reconstructed in the frequency range of $(-0.2,0.2)f_s$; this is due to the noise coloration, as already mentioned for the MRKLT reconstruction. 

\begin{figure}
\begin{multicols}{2}
        \includegraphics[width= \columnwidth]{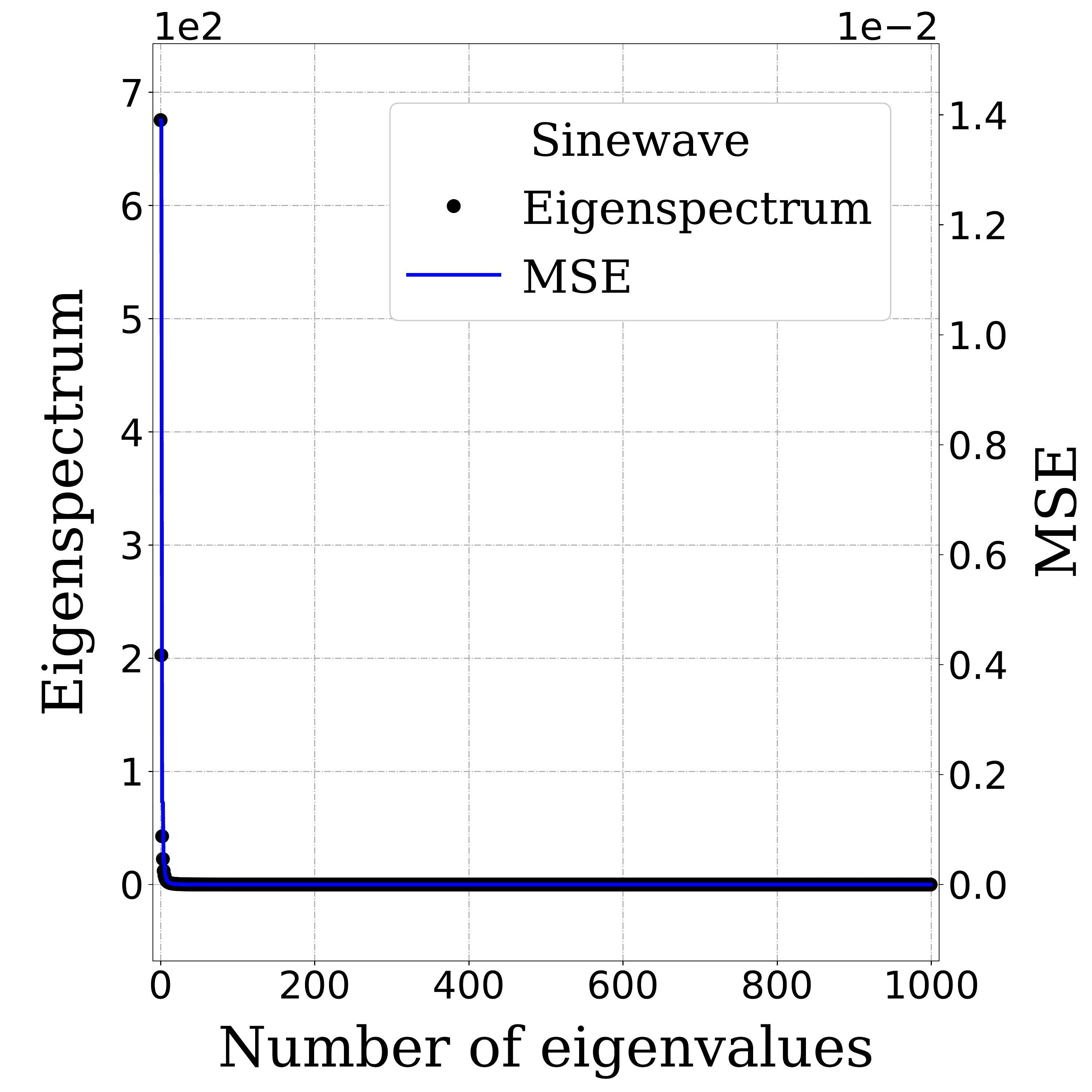}

        \includegraphics[width= \columnwidth]{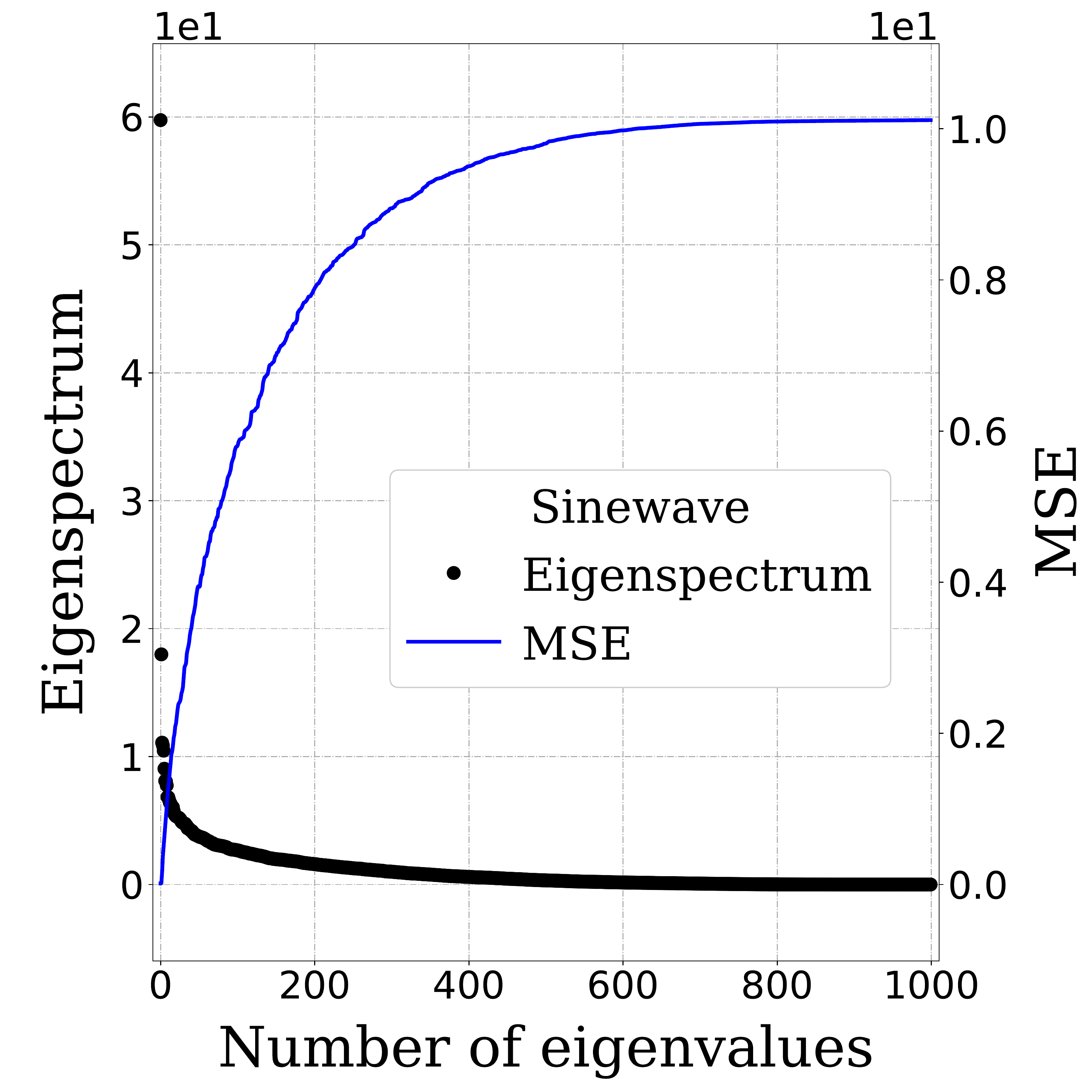}
        
\end{multicols}        
\begin{multicols}{2}
        \includegraphics[width= \columnwidth]{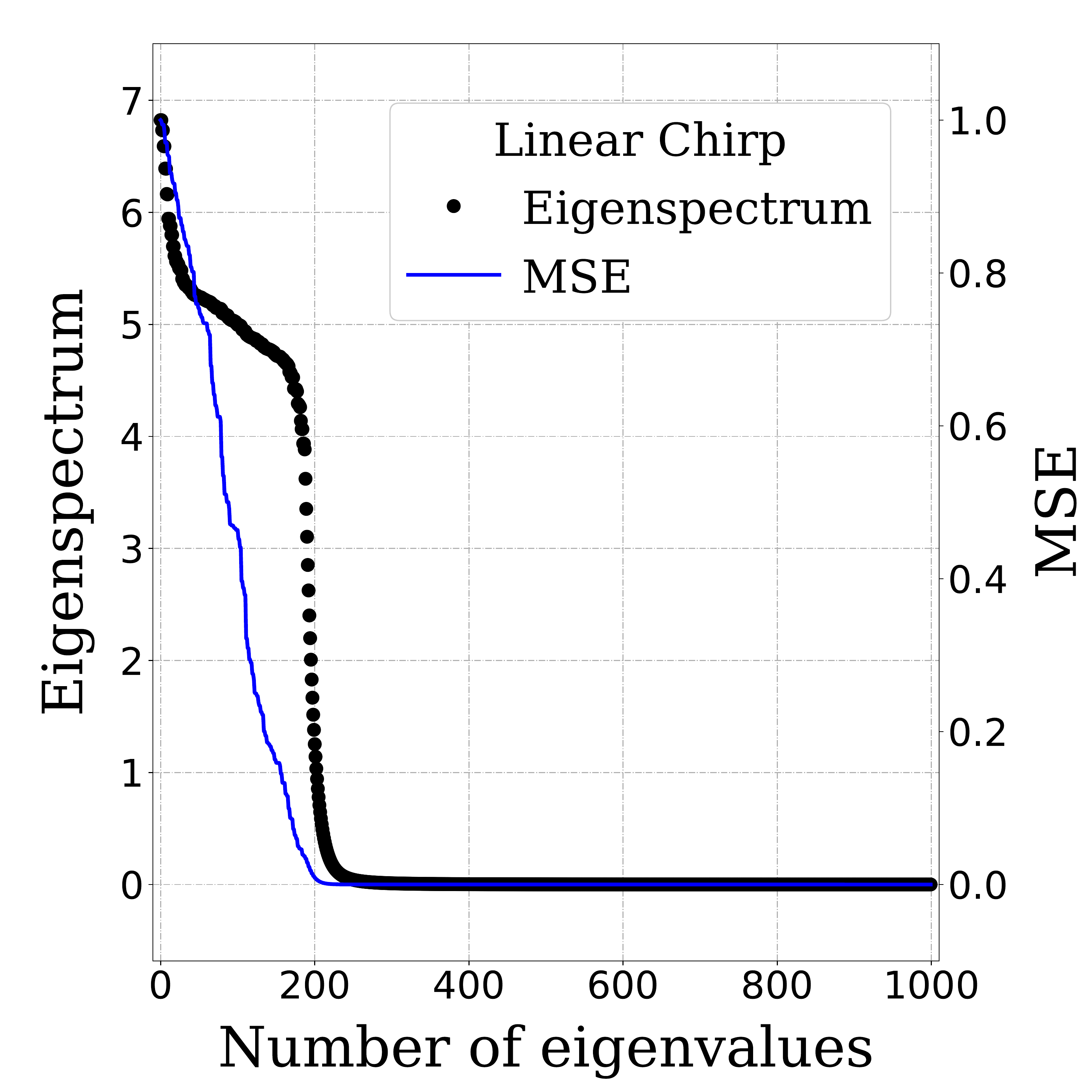}

        \includegraphics[width= \columnwidth]{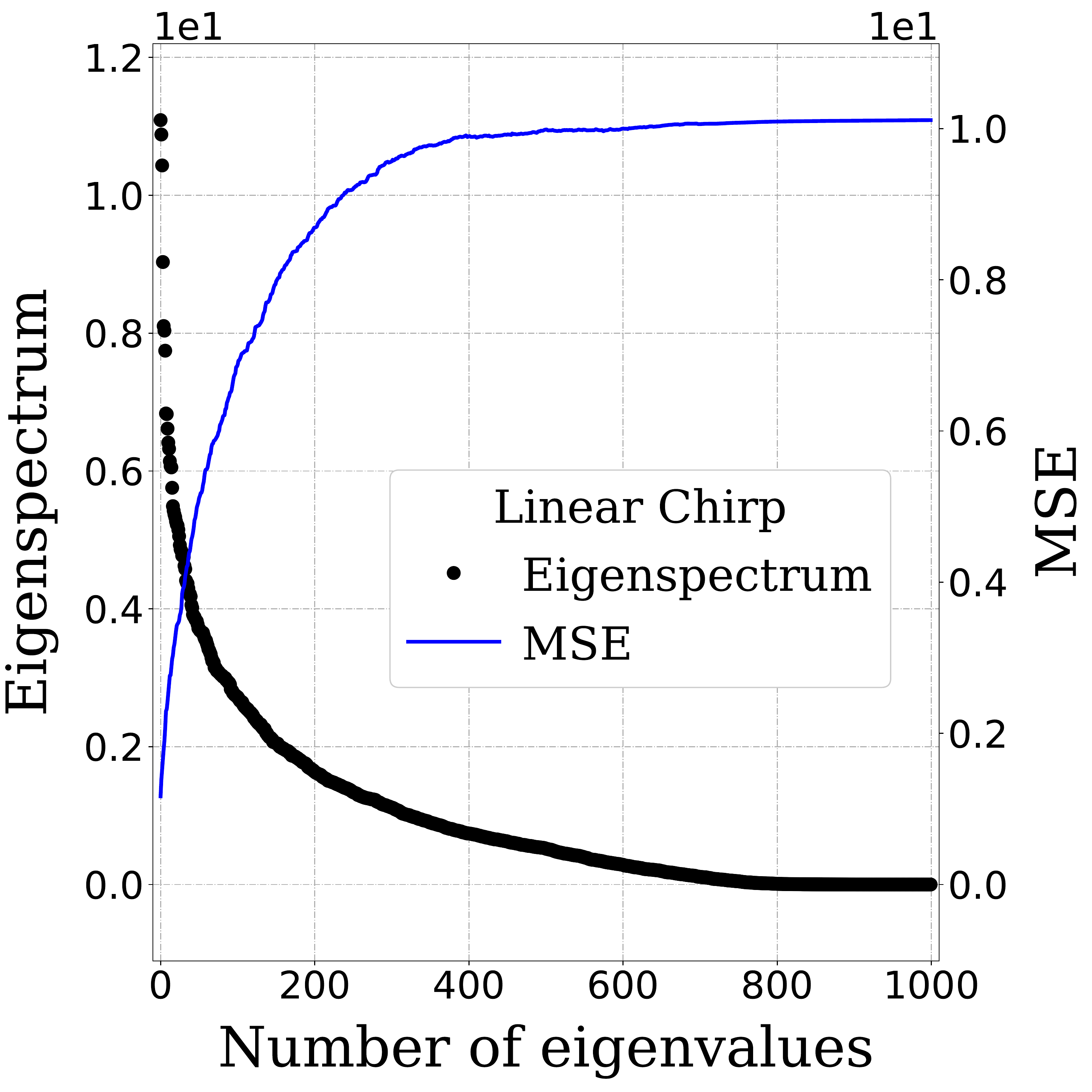}
        
\end{multicols} 
\begin{multicols}{2}
        \includegraphics[width= \columnwidth]{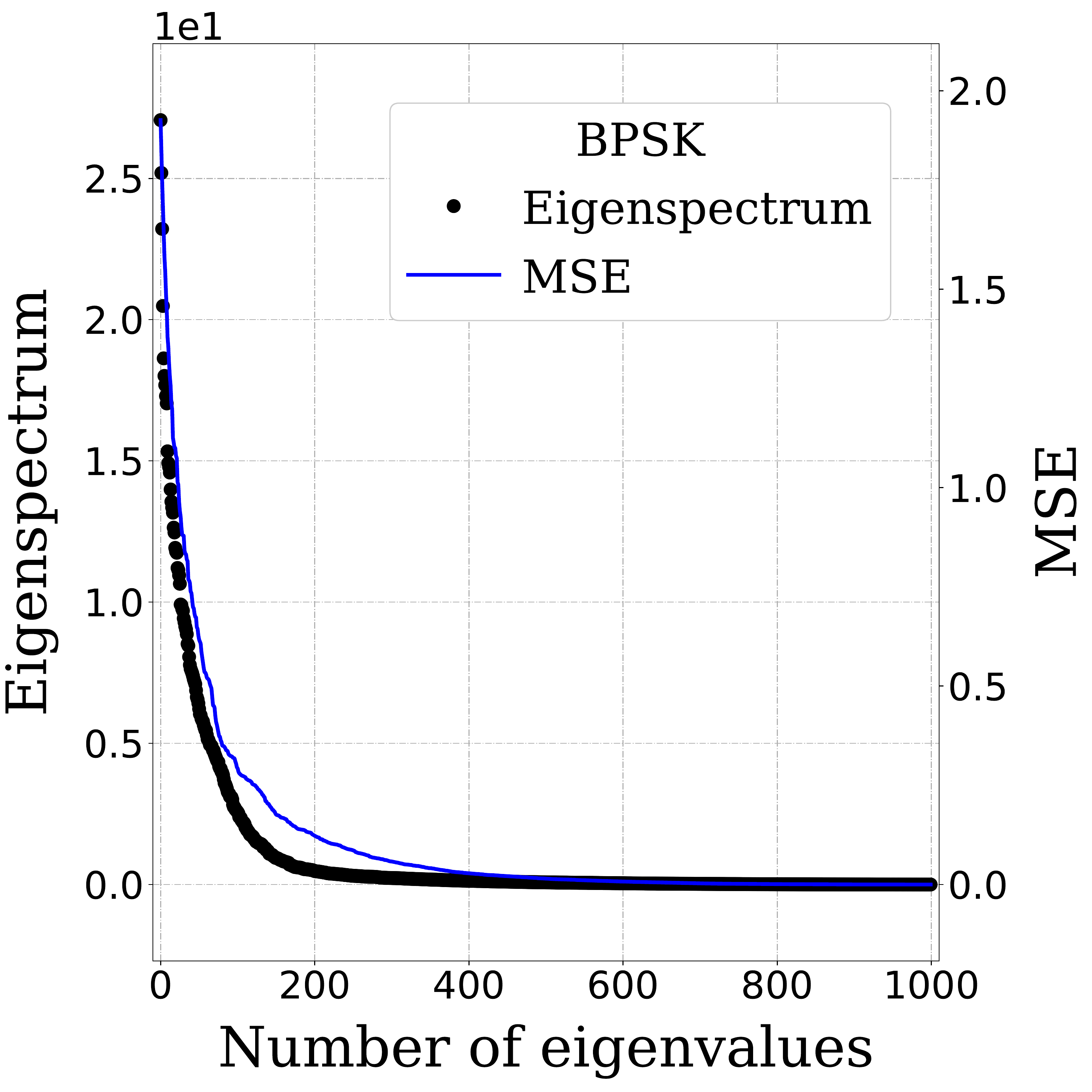}

        \includegraphics[width= \columnwidth]{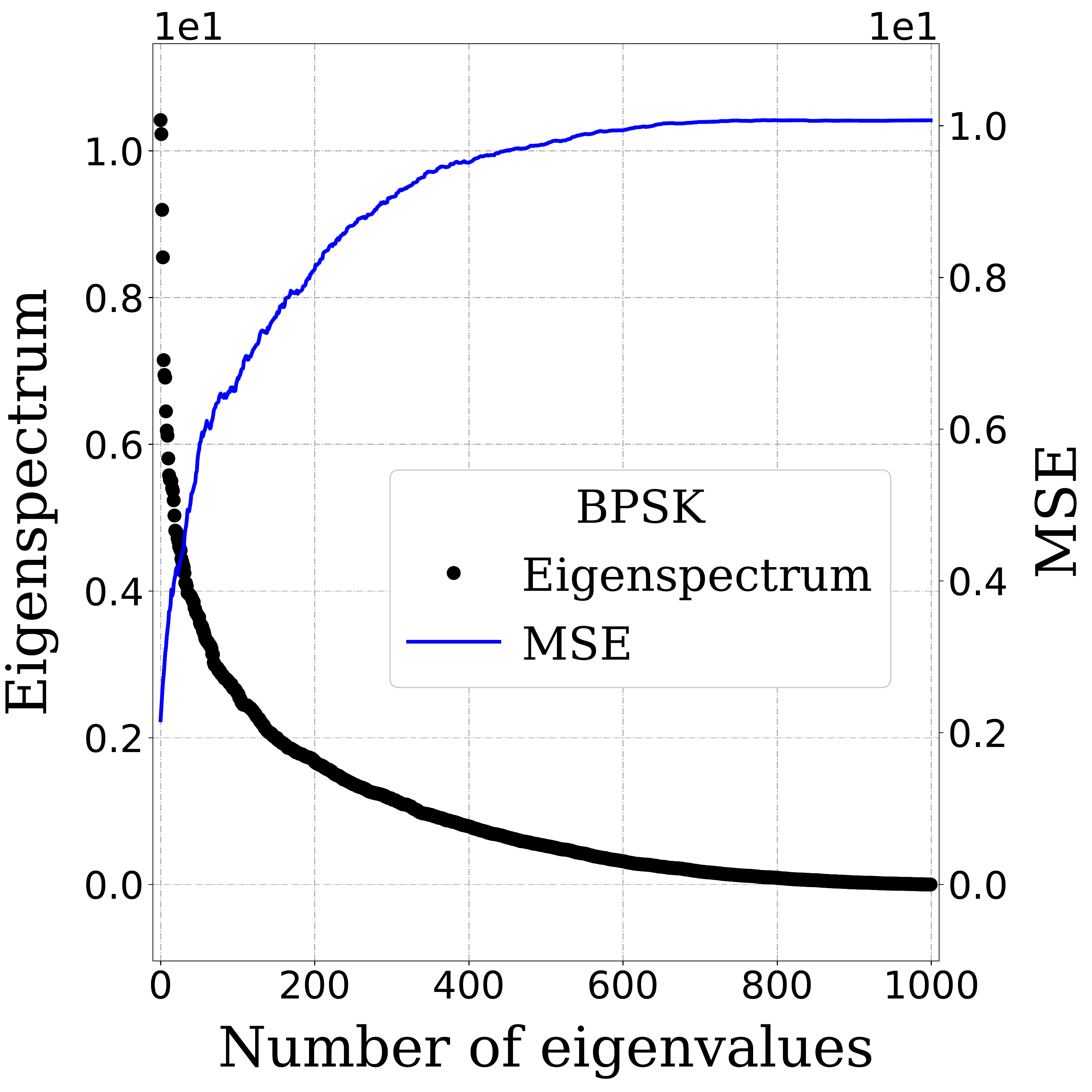}
        
\end{multicols} 
    \caption{Eigenspectrum (left y axis) and MSE (right y axis) as a function of the number of eigenvalues used for the TKLT reconstruction for different SOIs. (Top, middle, bottom)-left panels represent the noiseless case; (top, middle, bottom)-right panels represent the SOI buried in coloured noise with a SNR of -10 dB.}
\label{fig:mse_tklt}

\end{figure}

\begin{figure}
\begin{multicols}{2}
        \includegraphics[width= \columnwidth]{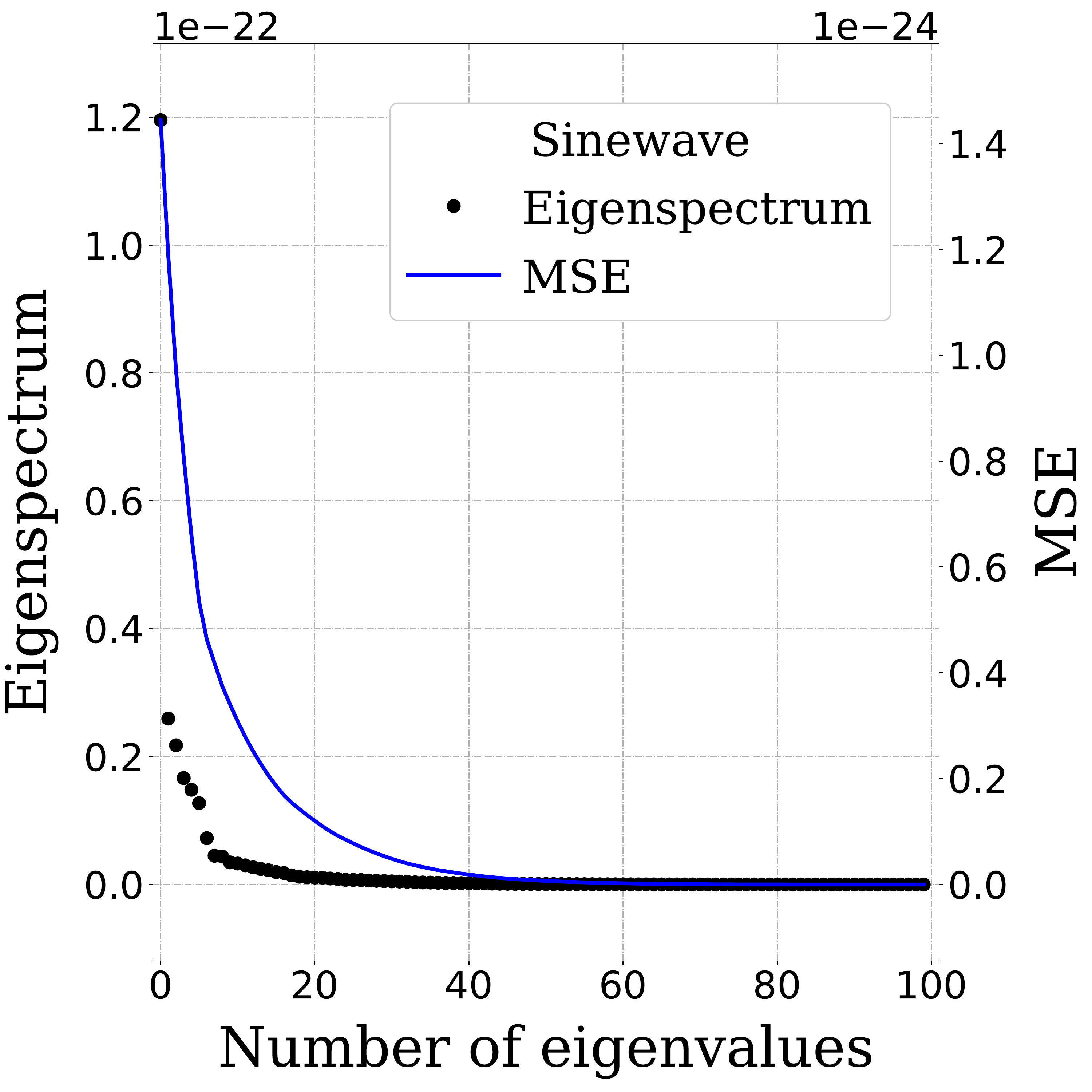}

        \includegraphics[width= \columnwidth]{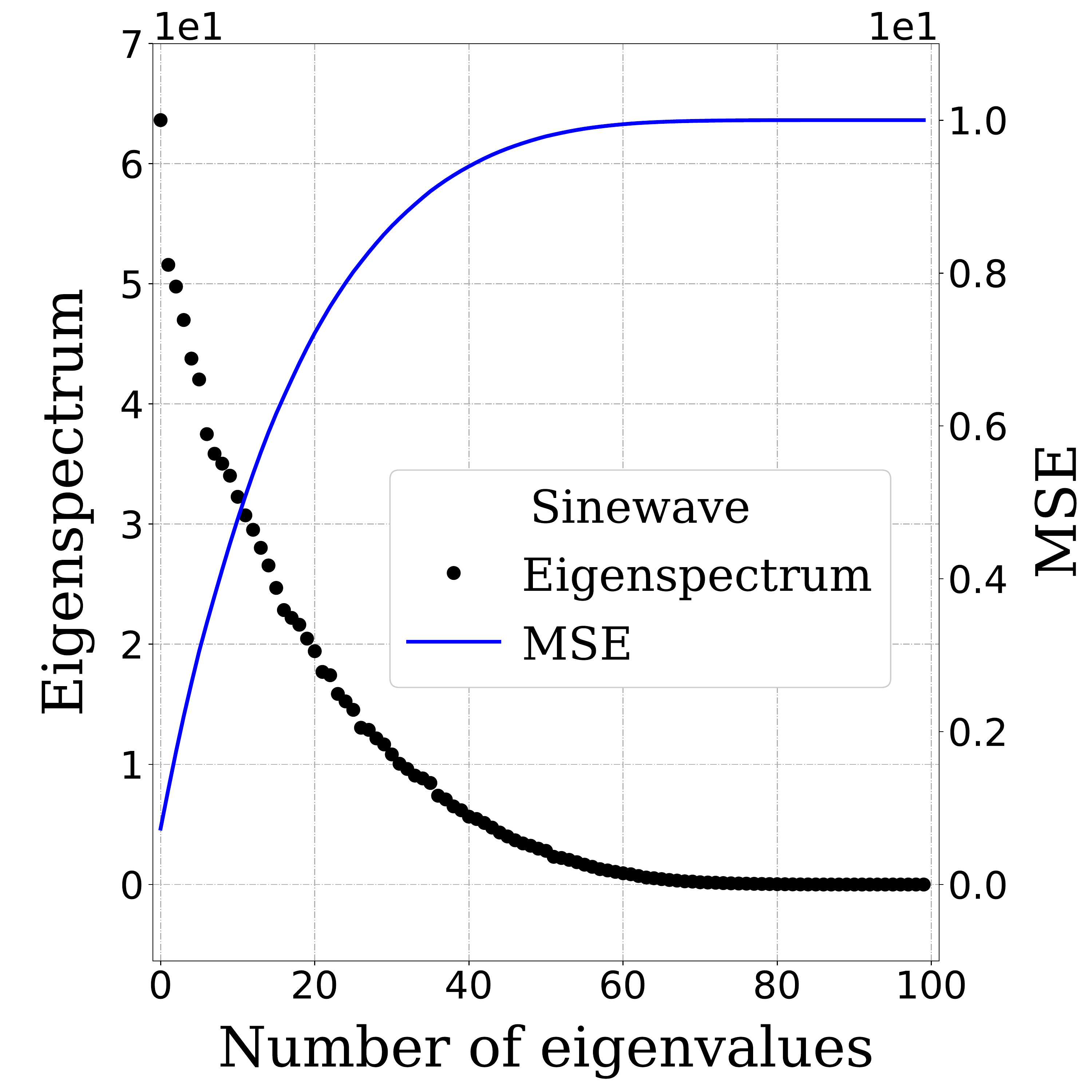}
        
\end{multicols}        
\begin{multicols}{2}
        \includegraphics[width= \columnwidth]{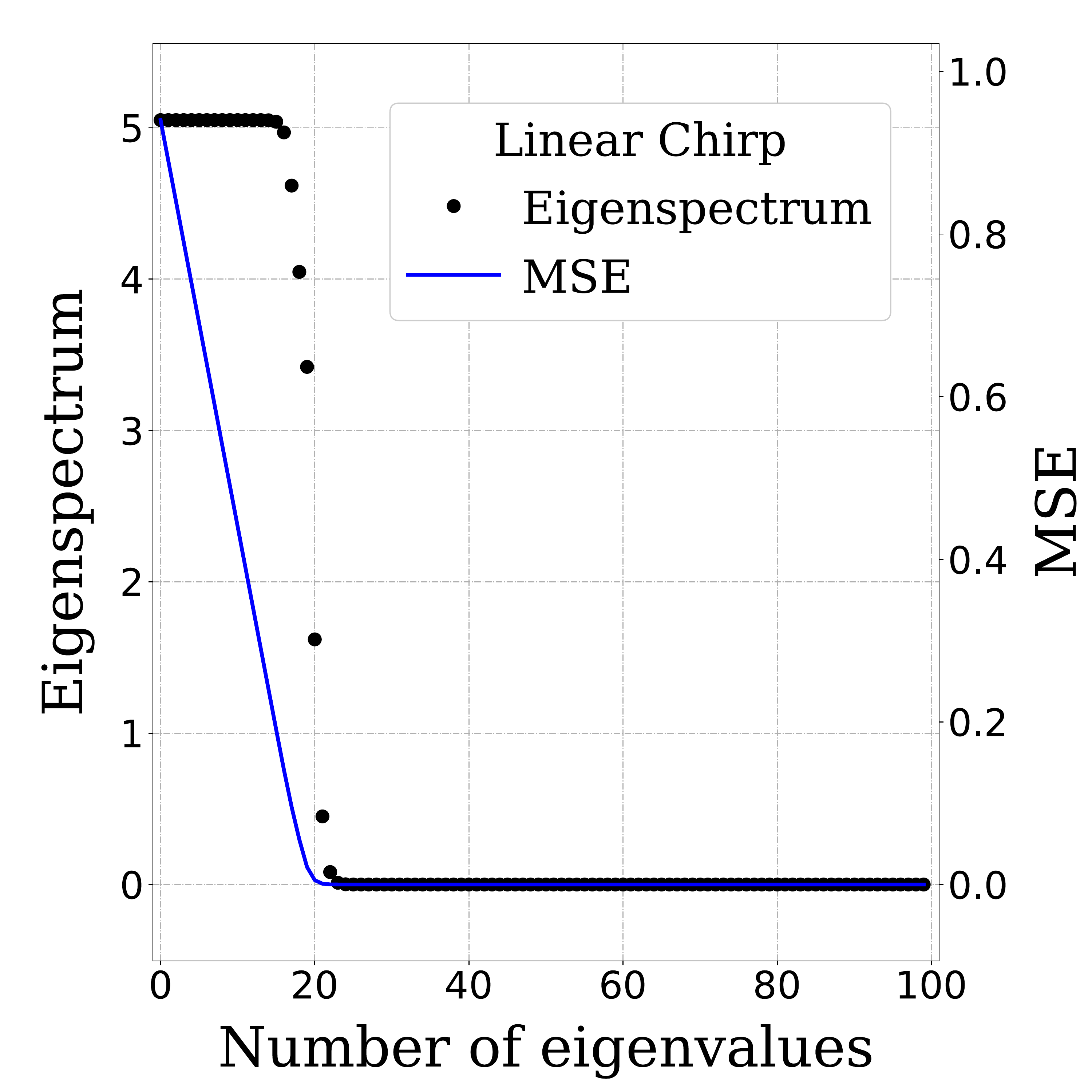}

        \includegraphics[width= \columnwidth]{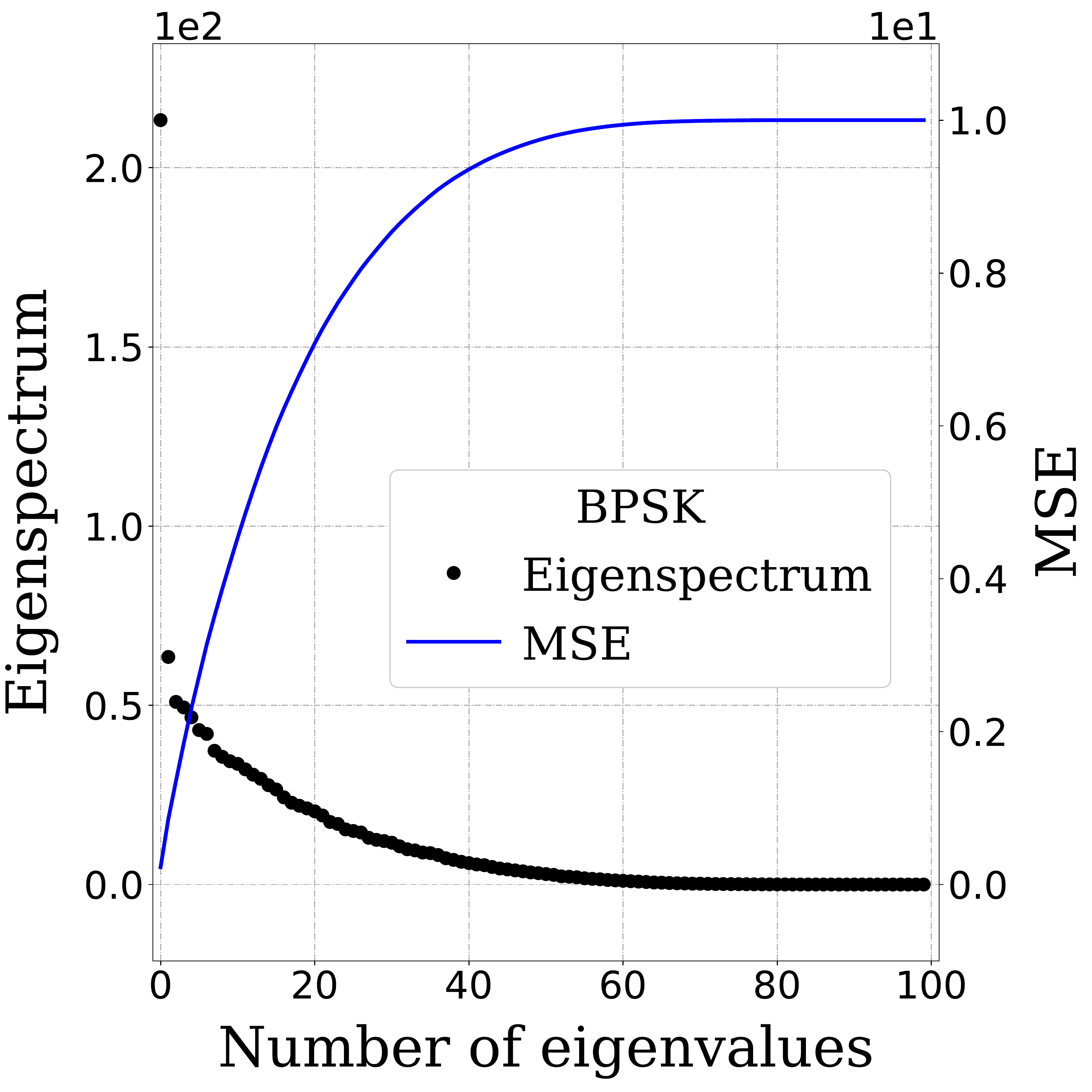}
        
\end{multicols} 
\begin{multicols}{2}
        \includegraphics[width= \columnwidth]{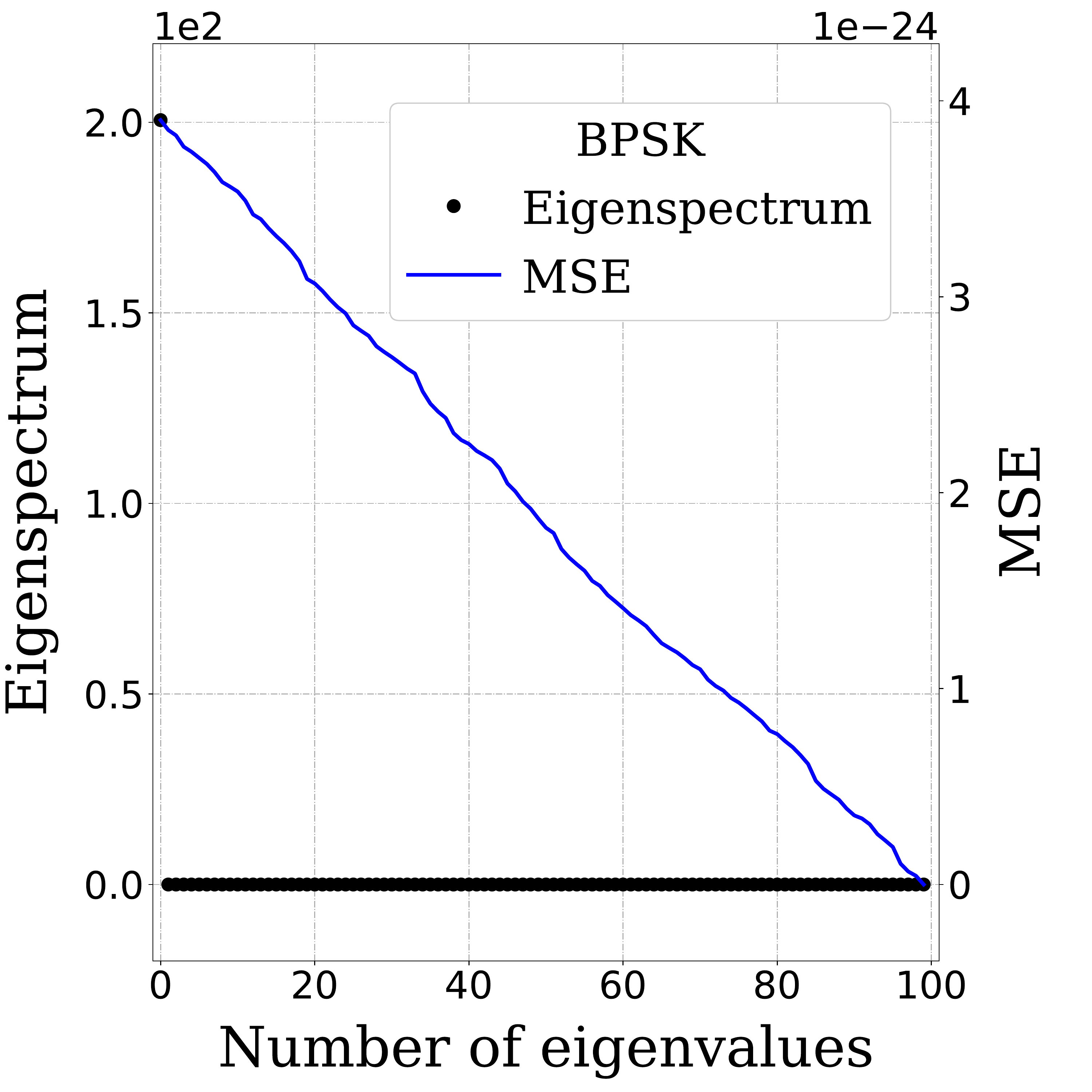}

        \includegraphics[width= \columnwidth]{Plots/Section4/CKLT/MSE/CKLT_BPSK_Noise.pdf}
        
\end{multicols} 

    \caption{Eigenspectrum (left y axis) and MSE (right y axis) as a function of the number of eigenvalues used for the CKLT reconstruction for different SOIs. (Top, middle, bottom)-left panels represent the noiseless case; (top, middle, bottom)-right panels represent the SOI buried in coloured noise with a SNR of -10 dB.}
\label{fig:mse_cklt}
\end{figure}

In figures \ref{fig:mse_tklt} and \ref{fig:mse_cklt} the black dots represent the eigenspectra for each SOI, computed using the TKLT and the CKLT respectively. The plots on the left show the case of the SOI without noise; the plots on the right show the case of the SOI buried in coloured noise with SNR$=-10$ dB. The blue lines in the same plots represent the mean square errors (MSE) between the SOI $s_i$ and the reconstructed signal $\tilde{x}_i$:
\begin{equation}
    \label{3.2.3}
    \text{MSE} = \frac{1}{N} \sum_{i=0}^{N-1} \left|s_i-\tilde{x}_i \right|^2 \ .
\end{equation}
MSE were computed by varying the number of eigenvalues for each point, starting from 1 to N for the TKLT and from 1 to W for the CKLT. For the TKLT, we considered an input vector of $N=10^3$ samples, while for the CKLT we considered an input vector with $N = 10^4$ samples and a KLT Window with $W=10^2$ samples. 

In the case of the TKLT, when the SOIs are not corrupted by the noise, only the eigenspectra of the sinewave and the chirp show a trend separating the signal sub-space from the noise sub-space. As already noticed, for the sinewave, only two eigenvalues are dominant. For the chirp, a change of slope followed by a concavity seem to delimit the signal space. The number of meaningful eigenvalues is 200, which is exactly the length of the TKLT eigenspectrum times the considered chirp drift rate. 
Lastly, the eigenspectrum of the BPSK shows a monotone decreasing behaviour with no particular feature.

The MSE curves of the three SOIs behave similarly to the eigenspectra. In the sine case, when considering two eigenvalues, the MSE rapidly converges to zero: the first two coefficients of the expansion are enough to reconstruct the SOI. For the chirp, the MSE consistently reaches zero near the second point where the eigenspectrum changes slope. When the SOIs are corrupted by the noise, only the sinewave eigenspectrum maintains the same appearance as the noiseless case. The other two eigenspectra are very similar to each other, suggesting that the TKLT did not manage to filter the noise out. The MSE curves correctly grow when we consider more eigenvalues and they all saturate at 10. This result is consistent since it corresponds to the variance of the noise for a SNR$=-10$ dB.

In the case of the CKLT, when the SOIs are not corrupted by the noise, the sinewave eigenspectrum has only one dominant eigenvalue (as opposed to the two dominant in the TKLT) and a cluster of sub-dominant eigenvalues appears. The chirp eigenspectrum shows similar features to the one reconstructed by the TKLT. Also in this case the number of meaningful eigenvalues corresponds to the chirp drift rate times the length of the eigenspectrum considered, which is the length of the KLT Window for the CKLT. Contrary to the TKLT case, the meaningful eigenvalues remain constant in value and there is no change in slope. The BPSK eigenspectrum shows only one significant eigenvalue and all the remaining eigenvalues converge rapidly to zero. All the MSE curves correctly decrease considering more eigenvalues. 

When the SOIs are corrupted by noise only the BPSK eigenspectrum shows a clear sub-space division. The sinewave and chirp eigenspectra are very similar to the ones obtained using the TKLT. As for the TKLT, also the CKLT MSE curves increase considering more eigenvalues and saturate at the value corresponding to the variance of the noise.

\section{Monte Carlo Simulations for CKLT}
\label{sec:montecarlo}
Given the computation time results summarised in Fig.\ref{fig:time_analysis}, we chose the CKLT as the most suitable algorithm to perform  Monte Carlo (MC) simulations for the evaluation of both reconstruction and detection performance for each SOI. In the next two sections we will discuss reconstruction and detection separately. We will present the different setups for the simulations, the metrics used in order to evaluate CKLT performance and the obtained results.

For our simulations we considered five SOIs: a sinewave, a linear chirp and a BPSK as models for typical interstellar telecommunication signals; a synthetic spectral line and a synthetic pulsar as models for signals of astrophysical origin. 
In the MC simulation, some parameters were randomly generated at each MC trial. For other parameters some discrete values were selected to evaluate the MC outcome based on their variation.
The parameters selected for the MC simulations were:
\begin{itemize}
    \item SNR. This parameter is used to deduce the level of SNR required to start to recover the SOIs buried in noise for the case of reconstruction. It is also an indicator of when CKLT starts to be a good detector, in the case of detection.
    \item Length of the KLT window. This parameter plays a key role in CKLT. The size of the covariance matrix computed and its consequential eigenspectrum depend on it. The simulation of several KLT windows values should prompt the optimal KLT window length for each case, which may vary with the SOI type.

\end{itemize}
The parameters randomly generated for each SOI were:
\begin{itemize}
    \item Sinewave. A normalised frequency is generated with uniform probability distribution $\mathcal{U}(0,1)f_s$, while the phase is generated with a uniform probability distribution $\mathcal{U}(0,2 \pi)$.
    \item Linear Chirp. The distributions of the normalised starting frequency and the phase are the same used for the sinewave, the drift rate is generated with a uniform distribution $\mathcal{U}(0,1)f_s/N$.
    \item Synthetic Pulsar. Randomness is ensured by generating the complex white noise used to build the SOI in each trial, and by considering, for each single pulse, a different amplitude with normal distribution $\mathcal{N}(1,0.5)$.
    \item BPSK. The parameters were uniformly distributed like in the sinewave, for the sinewave transmission signal, while the bits for the message signal were generated by randomly choosing between -1 and +1 (with probability $p=0.5$ each).
    \item Synthetic Spectral Line: the randomness of the experiment was ensured with the complex white noise used to generate the SOI itself.
\end{itemize}
In both MC simulations the input vector was a complex vector of $10^4$ samples. The length of the filter window used to generate the synthetic spectral line and the bit-period for the BPSK were both $10^2$ samples. The synthetic pulsar signal consists in $10^2$ pulses of $10^2$ samples each. The noise considered is the coloured noise described in section \ref{sec:mrklt}.

\subsection{Reconstruction}
For this analysis, we used the MSE between the SOI and the CKLT reconstructed signal as a metric and we studied how the MSE changes as a function of the SNR and of the KLT Window. We performed $10^3$ trials for the MC simulation. Here we follow the most conservative approach by considering the first dominant eigenvalue for the expansion as  we lack an a-priori closed-form expression to identify the meaningful number of eigenvalues which define a specific SOI.

\begin{figure}
\centering
        \includegraphics[width= 0.82 \columnwidth]{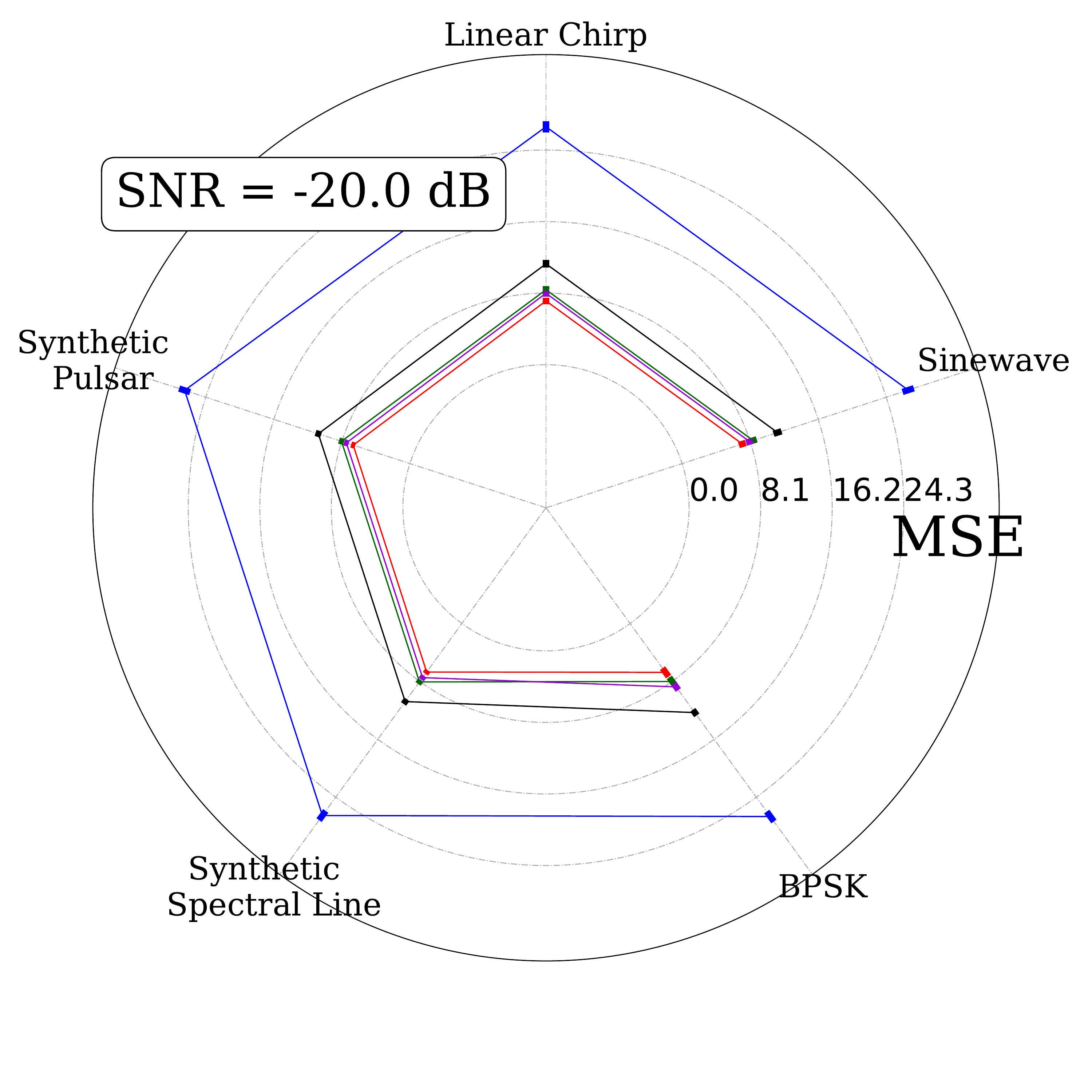}

        \includegraphics[width= 0.82  \columnwidth]{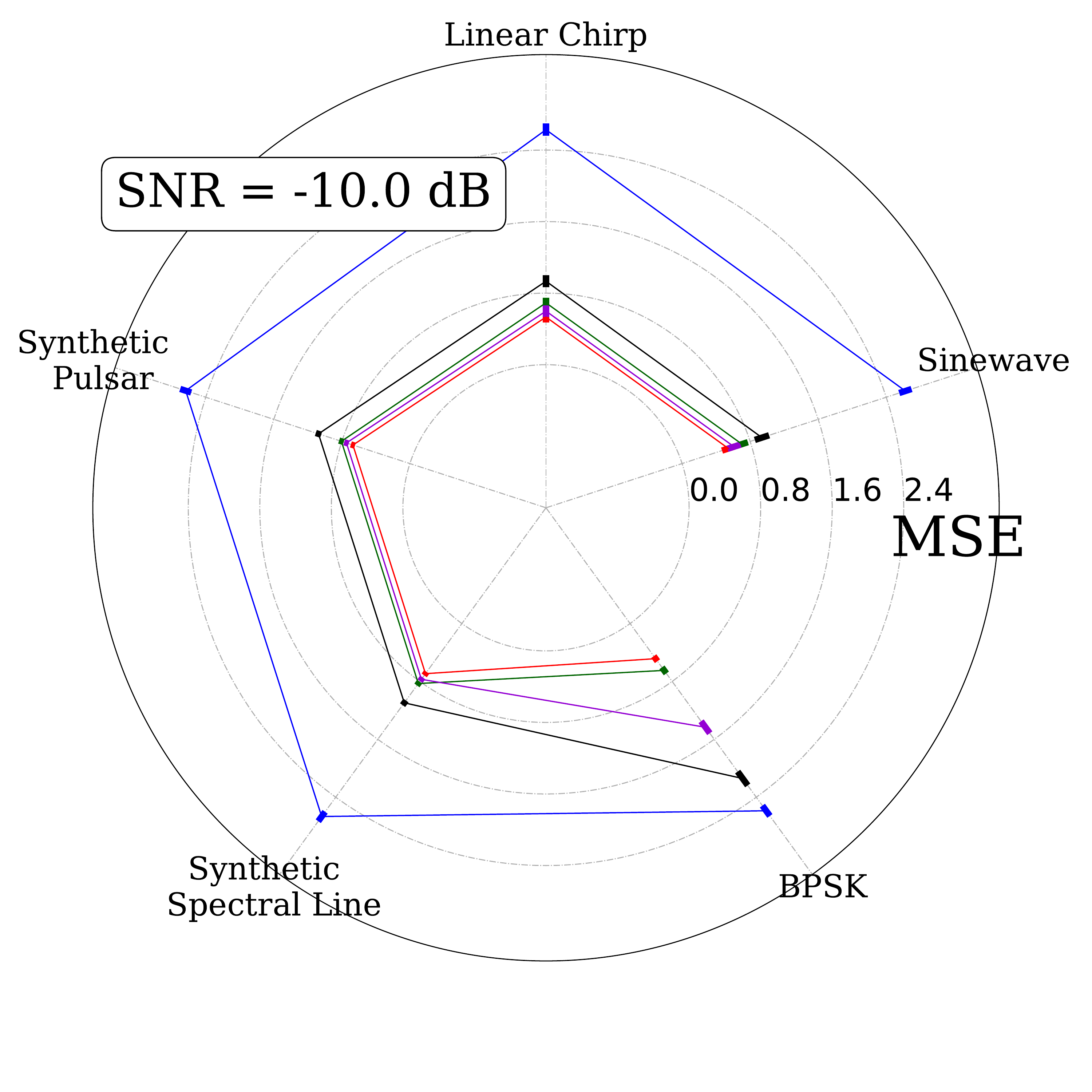}

        \includegraphics[width= 0.82 \columnwidth]{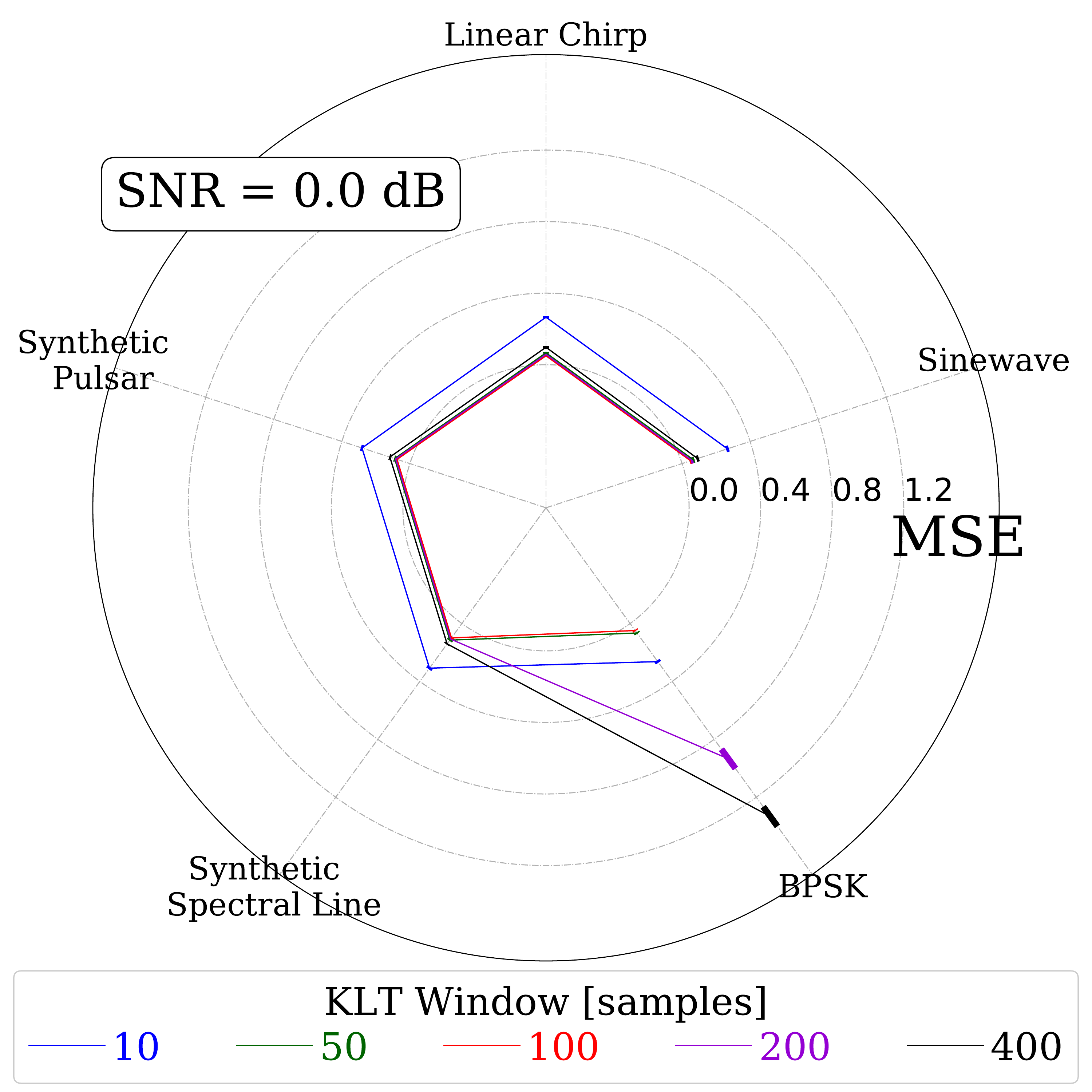}

    \caption{Radar charts for the reconstruction MC simulation. Each plot shows the average value of the MSE between the SOI and the reconstructed CKLT, with its respective error. This value is computed for each SOI (different axes of the radar chart) and for each KLT Window (different colours) at different SNRs: -20 dB (top panel), -10 dB (middle panel) and 0 dB (bottom panel).  }
    \label{fig:mse_radars}
\end{figure}

Figure \ref{fig:mse_radars} shows the results for the reconstruction. The radar charts show that, when the SNR increases, the MSE consistently decreases. The associated error decreases as well, as the initial input becomes more deterministic. It is apparent that the MSE is not influenced by the type of SOI considered. The fact that there is no significant dependence on the SOI is due to the presence of the noise as we can see from Fig. \ref{fig:mse_radars} (top panel) where SNR$=-20$ dB. Only the BPSK case, down to SNR$=-10$ dB, shows a consistent difference from the other SOIs, in particular when we consider a KLT window of 200 or 400 samples. This means that the BPSK is more sensitive to the choice of the KLT window. In fact, as opposed to the sinwave, the BPSK also contains the message signal and we cannot consider KLT windows longer than the bit-period. 

In general, there is a clear dependence on the choice of the KLT window. The MSE reaches a minimum value at 100 samples: this value is the square root of the length of the input vector. This, in turn, means that, when the noise is the dominant term, the optimal choice is to consider K = W. 
When noise and SOI are comparable in power the dependence on the KLT window is less significant except for the case of the BSPK, as we already discussed. 

\subsection{Detection}
For the detection analysis the MC outcomes in both hypotheses were the decision parameters from equations (\ref{3.3.2},\ref{3.3.3},\ref{3.3.4},\ref{3.3.6}). Because of the low computational burden, the number of trials was $10^4$. Since we are particularly interested in detection at low SNR, for this simulation we considered only the optimal KLT window of 100 samples, as reconstruction MC simulations suggested. In order to compare the four detectors, we evaluated the Area Under Curve (AUC) of the Receiver Operating Characteristic (ROC) \citep{schreier2010statistical} curves for the detectors at various SNR. The ROC curves are generated by calculating the detection probability $P_d$ and the false alarm probability $P_{fa}$ which are obtained by integrating the MC outcome histograms binned with $10^3$ bins.  

\begin{figure}
\begin{multicols}{2}
        \includegraphics[width= \columnwidth]{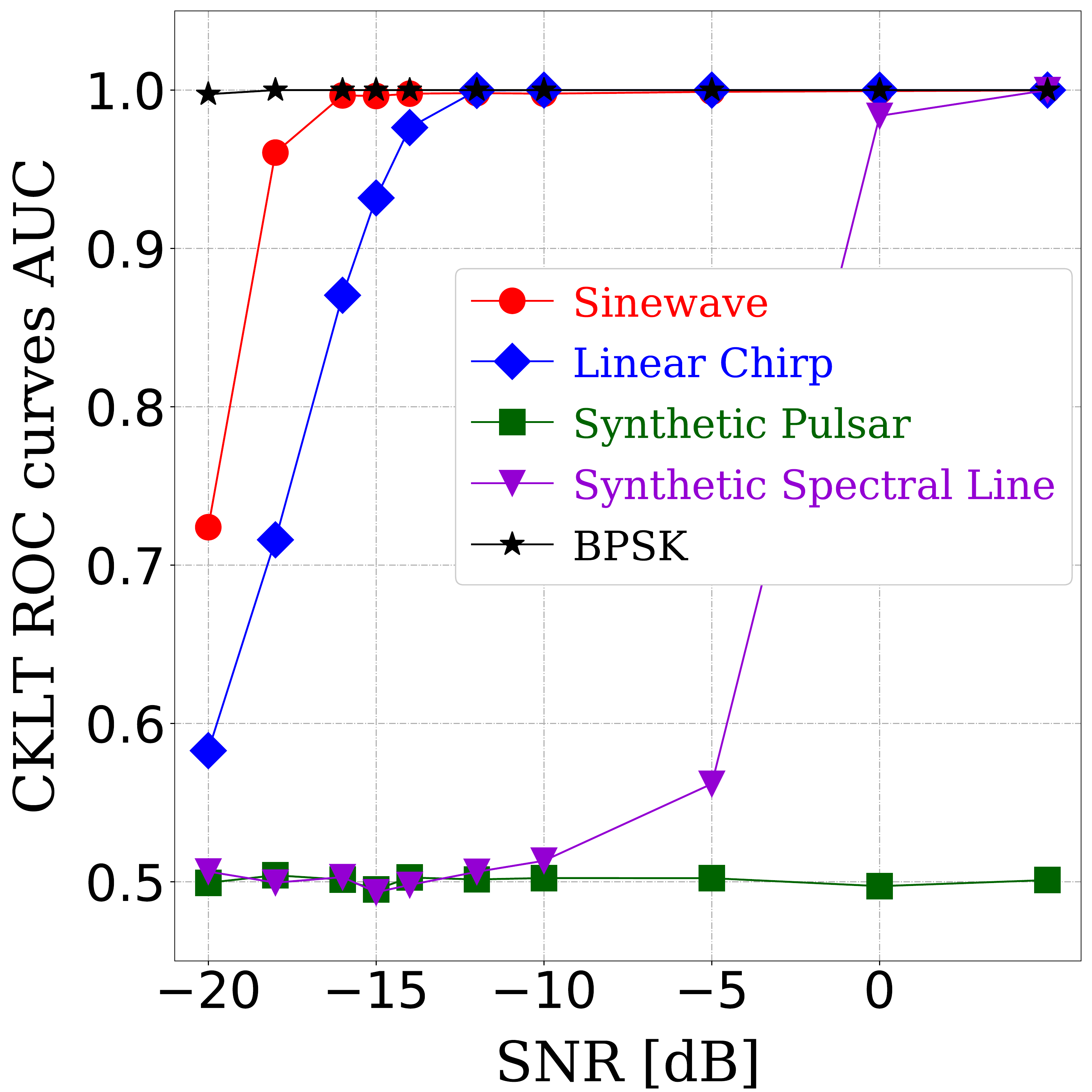}

        \includegraphics[width= \columnwidth]{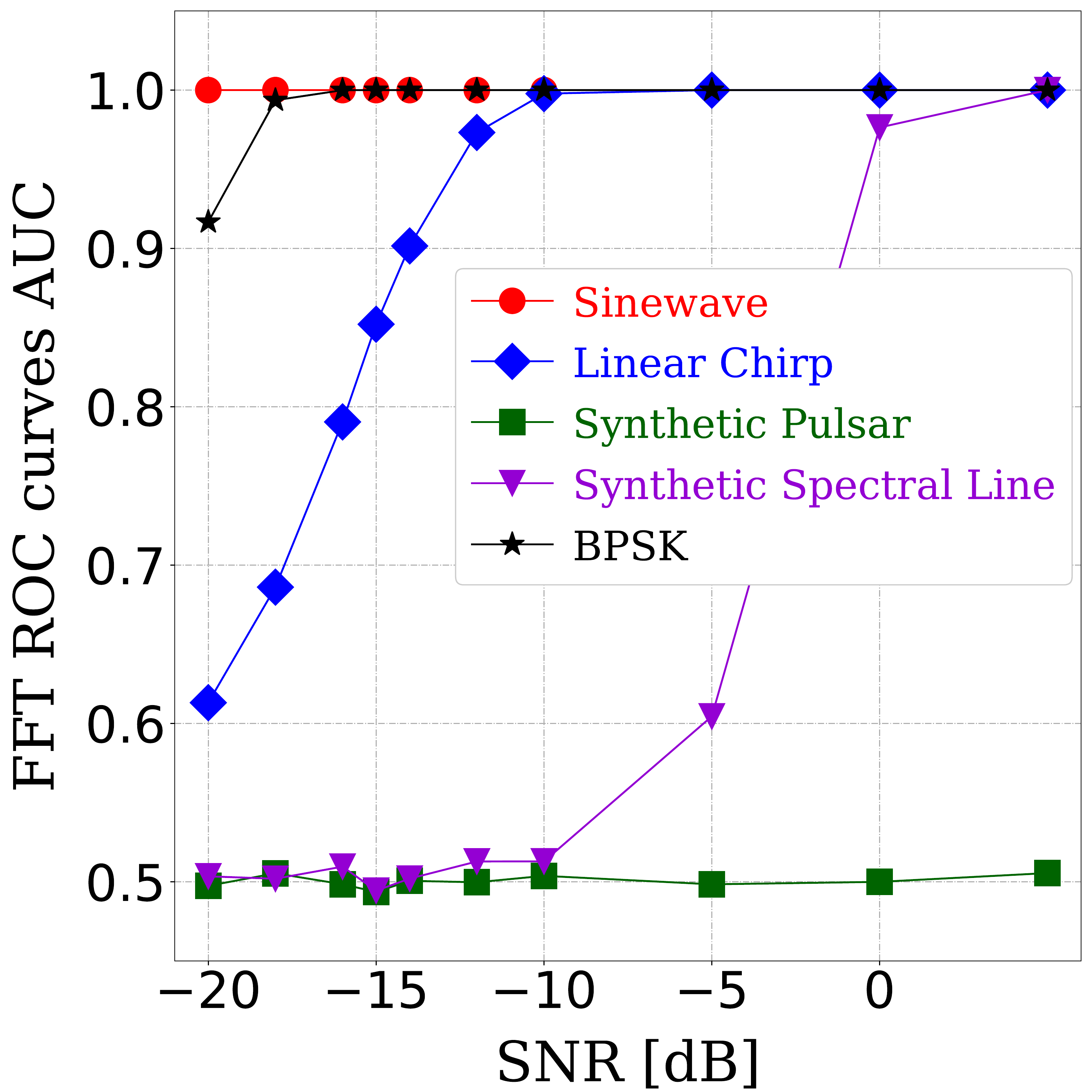}
        
\end{multicols}
\begin{multicols}{2} 
        \includegraphics[width= \columnwidth]{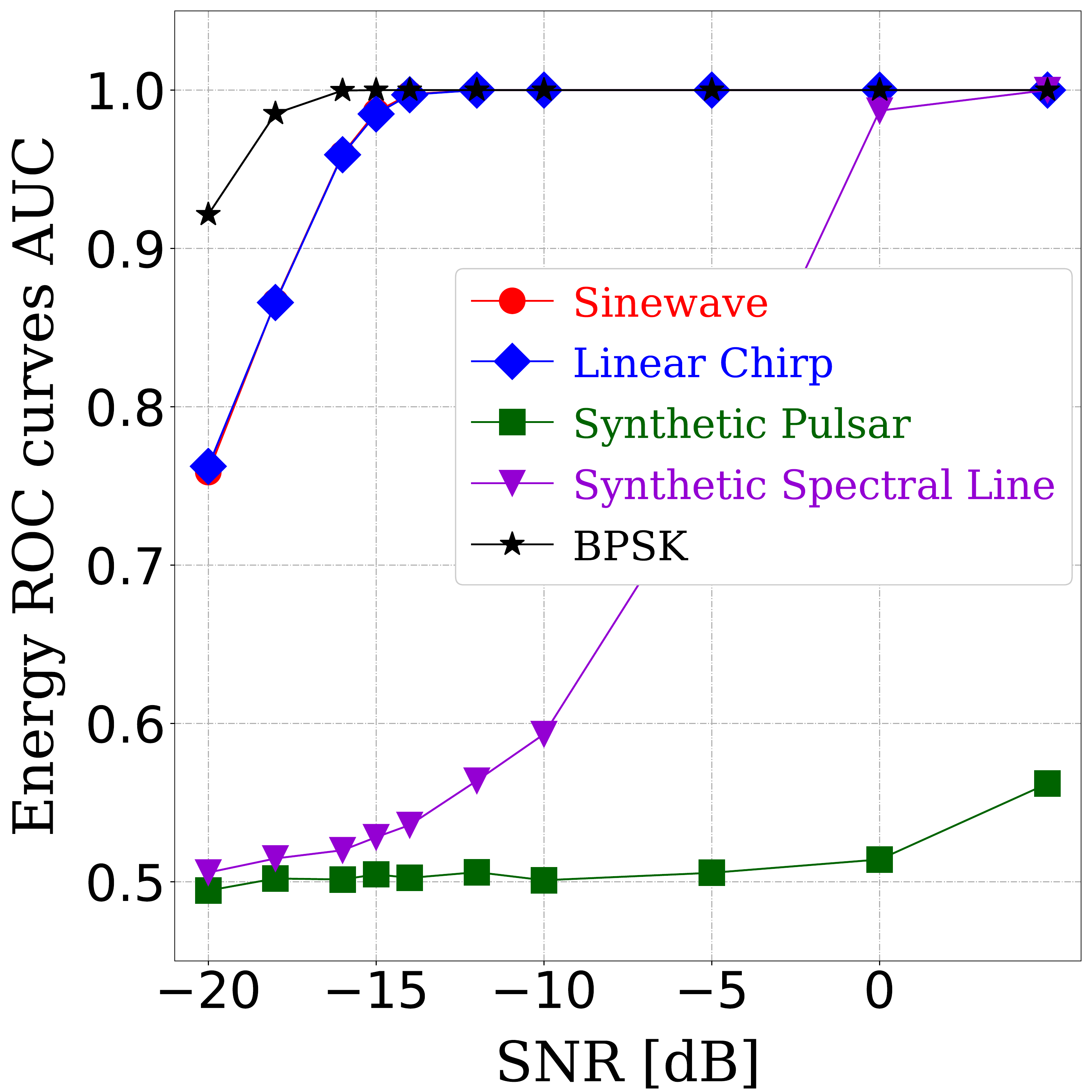}

        \includegraphics[width= \columnwidth]{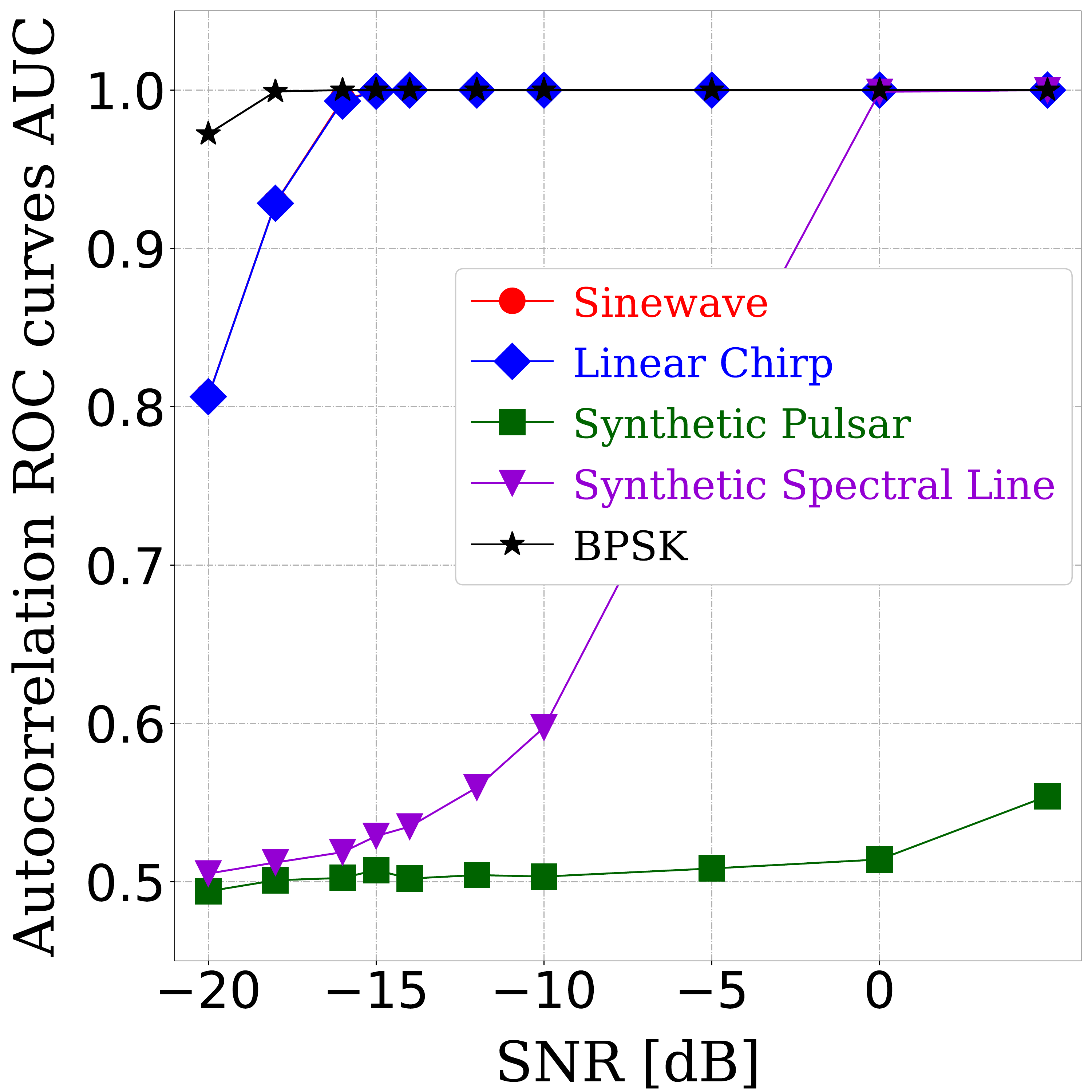}
        
\end{multicols}  
    \caption{AUC as a function of the SNR of the ROC curves for the CKLT detector (top-left panel), the FFT detector (top-right panel), the Energy detector (bottom-left panel) and the Autocorrelation detector (bottom-right panel). }
    \label{fig:auc_detectors}
\end{figure}

Figure \ref{fig:auc_detectors} shows the AUC of the ROC curves as a function of increasing SNR for the four detectors we considered. For the sinewave, all the detectors work well at SNRs whitin -16 dB, while for very low SNRs only the FFT detector does not lose performance power (Fig.\ref{fig:auc_detectors}, top-right panel). This result is expected, since the best eigenbases for a monochromatic signal are sines and cosines. For the linear chirp, the detector based on the autocorrelation (Fig.\ref{fig:auc_detectors}, bottom-right panel) is the most efficient at low SNR, and all the four detectors start to perform equally from SNR$=-5$ dB. The BPSK case shows analogous results to the sinewave except at low SNR where the CKLT performs best (Fig.\ref{fig:auc_detectors}, top-left panel). In the case of the synthetic pulsar, only the energy and autocorrelation detectors can discriminate the SOI at positive SNR, and none of them can when it is negative. Similarly to the pulsar case, also for the spectral line no detector is efficient at low SNR. Also in this case only the energy and autocorrelation detectors begin to detect the SOI starting from SNR$=-5$ dB. Conversely, when the SNR is positive all the detectors have good performance.

\section{Real Data: Voyager 1}

\begin{figure}
\begin{multicols}{3}
         \includegraphics[width= \columnwidth]{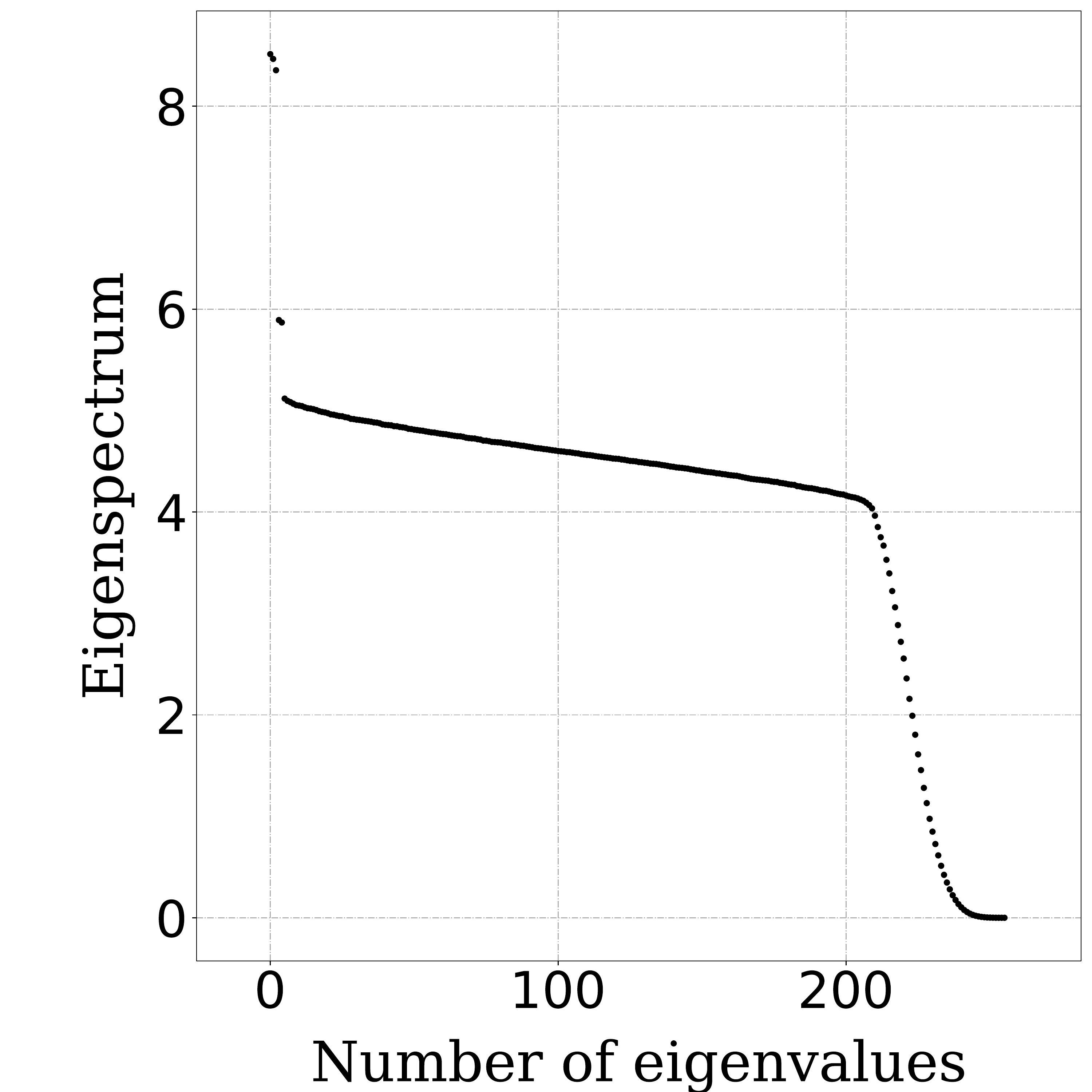}

        \includegraphics[width= \columnwidth]{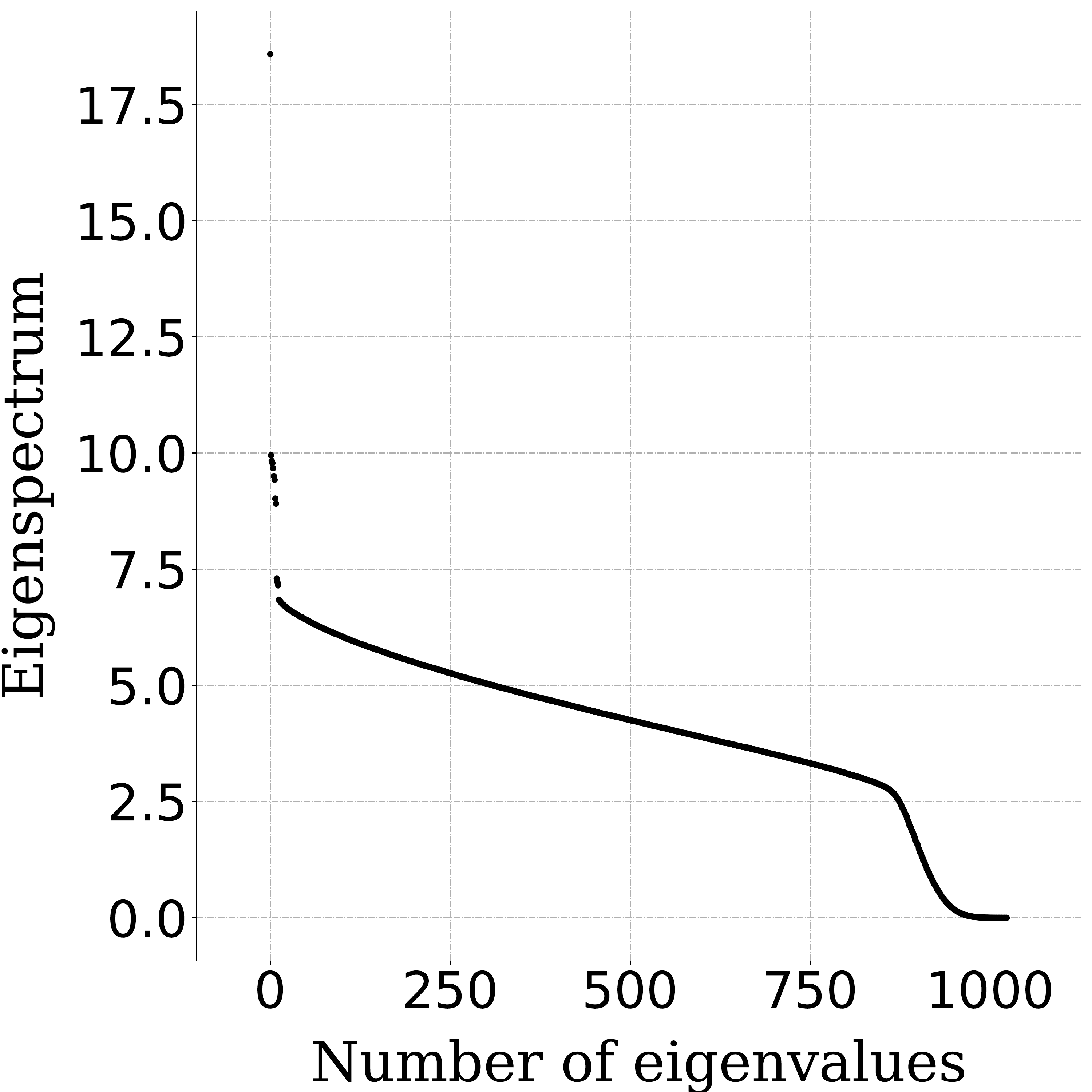}

        \includegraphics[width= \columnwidth]{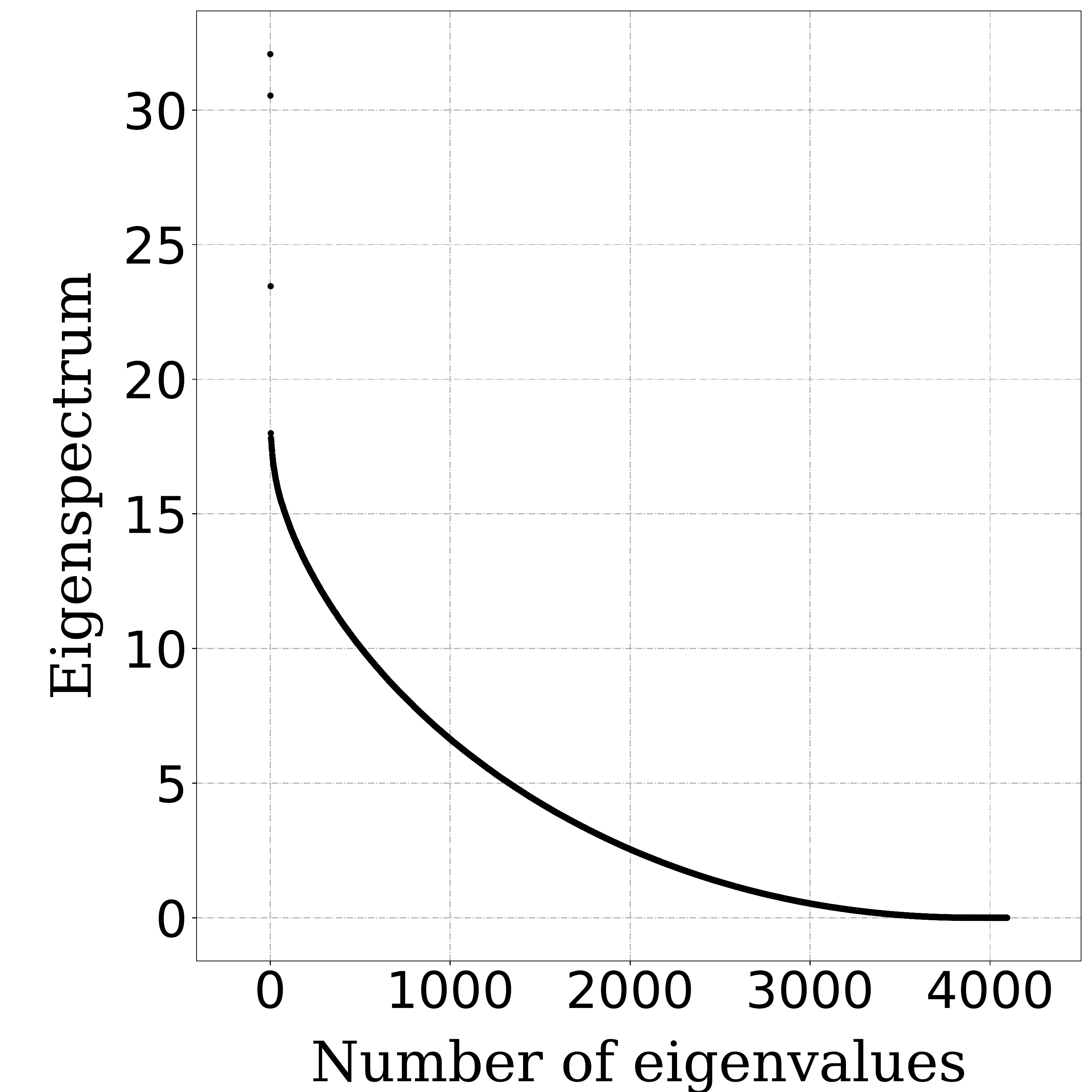}
        
\end{multicols}   
    \caption{CKLT eigenspectra for Voyager 1 data at different KLT Windows: 256 samples (left panel), 1024 samples (central panel), 4096 samples (right panel).  }
    \label{fig:voyager_eigenspectra}
\end{figure}

\begin{figure}
    \centering
        \includegraphics[width= 0.85 \columnwidth]{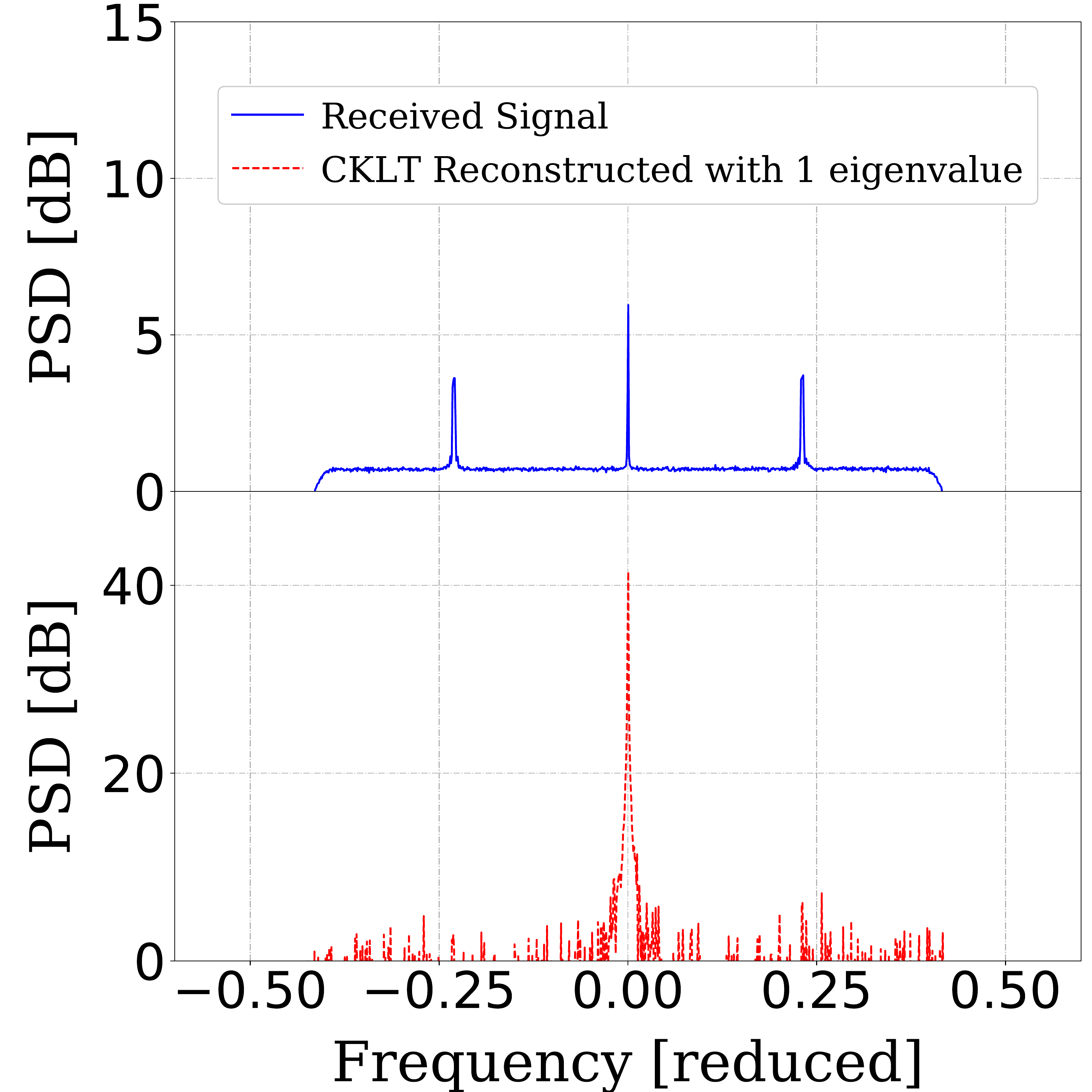}
        
        \includegraphics[width= 0.85 \columnwidth]{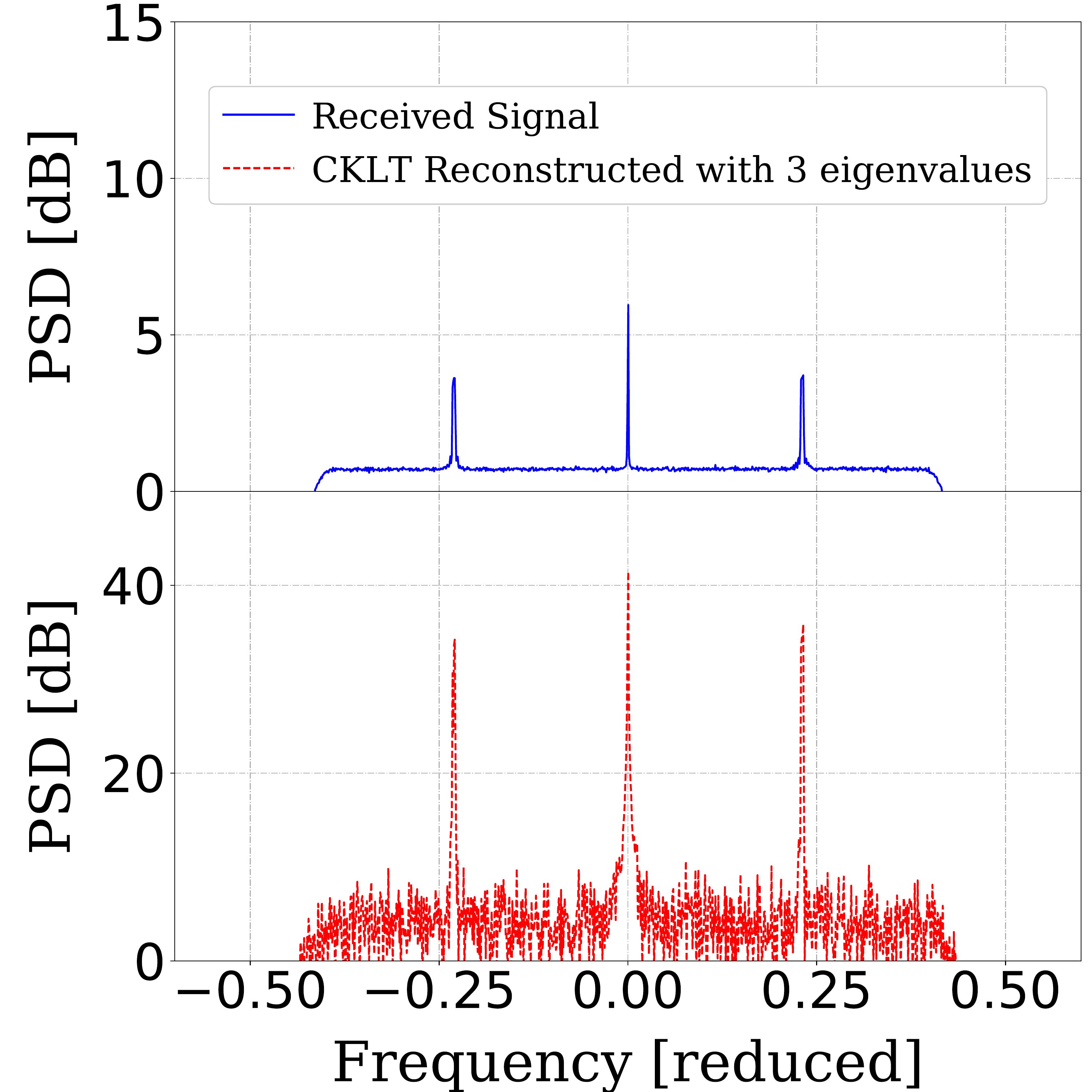}
        
        \includegraphics[width= 0.85 \columnwidth]{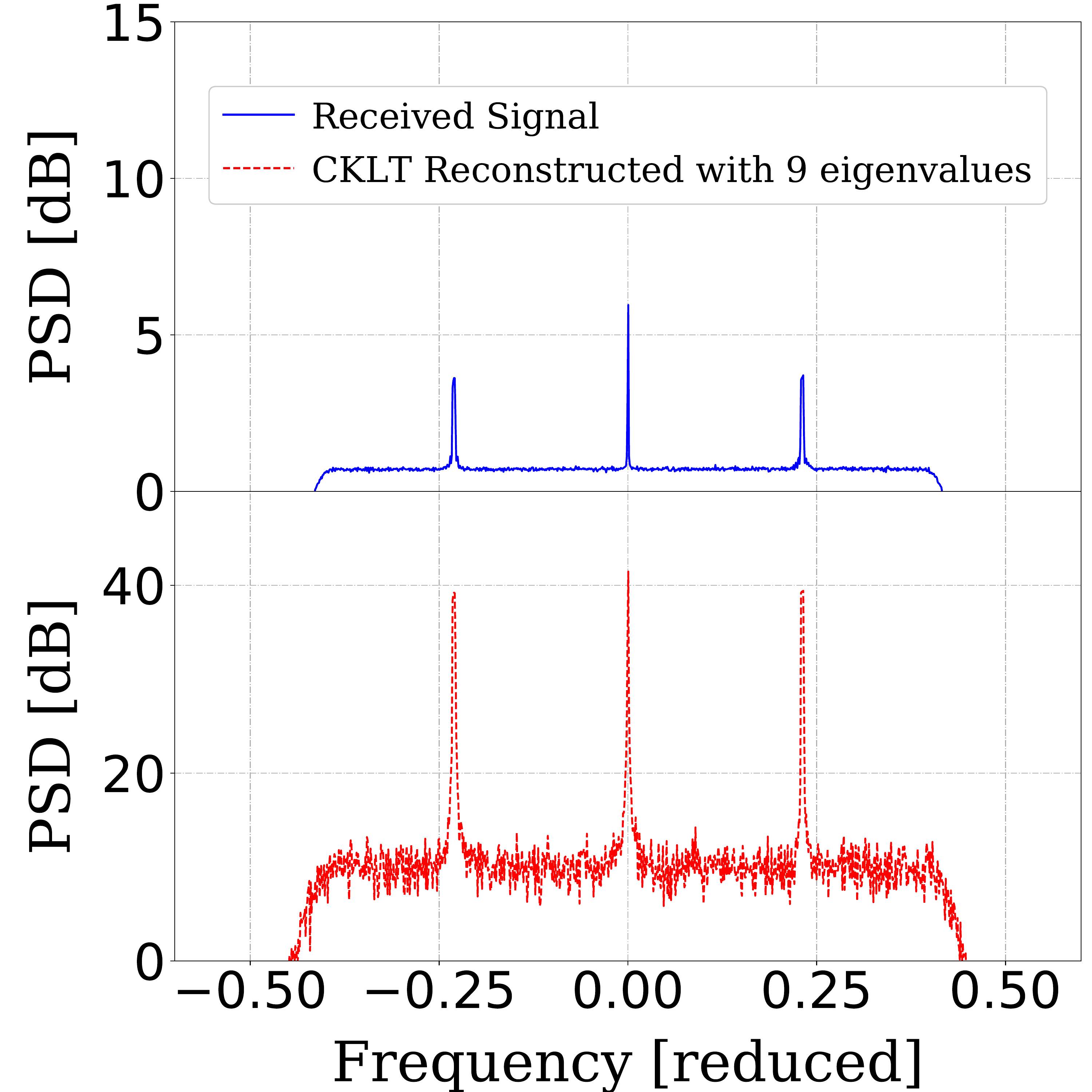}

    \caption{CKLT reconstruction for Voyager 1 data using a KLT window of 1024 samples considering different eigenvalues: 1 eigenvalue (top panel), 3 eigenvalues (middle panel), eigenvalues (bottom panel).}
    \label{fig:voyager_reconstruction}
\end{figure}

\label{sec:voyager}
We performed CKLT analysis of the Voyager 1 \citep{doi:10.1002/9781119169079.ch3} telemetry signal collected by the Breakthrough Listen group (UC Berkeley) with the Green Bank Telescope. The Voyager 1 telemetry signal consists in general into a bi-phase modulation BPSK where the central carrier is emitted by the spacecraft a 8415 MHz. This value does not consider the Doppler shift due to the relative motion between the telescope and Voyager 1. The sub-carrier is, in turn, modulated to carry individual phase shifts that are designed to represent groups of binary 1s and 0s. The received signal is a complex vector of $2^{24}$ samples.

Figure \ref{fig:voyager_eigenspectra} shows three eigenspectra computed using different KLT windows.  Fig.\ref{fig:voyager_eigenspectra} (right panel) shows that the noise component starts to dominate for larger samples (4096 samples is exactly the square root of the length of the signal). Fig.\ref{fig:voyager_eigenspectra} (left panel) shows that the sub-dominant components are very close in value to the dominant ones for lower samples. Only the eigenspectrum in Fig.\ref{fig:voyager_eigenspectra} (central panel) shows a clear division between the SOI and noise sub-spaces. In this case there is a single dominant eigenvalue and a cluster of 8 sub-dominant eigenvalues. After this cluster, the eigenspectrum decreases constantly and it rapidly converges to zero at approximately 80 \% of the KLT window. This behaviour is similar to the case shown in Fig.\ref{fig:voyager_eigenspectra} (left panel). It is not clear why this happens, and it will be the subject of further studies. The Voyager 1 study-case shows that it is very hard to interpret the eigenspectra and have a complete understanding of all the meaningful eigenvalues. Furthermore, at odds with the MC reconstruction results, the Voyager 1 signal shows that the optimal KLT Window is not necessarily the square root of the total length of the signal. We point out that, because of the high computation burden of our MC simulations, it was not possible to explore the high number of samples needed to investigate long signals as this. 

Figure \ref{fig:voyager_reconstruction} shows the comparison between the received and CKLT reconstructed average PSDs computed with a resolution of 1024 samples. When we consider only 1 eigenvalue (Fig.\ref{fig:voyager_reconstruction}, top panel), the CKLT PSD shows only the main carrier, though with a considerable power enhancement relative to the PSD of the received data.  
In order to see the sub-carriers we need at least three eigenvalues (Fig.\ref{fig:voyager_reconstruction}, middle panel). Finally, when we select the dominant eigenvalue plus the cluster of 8 we reconstruct the PSD with the minimum amount of noise (Fig.\ref{fig:voyager_reconstruction}, bottom panel). In this case the gain in power is approximately 25 dB for the main carrier and the sub-carriers.

\section{Conclusions}
\label{sec:conclusion}
In this work, we have presented the use of the KLT as a noise filter for signal processing in astronomy. 
As a first approach we developed a KLT based on multiple realisations of the same signal (MRKLT), which was extremely successful in reconstructing all the examined SOIs down to low SNR (-20 dB). While this method shows promising results, it is not suitable for single-receiver radio-telescopes providing a unique signal realisation at a time. Phased array radio telescopes, however, do provide multiple realisations of the same signal and could, as such, take advantage of the MRKLT for signal de-noising and recovery. This application falls outside the scope of the present paper and has not been addressed.

We then compared standard KLT techniques based on the Toeplitz matrix (TKLT) \citep{1993ASPC...47..129D,2010AcAau..67.1427M} for the KLT kernel with a new method (CKLT), which provides a significant improvement in computation time. Both techniques show good performance for narrow-band signals, while they show limitations for wide-band signals, as highlighted by the case of the linear chirp. For SOIs of this kind, further studies are needed in order to identify a closed-form expression for the choice of a meaningful number of eigenvalues. 
We considered several models for typical astrophysical and interstellar-telecommunication SOIs, and performed a Monte Carlo analysis for the CKLT in order to study its reconstruction and detection performance. SOI reconstruction simulations show good results starting from as low as SNR$=-10$ dB. SOI detection simulations, on the other hand, show comparable results with standard detection techniques. Finally, we provided a real data application by reconstructing the Voyager 1 telemetry signal. The signal displays a significant gain in power after the CKLT application on the collected data. These first promising results obtained with Voyager 1 suggest that the KLT might be an extremely powerful instrument for interstellar-telecommunication. For astrophysical signals such as spectral lines, or transients (like pulsars or FRBs) the KLT applied to single complex voltage data does not appear as a viable substitute for most commonly used detectors, since priors regarding the SOI are rarely available.

\section*{Acknowledgements}

MT, MP and AT acknowledge support from the Regione Autonoma della Sardegna through project funding "Development and implementation of innovative mathematical algorithms for the study of FRBs", C.R.P. 127, Ob. Fu. 1.05.01.18.31. 

The authors thank the Berkeley SETI Research Center and the Breakthrough Listen group at UC Berkeley for providing the Voyager I data.

The authors appreciate the unknown referee’s comments which significantly contributed to improving the quality of the publication.

\bibliographystyle{mnras}
\bibliography{Bibliography} % if your bibtex file is called example.bib

\appendix

\section{Matrix Diagonalisation}

\label{app:matrix}

In the most general case, the KLT has a major drawback: it demands for high computational power. This is mainly due to the  operation of diagonalising the autocovariance matrix.
We will here attempt to assess the number of operations required to diagonalise a matrix.
We will limit our analysis to multiplications, exponents, divisions and square roots operations, and will not consider the additions as they are negligible in terms of computational requirements. Mathematically, given the computational complexity of a multiplication $M(n)$, and the computational complexity of an addition $D(n)$, $n$ being an arbitrary platform bit-depth, then $D(n) = o (M(n))$.

The other operations (exponent, division and square root) have computational complexity comparable to the multiplication; if we, for instance, use Newton's algorithm \citep{NewRapSqRoot}, their complexity $C(n)=O(M(n))$, that is a positive real number $k$\ and a real value $x_0$ exist such that $|C(n)| \leq kM(n)$\ for all $x_0 \geq n$.

We start from the general case of an autocovariance matrix given by (\ref{2.8}). This matrix is
Hermitian.
According to the Jacobi algorithm \citep{GolubMatComp} (see pseudocode in algorithm \ref{JacAlg}), given the Hermitian $N$-square matrix $A$, all the
off-diagonal elements must be reduced to zero by means of appropriate matrix rotations. 
\begin{algorithm}
\caption{Jacobi Algorithm}\label{JacAlg}
\begin{algorithmic}
\Procedure{}{Jacobi Algorithm}
\State $\textbf{for }  p = 1:n-1$
\State \quad $\textbf{for }  q = p+1:n-1$
\State \qquad $(c,s) = \textbf{sym.schur2}(A,p,q)$
\State \qquad $A = J(p,q,\theta)^T A J(p,q,\theta)$
\State \qquad $V = V J(p,q,\theta)$
\State \quad $\textbf{end}$
\State $\textbf{end}$
\EndProcedure
\end{algorithmic}
\end{algorithm}
In the algorithm \ref{JacAlg}, a loop is iterated on all the $(N^2-N)/2$\ upper off-diagonal elements (the lower ones are obtained using symmetry properties). Another algorithm is nested and is the Schur algorithm \ref{SchAlg}.
In the Schur algorithm (called within the Jacobi algorithm), a division is carried on (to calculate $\tau$) plus (in the \textbf{if}\ condition) a division, a square root and an exponent, for a total of 4 operations.
\begin{algorithm}
\caption{Schur Algorithm}\label{SchAlg}
\begin{algorithmic}
\Procedure{$[c,s]=\textbf{sym.schur2}(A,p,q)$}{Schur Algorithm}
\State $\textbf{if }  A(p,q) \neq 0$
\State \quad $\tau = (A(q,q)-A(p,p))/(2A(p,q))$ 
\State \quad $\textbf{if } \tau \geq 0$
\State \qquad $t = 1/(\tau \sqrt{1+\tau^2})$
\State \quad $\textbf{else }$
\State \qquad $t = -1/(-\tau \sqrt{1+\tau^2})$
\State \quad $\textbf{end }$
\State \quad $c = 1/ \sqrt{1+t^2})$
\State \quad $s=tc$
\State $\textbf{else }$
\State \quad $c=1$
\State \quad $s=0$
\State $\textbf{end}$
\EndProcedure
\end{algorithmic}
\end{algorithm}
The Schur algorithm returns a $N \times N$ rotation matrix that is 0 everywhere but the diagonal elements and, in particular, the following elements:
\begin{enumerate}
    \item $J(p,p, \theta) = J(q,q, \theta) = c$
    \item $J(p,q, \theta) = s$
    \item $J(p,q, \theta) = s^*$
\end{enumerate} Once the $J$\ matrix is returned, the autocovariance matrix $A$ and the eigenvector matrix $V$\ can be updated by means of $6N$\ and $3N$\ operations respectively, if the symmetry is exploited.
In the end, $(4+6N+3N)$\ operations are repeated $(N^2-N)$\ times. Therefore the computational complexity of the diagonalisation is $O(N^3)$. 
The Jacobi algorithm holds true for Hermitian matrices. Whenever the process described is simpler, further properties can be applied to the autocovariance matrix. This is the case, for instance, of a stationary process, where the autocovariance matrix is a Toeplitz matrix; the Arnoldi algorithm \citep{ArnoldiAlgorithm} can then be used.

The Arnoldi algorithm produces a set of orthogonal vectors, obtained by an iterative projection on a given matrix A. The subspace generated from this projection is called Krilov subspace \citep{GolubMatComp} and the iterations for this method are given by the following equation:
\begin{equation*}
AV_{k}=V_{k}H_{k}+f_{k}e_{k}^{T},\;\;\;V_{k}^{H}V_{k}=I_{k},\;\;\;V_{k}^{T}f_{k}=0.
\end{equation*}
where $A$\ is the autocovariance or covariance matrix obtained from the signal
samples, $V_{k}$ are the projections and $H_{k}$\ is the representation of the projection of $A$ on $K\left(A,v_{0};k\right)$.

The advantage of this algorithm for our purposes is twofold: first it does not change the properties of the matrix $A$, so that the Hermitian structure can be exploited and is stored efficiently; second, it returns a $k$-rank matrix $H$, with $k$\ the amount of eigenvalues of interest relating the largest eigenvalues of $A$. 

The routine consists of the following steps:
\begin{enumerate}
    \item build the autocovariance matrix
    \item run the Arnoldi algorithm iterations
    \item compute Givens rotations \citep{GolubMatComp}
\end{enumerate}

\begin{figure}

    \centering

        \includegraphics[width=0.85\columnwidth]{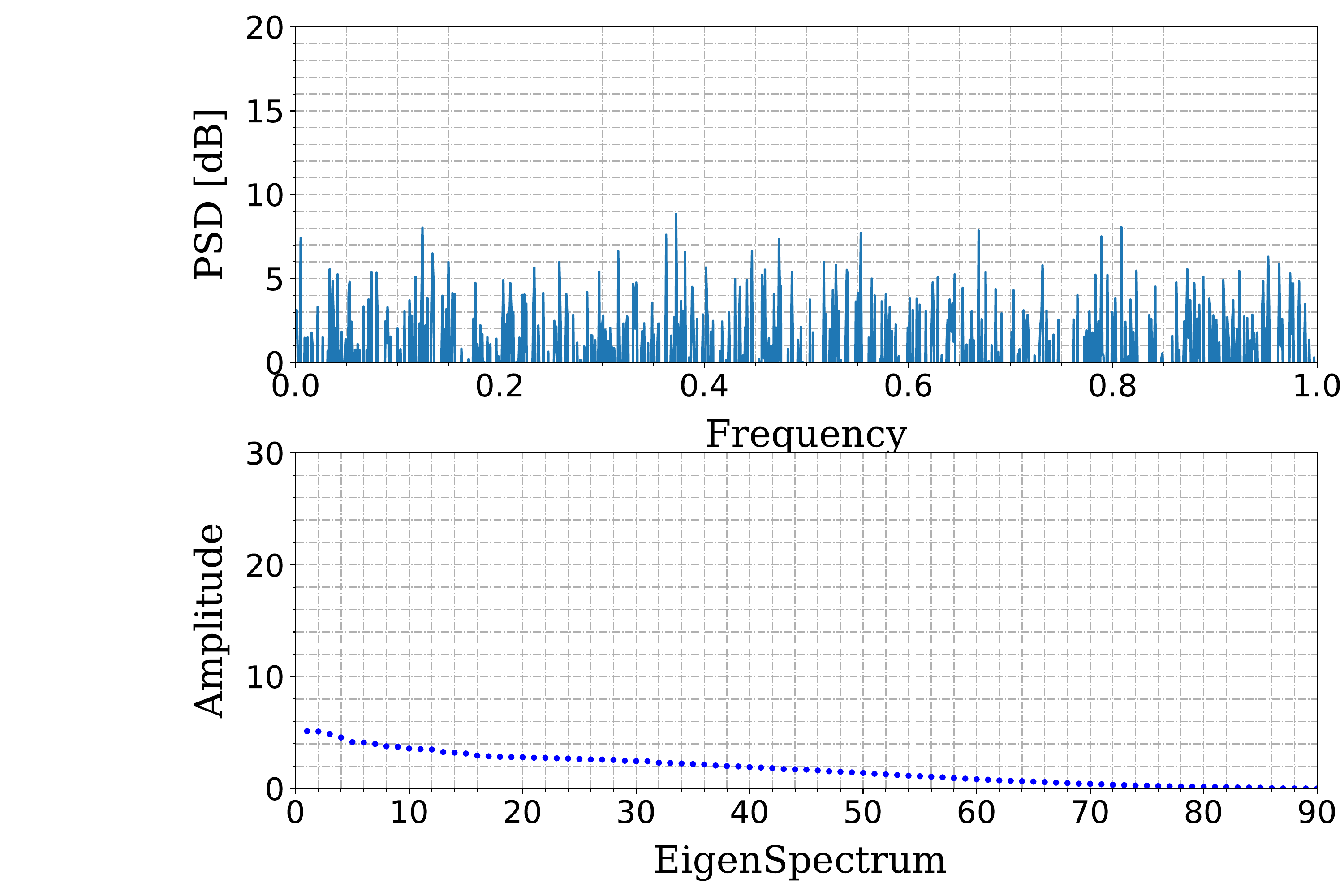}
     
        \includegraphics[width=0.85\columnwidth]{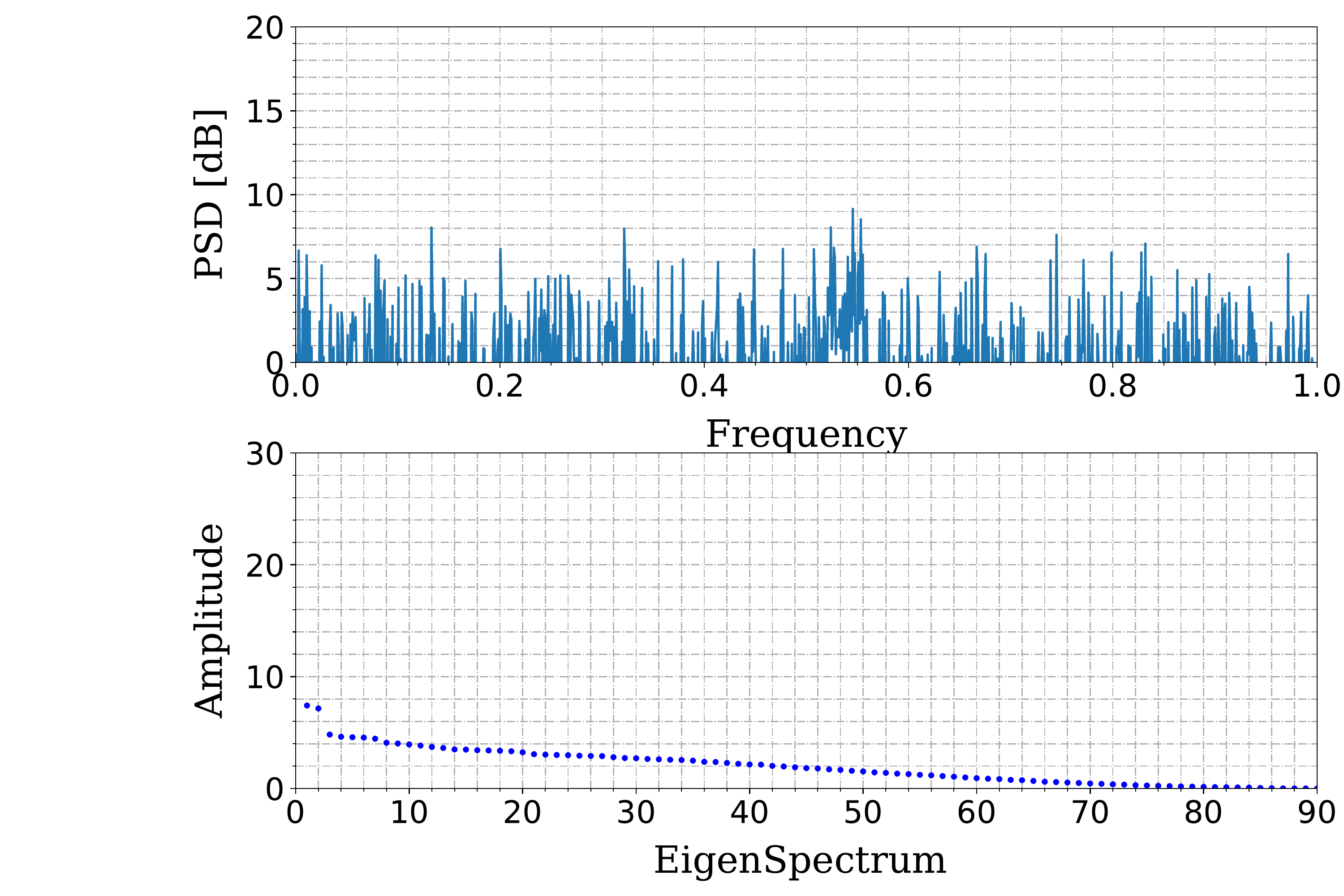}

        \includegraphics[width=0.85\columnwidth]{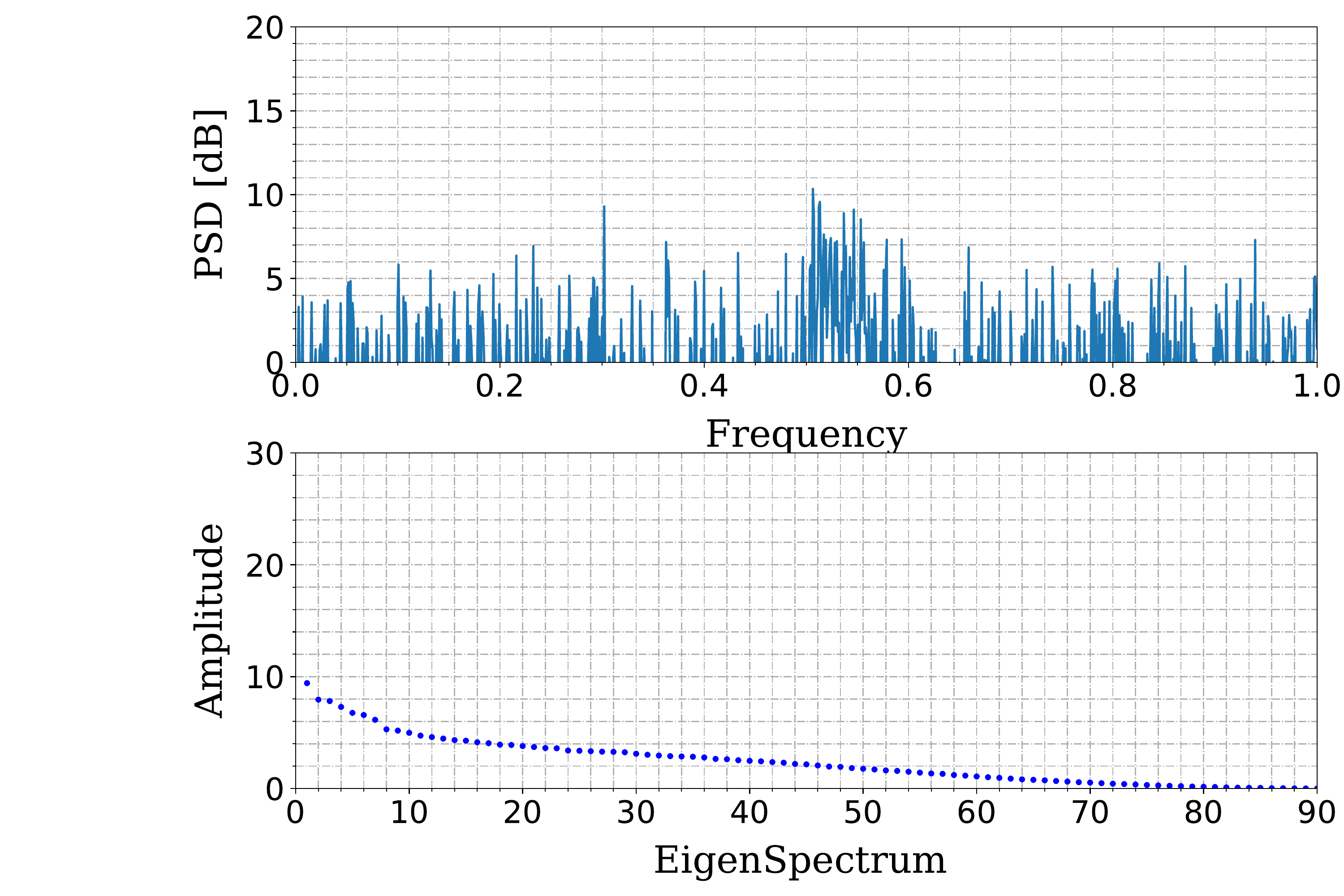}

        \includegraphics[width=0.85\columnwidth]{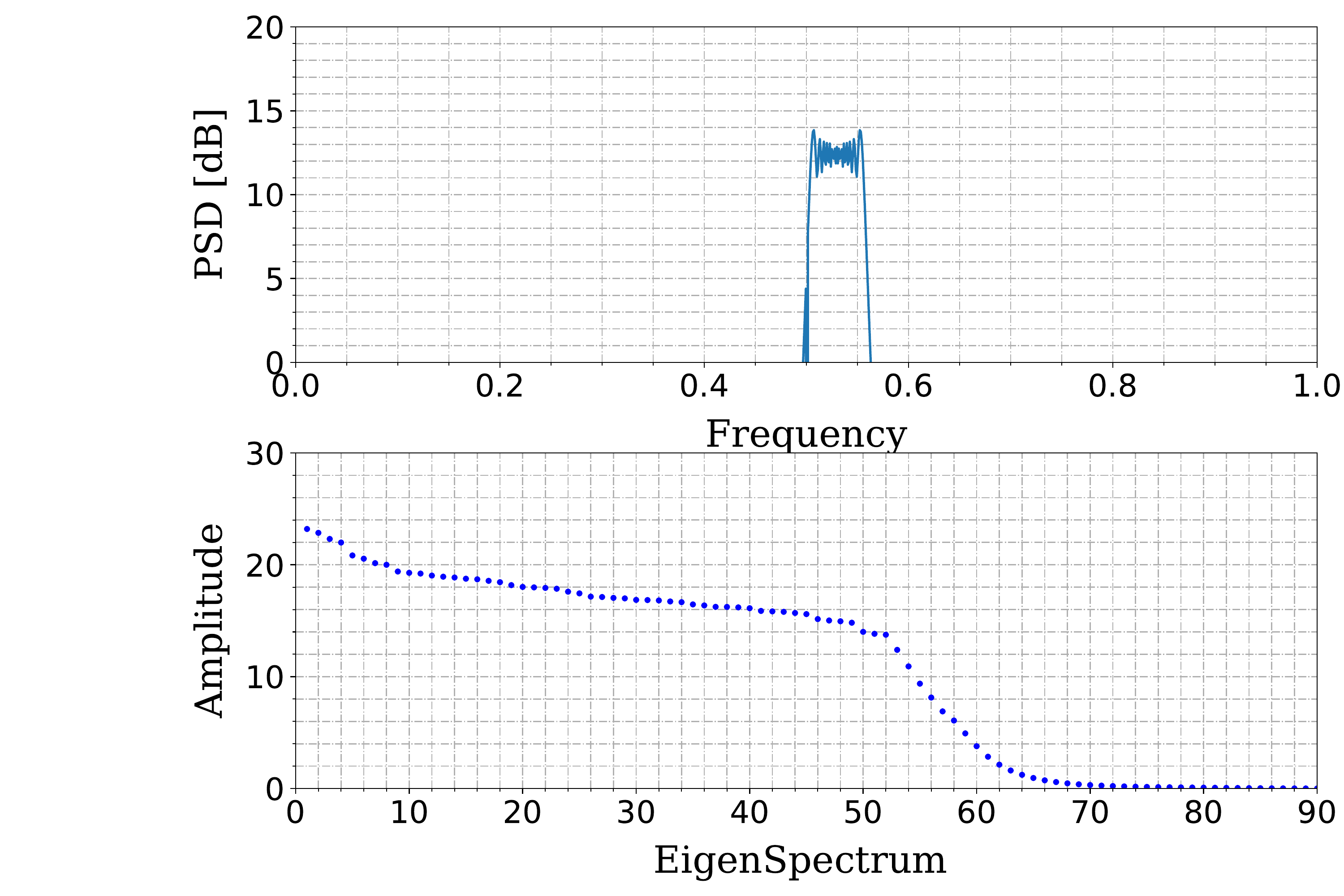}

    \caption{ Test simulation using a linear chirp with 1024 samples and 90 eigenvalues. For each panel the upper plot represents the PSD of the reconstructed signal, the lower plot shows the corresponding eigenspectrum. The top panel shows the PSD and eigenspectrum for only white noise. The upper-middle and the lower-middle panels show the same as the top panel but for a chirp buried in white noise with a SNR of -18 dB and -13 dB respectively. The bottom panel shows the same results for a chirp without noise.}
    \label{fig:tklt_detection}
\end{figure}

\begin{algorithm}
\caption{Arnoldi Truncated Algorithm}\label{ArnAlg}
Input: (A,$v_{0}$) 

Output:$(V_{k},H_{k},f_{k})$ such that $AV_{k}=V_{k}H_{k}+f_{k}e_{k}^{T}$,
$V_{k}^{T}V_{k}=I_{k}$and $V_{k}^{T}f_{k}=0$

1. $v\leftarrow v_{1}/\left\Vert v_{1}\right\Vert $

2. $w\leftarrow Av;
$

3. $H_{1}=(\alpha_{1});V_{1}=(v_{1});f_{1}=w-v_{1}\alpha_{1};$

4. for j=1,2,3,...,k-1

4.1 \quad $\beta_{j}=\left\Vert f\right\Vert ;v_{j+1}=f/\beta_{j};$

4.2 \quad $V_{j+1}=(V_{j},v_{j+1});H_{j}=\dbinom{H}{\beta_{j}e_{j}^{T}}$

4.3 \quad $z=Av_{j+1};$

4.4 \quad $h=V_{j+1}^{H}z;H_{j+1}=(H_{j},h);$

4.5 \quad $f_{j}=z-V_{j+1}h;$

5. end
\end{algorithm}
 The computational cost of Arnoldi implementation is $O(kNlogN)$, where $k$ are the eigenvalues of interest. The procedure returns a set of eigenpairs of the matrix $H$ that are an approximation of those of matrix $A$.
In order to show the responsiveness of the eigenvalue spectrum, a linear chirp at
different SNR ratios was analyzed.  
Figure \ref{fig:tklt_detection} (top panel) shows the pure noise case. The upper plot of the panel is its PSD, while the plot on the bottom  is the eigenspectrum. The following figures \ref{fig:tklt_detection} (upper-middle panel), \ref{fig:tklt_detection} (lower-middle panel) and \ref{fig:tklt_detection} (bottom panel) show the chirp with a SNR equal to, respectively, -18 dB, -13 dB and noiseless. Note the behaviour of the largest eigenvalue in each plot: as the SNR increases, the largest eigenvalue increases in turn. 

Also, in the particular case of the Arnoldi algorithm, the most time-consuming computation is at step 4.3, where a matrix-vector multiplication is run, thus with a complexity $O(N^2)$.
By exploiting the particular Hermitian structure of $A$, it is possible to optimise this step: by virtue of the FFT, the matrix multiplication is a convolution in the time domain, equal to a multiplication in the frequency domain, at a cost of $N\log(N)$.
The table \ref{tab:timingtable} below shows results obtained from the same algorithm implemented both in Matlab and C-CUDA code. Each row shows time consumption and the first eigenvalue given by Matlab and C-CUDA.
If we compare the last two columns of table \ref{tab:timingtable}, we notice that the two methods differ by an absolute error less than $4\cdot10^{-3}$. The first 45 eigenvalues obtained with the two different methods give a root mean square error of 0.3. Absence of convergence can be experienced, that translates into as a loss of orthogonality between the extracted eigenvectors. A restart of the algorithm is then required. It turns out that, for the few eigenvalues we used, that was not necessary.
The results show also that for $N$ less than 4096 samples the CPU is faster than the GPU due to low speed transfer between host and GPU-device.
The tests presented here were obtained using a linux platform Ubuntu 18.04 with CUDA 10.0 running on a  Intel\textsuperscript{\textregistered} Core\textsuperscript{TM}  i7 CPU @ 2.50GHz.The GPU device is GeForce GTX 850M @ 902Mhz, onboard memory 2GiB DDR3, 640 CUDA cores.

\begin{table*}
	\centering
	\caption{Computation time (in seconds) and value of the first eigenvalue computed using the Arnoldi algorithm, implemented both with Matlab and C-CUDA. N refers to the dimension of the input matrix. }
	\label{tab:timingtable}
	\begin{tabular}{lcccr} 
		\hline
		N (samples) & Matlab Computation Time (s)  & C-CUDA Computation Time (s) & Matlab First Eigenvalue & C-CUDA First Eigenvalue\\
		\hline
		1024   &    0.12204   &   0.46998  &     23.20282    &       23.20281   \\
        2048   &    0.14946   &   0.49409  &     46.41253    &       46.41252   \\
        4096   &    0.33667   &   0.54417  &     92.83205    &       92.83202   \\
        10240  &    0.82349   &   0.66072  &    232.09068    &      232.09063   \\
        20480  &    2.07194   &   0.80144  &    464.18832    &      464.18832   \\
        40960  &    3.17558   &   1.12395  &    928.38393    &      928.38372   \\
        51200  &    3.35672   &   1.18899  &   1160.48160    &     1160.48130   \\
        102400 &    8.98926   &   2.79724  &   2320.97040    &     2320.96980   \\
        512000 &   71.05971   &  13.78155  &  11604.88000    &    11604.87800   \\
        819200 &   78.51755   &  22.65362  &  18567.81300    &    18567.80800   \\
		\hline
	\end{tabular}
\end{table*}

\bsp	
\label{lastpage}
\end{document}